\documentclass[12pt,letterpaper]{article}
\pdfoutput=1
\usepackage{graphicx,array}
\usepackage{color}
\usepackage{latexsym}
\usepackage{amsthm}
\usepackage{amsmath}
\usepackage{amssymb}
\usepackage{hyperref}
\usepackage{bbold}
\usepackage{subfig}
\usepackage{cite}
\usepackage{cancel}
\usepackage{enumitem}

\def\simgt{\mathrel{\lower2.5pt\vbox{\lineskip=0pt\baselineskip=0pt
           \hbox{$>$}\hbox{$\sim$}}}}
\def\simlt{\mathrel{\lower2.5pt\vbox{\lineskip=0pt\baselineskip=0pt
           \hbox{$<$}\hbox{$\sim$}}}}

\newcommand{\be}{\begin{equation}}
\newcommand{\ee}{\end{equation}}
\newcommand{\bea}{\begin{eqnarray}}
\newcommand{\eea}{\end{eqnarray}}
\newcommand{\Eq}[1]{Eq.~(\ref{#1})}
\newcommand{\Eqs}[2]{Eqs.~(\ref{#1}) and (\ref{#2})}
\newcommand{\Sec}[1]{Sec.~\ref{#1}}
\newcommand{\Secs}[2]{Secs.~\ref{#1} and \ref{#2}}
\newcommand{\Fig}[1]{Fig.~\ref{#1}}
\newcommand{\Figs}[2]{Figs.~\ref{#1} and \ref{#2}}
\newcommand{\App}[1]{App.~\ref{#1}}

\newcommand{\GeV}{\textrm{ GeV}}
\newcommand{\TeV}{\textrm{ TeV}}
\newcommand{\gsim}{\lower.7ex\hbox{$\;\stackrel{\textstyle>}{\sim}\;$}}
\newcommand{\lsim}{\lower.7ex\hbox{$\;\stackrel{\textstyle<}{\sim}\;$}}

\newcommand{\squared}[1]{\left| #1 \right|^2}

\def\sigmabar{\overline\sigma}

\hypersetup{colorlinks,citecolor= blue,linkcolor= black}

\numberwithin{equation}{section}

\begin{document}

SCIPP 16/13 \\


\hfill


\begin{center}
{\LARGE\bf
 Gravitino or Axino Dark Matter \\  \vspace{0.1in}
 with \\   \vspace{0.2in}
 Reheat Temperature as high as $10^{16}$ GeV
}
\\ \vspace*{0.5cm}

\bigskip\vspace{1cm}{
{\large \mbox{Raymond T. Co$^{1,2}$, Francesco D'Eramo$^{3,4}$ and Lawrence J. Hall$^{1,2}$} }
} \\[7mm]
{\it 
$^1$Berkeley Center for Theoretical Physics, Department of Physics, \\ University of California, Berkeley, CA 94720, USA \\
$^2$Theoretical Physics Group, Lawrence Berkeley National Laboratory, Berkeley, CA 94720, USA \\
$^3$Department of Physics, University of California Santa Cruz, Santa Cruz, CA 95064, USA \\
$^4$Santa Cruz Institute for Particle Physics, Santa Cruz, CA 95064, USA} 
\end{center}

\bigskip
\centerline{\large\bf Abstract}

\vspace{0.3cm}

\begin{quote} \small


A new scheme for lightest supersymmetric particle (LSP) dark matter is introduced and studied in theories of TeV supersymmetry with a QCD axion, $a$, and a high reheat temperature after inflation, $T_R$.  A large overproduction of axinos ($\tilde{a}$) and gravitinos ($\tilde{G}$) from scattering at $T_R$, and from freeze-in at the TeV scale, is diluted by the late decay of a saxion condensate that arises from inflation.   The two lightest superpartners are $\tilde{a}$, with mass of order the TeV scale, and $\tilde{G}$ with mass $m_{3/2}$ anywhere between the keV and TeV scales, depending on the mediation scale of supersymmetry breaking.  Dark matter contains both warm and cold components: for $\tilde{G}$ LSP the warm component arises from $\tilde{a} \rightarrow \tilde{G}a$, while for $\tilde{a}$ LSP the warm component arises from $\tilde{G} \rightarrow \tilde{a}a$.   The free-streaming scale for the warm component is predicted to be of order 1 Mpc (and independent of $m_{3/2}$ in the case of $\tilde{G}$ LSP).  $T_R$ can be as high as $10^{16}$ GeV, for any value of $m_{3/2}$, solving the gravitino problem.  The PQ symmetry breaking scale $V_{PQ}$ depends on $T_R$ and $m_{3/2}$ and can be anywhere in the range $(10^{10} - 10^{16})$ GeV.  Detailed predictions are made for the lifetime of the neutralino LOSP decaying to $\tilde{a}+ h/Z$ and $\tilde{G}+h/Z/\gamma$, which is in the range of $(10^{-1}-10^6)$m over much of parameter space.   For an axion misalignment angle of order unity, the axion contribution to dark matter is sub-dominant, except when $V_{PQ}$ approaches $10^{16}$ GeV.   
\end{quote}

\setcounter{tocdepth}{1}

\newpage
\tableofcontents

\section{Introduction}
\label{sec:intro}

Perturbative theories with supersymmetry broken at the TeV scale are well-motivated by the hierarchy problem, even if they do not completely solve it, and lead to Higgs boson masses in the region discovered at the LHC.    Dark matter could be the lightest superpartner (LSP), cosmologically produced by the freeze-out mechanism.  On the other hand, the strong CP problem is elegantly solved by introducing a Peccei-Quinn (PQ) symmetry \cite{Peccei:1977hh,Peccei:1977ur} broken at scale $V_{PQ}$,\footnote{In this work we use the PQ breaking scale $V_{\rm PQ}$ defined in \Eq{eq:VPQdef}, instead of the axion decay constant $f_a$. These two quantities are connected by a color anomaly coefficient, as shown explicitly in \Eq{eq:VPQdef2} of App.~\ref{app:Interactions}.} leading to a light axion degree of freedom $a$ \cite{Weinberg:1977ma, Wilczek:1977pj} that relaxes the $CP$-violating phase $\bar{\theta}$ to zero. In this case dark matter could be axions produced by the misalignment mechanism, with $V_{PQ}$ of order $10^{12}$ GeV a motivated possibility.

However, the cosmology of these two leading candidates for dark matter, LSPs and axions, is changed enormously in theories that have {\it both} weak scale supersymmetry and axions.  The axion, $a$, must be promoted to a superfield
 \be
 A=\frac{s+ia}{\sqrt 2}+\sqrt 2\theta \tilde a+\theta^2 F
 \label{eq:axionsupermultiplet}
 \ee
and the saxion, $s$, and the axino, $\tilde a$, both play central roles in cosmology. 

In this work, for reasons discussed below, we focus on DFSZ theories~\cite{Dine:1981rt,Zhitnitsky:1980tq}, where the PQ symmetry forbids the $\mu$ term of the minimal supersymmetric standard model (MSSM). At the scale $V_{\rm PQ}$, PQ breaking induces the $\mu$ term as well as a coupling of the axion supermultiplet with the MSSM Higgs superfields 
\be
W_{\rm DFSZ} = \mu H_u H_d  + q_\mu \frac{\mu}{V_{PQ}} \, A \, H_u H_d  + \ldots \ ,
\ee
with $q_\mu$ a model dependent parameter defined in App.~\ref{app:Interactions}. The superpotential cubic coupling is responsible for axino production in the early universe~\cite{Chun:2011zd, Co:2015pka}, either through decays or inverse decays of charginos and neutralinos, $\tilde{\chi} \rightarrow \tilde{a}$. This axino production by the freeze-in (FI) mechanism is IR dominated~\cite{Hall:2009bx}, namely most of the axinos are produced at temperatures around the TeV scale. Depending on the fermion content of the PQ breaking sector, a large abundance of axinos can also be produced in the UV, at the temperature $T_R$ at the end of inflationary reheating~\cite{Covi:2001nw, Strumia:2010aa}, analogous to UV production of gravitinos~\cite{Nanopoulos:1983up}.

In order to make this distinction sharper, we define two different types of theories:
\begin{description}
\item[DFSZ$_0$:] the heaviest colored fermion carrying PQ charge is the top quark, so the only source for axino production is the IR dominated freeze-in; 
\item[DFSZ$_+$:] there is at least one heavy (with mass of order $V_{\rm PQ}$) colored fermion carrying PQ charge, and thus we also have UV dominated production at $T_R$ from gluino scattering off quarks and gluons.
\end{description}
In DFSZ$_0$, an IR contribution to axino production via scattering also arises from the supersymmetrized $a G \tilde{G}$ operator generated when the top quark is integrated out~\cite{Bae:2011jb}, but it is suppressed compared to the one from decays and we neglect it. In theories with a low gravitino mass, the decay of neutralinos can also populate gravitinos by the FI mechanism but we find this contribution sub-dominant to the ones mentioned above.

\begin{figure} [t]
\begin{center}
\includegraphics[width=0.495\linewidth]{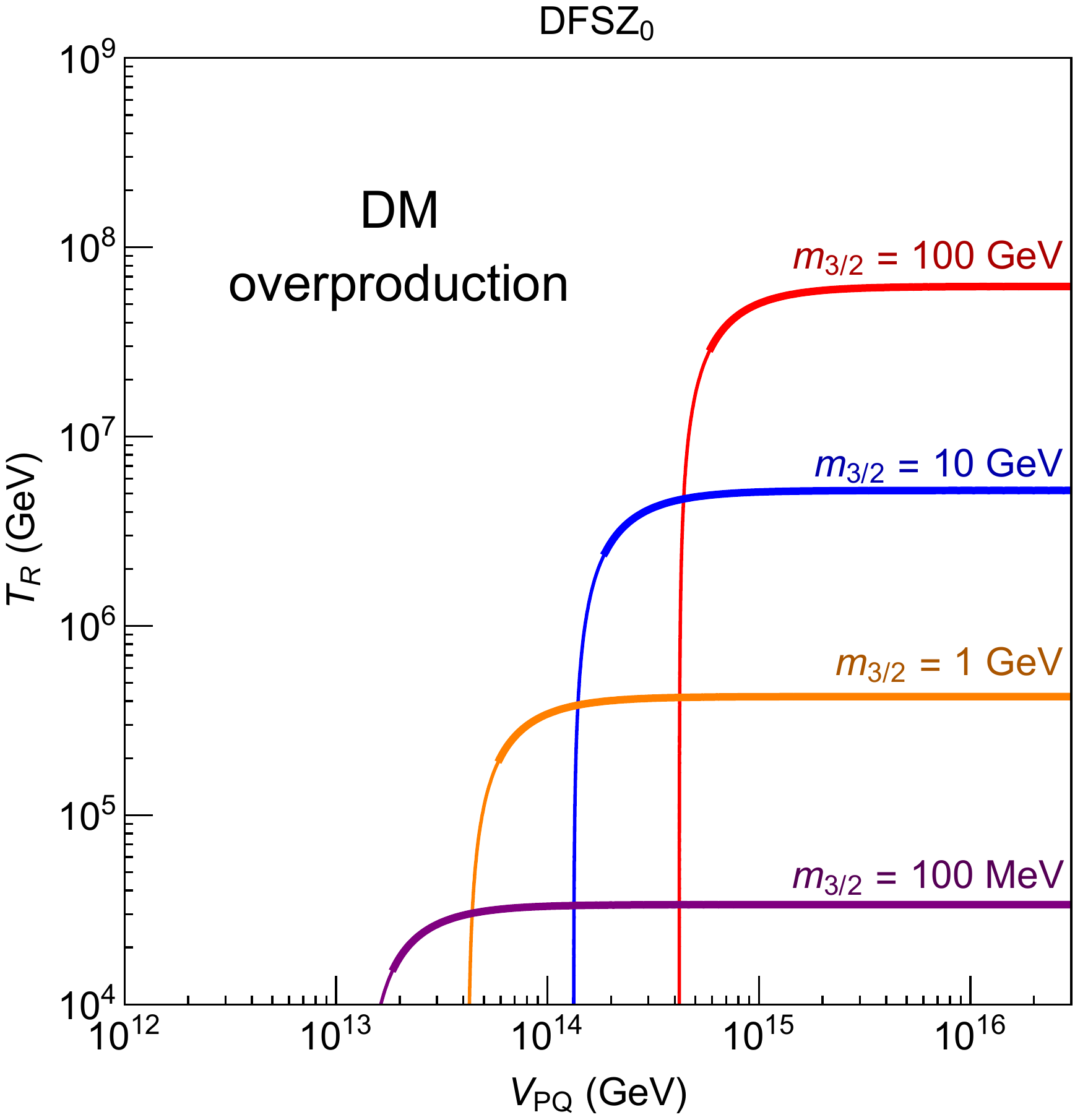} \includegraphics[width=0.495\linewidth]{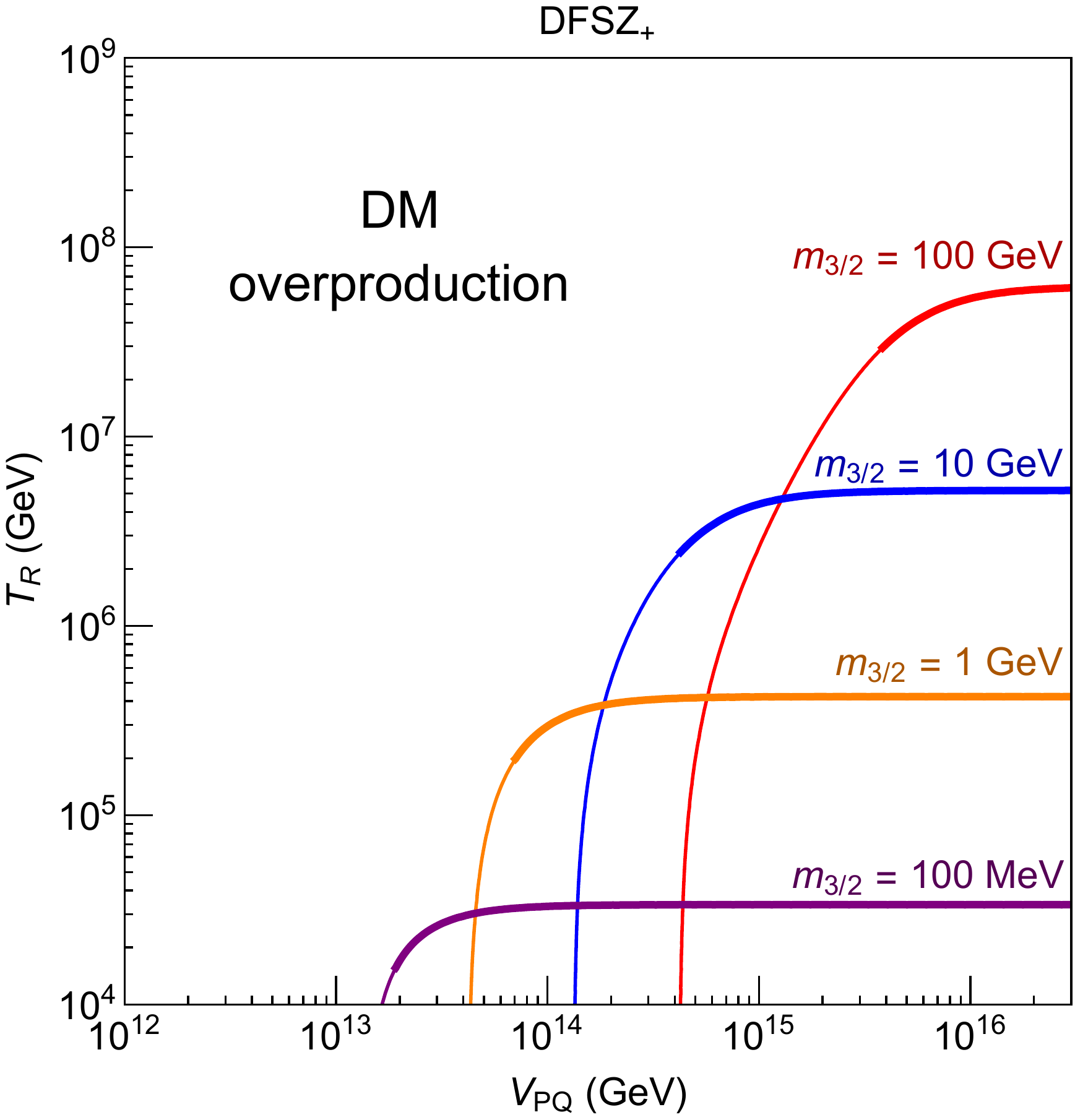}
\end{center}
\caption{Contours yielding the observed dark matter abundance for a gravitino LSP without a saxion condensate for DFSZ$_0$ (left panel) DFSZ$_+$ (right panel), with $M _1 = M_2/2 =  \mu =1$ TeV, $\tan \beta = 2$ and $q_\mu = 2$. The axino mass is in the range $m_{3/2} \lesssim m_{\tilde{a}} \lesssim \mu$. For UV production in DFSZ$_+$ we fix $N_{DW} = 6$. Vertical lines correspond to axino freeze-in via decays $\tilde{\chi} \rightarrow \tilde{a}$, followed by $\tilde{a} \rightarrow \tilde{G} \, a$, whereas horizontal lines correspond to UV gravitino production at $T_R$. The thick (thin) portions of the lines refer to a freeze-in contribution smaller (larger) than $50\%$.}
\label{fig:AxinoProblem}
\end{figure}

In the absence of a saxion condensate, UV production of both axinos and gravitinos puts a very powerful bound on $T_R$~\cite{Cheung:2011mg}. This is illustrated in \Fig{fig:AxinoProblem} both for DFSZ$_0$ (left panel) and DFSZ$_+$ (right panel) for gravitino LSP and other superpartners at the TeV scale.  Contours that yield the DM relic density are shown in the $(V_{\rm PQ}, T_R)$ plane for four values of the gravitino masses. Even for a gravitino with a weak scale mass, the reheat temperature after inflation is strongly bounded, $T_R \lesssim 10^8 \, {\rm GeV}$. Thus LSP dark matter is typically overproduced in the absence of a saxion condensate, unless $V_{\rm PQ}$ is very large and $T_R$ is very low. However, if $V_{PQ}$ is very large so that axino freeze-in is significantly suppressed, the universe is typically overclosed by axions, unless a low value of the axion misalignment angle is selected by an anthropic requirement. 

Saxion cosmology can greatly change this conclusion. Axion theories typically have a domain wall problem, which we assume is solved by breaking the PQ symmetry before inflation, and not restoring it afterwards. We define the saxion field so that today the saxion vev is zero.  During inflation supersymmetry breaking yields a potential for the saxion, displacing the vev away from today's value. Depending on the sign of the quadratic term, the vacuum value $s_I$ is either $V_{PQ}$ or of order the cutoff of the field theory, $M_* \gg V_{PQ}$ ~\cite{Randall:1994fr, Hashimoto:1998ua}.\footnote{The possibility $s_I=0$ requires a special symmetry structure that we do not consider in this paper.} 
If $s_I \gsim 10^{13}$ GeV or $s_I \sim M_*$, 
this saxion condensate comes to dominates the energy density of the universe, producing an early matter-dominated (MD) era.  When the saxion condensate decays, large entropy is created that has a key effect on both LSP and axion contributions to dark matter.  The conventional picture of dark matter survives only for the restricted case of $s_I \sim V_{PQ} \lsim 10^{13}$ GeV, when the saxion condensate never dominates, allowing the usual favorite cases of LSP freeze-out or axion misalignment with $V_{PQ} \sim 10^{12}$ GeV. However, as we have already mentioned, this case of low $V_{PQ}$ has a serious cosmological problem. Even if we choose $T_R$ low enough to suppress UV production, there is still the IR contribution from freeze-in, which cannot be suppressed unless we select $T_R$ below the superpartner mass scale. Hence we consider larger $s_I$, giving a saxion condensate MD era that has important consequences for dark matter abundance.  

In KSVZ~\cite{Kim:1979if,Shifman:1979if} axion theories the MSSM $\mu$ term is not forbidden by PQ symmetry and the axion multiplet does not couple to the Higgs superfields. Thus the saxion typically has a large decay branching ratio to axions and the decay of the saxion condensate gives an axion contribution to dark radiation that is excluded~\cite{Graf:2013xpe}. 
Hence we focus on DFSZ theories.  

In our scheme the dominant saxion decay is to pairs of Higgs, $W$ or $Z$ bosons, giving a very low reheat temperature of the saxion\footnote{We define $T_R$ and $T_{Rs}$ as the reheat temperatures after inflation and saxion decays, respectively.} $T_{Rs}$;  for example, $T_{Rs} \sim$ GeV -- MeV for $V_{PQ} \sim (10^{13}-10^{16})$ GeV, respectively.   The decay of the saxion condensate has crucial implications for dark matter:  
\begin{itemize}
\item  The LSP abundance from freeze-out is greatly diluted~\cite{Chung:1998rq,Giudice:2000ex}. The freeze-out mechanism can only give the observed dark matter abundance if it first overproduces LSPs that are then diluted.  This could happen by raising superpartners well above the TeV scale.
\item  For $V_{PQ} > 10^{13}$ GeV, the entropy is released after the axion condensate starts to oscillate, diluting misalignment axions.  For a misalignment angle of order unity, the value of $f_a$ needed for axion dark matter is raised from  $\sim 10^{12}$ GeV to $\sim 10^{15}$ GeV \cite{Kawasaki:1995vt}, suggesting the possibility that PQ and grand unified gauge symmetries are broken together \cite{Co:2016vsi}. 
\item  The large abundances of gravitinos and axinos produced at $T_R$, and of axinos produced by freeze-in at the TeV scale, are diluted by saxion decays, greatly weakening the constraints of \Fig{fig:AxinoProblem} and allowing much higher $T_R$ and lower $V_{PQ}$.
\item  If kinematically allowed, the saxion condensate leads to a late production of superpartners, and therefore LSPs, via such decays $s \rightarrow \tilde{a} \tilde{a}, \tilde{a} \tilde{G}, \tilde{\chi} \tilde{\chi}$.
\end{itemize}
Given these crucial effects of the saxion condensate, what are the leading candidates and production mechanisms for dark matter?  Misalignment axions with $V_{PQ}$ the scale of grand unification becomes one attractive option \cite{Co:2016vsi}. In this paper we consider LSP possibility by assuming that the axion abundance after dilution is sub-dominant. A misalignment angle $\theta_i \sim \mathcal{O}(1)$ is sufficiently small for $V_{PQ} \lsim 10^{14}$ GeV, whereas $\theta_i \lsim \mathcal{O}(0.3)$ allows $V_{PQ}$ as high as $10^{16}$ GeV.  

We work in a generic framework of TeV scale supersymmetry, taking the saxion and axino masses of order the TeV scale, as well as all other superpartners except, perhaps, the gravitino which could be much lighter.   Freeze-out gives too low an abundance for a typical TeV-scale superpartner spectrum, although a bino-like LSP is a possibility providing the condensate is not too large. On the other hand, if $s$ decays to superpartners the late decay of the condensate typically overproduces LSP dark matter.  Hence we assume these decays are kinematically forbidden and focus on two mechanisms for LSP production: IR freeze-in of axinos from neutralinos and charginos, $\tilde{\chi} \rightarrow \tilde{a}$, and UV scattering at $T_R$ producing axinos and gravitinos, $g \tilde{g} \rightarrow \tilde{a}, \tilde{G}$.   We consider two possibilities for the LSP: $\tilde{a}$ and $\tilde{G}$, and perform a systematic analysis for the dark matter abundance over a very wide range of $T_R$ and $V_{PQ}$, identifying regions of both cold and warm dark matter.  A key feature of our work is to connect the dark matter cosmology with displaced vertex signals, arising from superpartner production and decay to gravitinos \cite{Dimopoulos:1996vz} and axinos \cite{Martin:2000eq}, at the LHC and future colliders.  

Another aspect of the gravitino problem is the powerful constraints from Big Bang nucleosynthesis (BBN) arising from decays between the lightest observable sector superpartner (LOSP) and the gravitino\cite{Weinberg:1982zq}.  We avoid this by making the axino and gravitino the two lightest superpartners.  The LOSP then decays before BBN and if the gravitino is the NLSP it decays to $\tilde{a}a$, which has no effect on BBN.

There is a considerable literature on the effects of a saxion condensate on LSP abundances.  The gravitino and axino problems were solved by the decay of a saxion condensate in Ref.~\cite{Kawasaki:2008jc}, which also considered the possibility of gravitino dark matter from inflaton decay.   This was further developed by identifying PQ-breaking fields as the waterfall fields of a hybrid inflation model with vevs of order $10^{15}$ GeV \cite{Kawasaki:2011ym}.  An alternative solution to the gravitino problem  involved a very light, keV axino  \cite{Asaka:2000ew,Baer:2010gr}.  In other work a saxion condensate was used to obtain a PQ breaking scale as large as the unification scale in theories with axino or neutralino LSP \cite{Baer:2011eca}, and further work considered mixed axion/neutralino dark matter \cite{Bae:2014rfa}.  
A KSVZ scheme with a saxion condensate relevant for gravitino dark matter, but at lower values of $V_{PQ}$ than considered in this paper, is given in Ref.~\cite{Hasenkamp:2010if}.

We describe the saxion condensate MD era in \Sec{sec:SaxiCosmo}, by identifying the characteristic temperatures associated to this epoch and giving analytical expressions for them. The production mechanisms for axino and gravitino are discussed in detail in \Sec{sec:axinoprod}, where we account for the dilution for the saxion condensate and give numerical results for the yields $Y_{\tilde a}$ and $Y_{3/2}$ as a function of $V_{PQ}$.    In Sec.~\ref{sec:AxGravLSP}, we discuss key consequences of our scheme with the axino and gravitino as the two lightest superpartners.  The NLSP decay leads to a component of dark matter that is warm, with free streaming length as illustrated in Fig.~\ref{fig:FSLength}. It is intriguing that this result can be consistent with Lyman-$\alpha$ forest observations while at the same time solving issues with cosmological structures at small scales.  Displaced signals at colliders resulting from the decay of the LOSP to axinos and gravitinos are discussed, as well as the axion dark radiation resulting from decay of the saxion condensate. 

In sections \ref{sec:HighScale} and \ref{sec:LowScale} we present our results for two classes of SUSY spectra: high scale (i.e. ``gravity'') and low scale (i.e. ``gauge'') mediation, respectively. The LSP relic density for high scale mediation is shown in Figs.~\ref{fig:HighDFSZ0} and \ref{fig:HighDFSZ+}, whereas the analogous results for low scale are in Figs.~\ref{fig:LowDFSZ0} and \ref{fig:LowDFSZ+}. A remarkable signal of our framework, which holds regardless of the mediation scale, is displaced events at colliders.  We study how these lifetimes vary with the supersymmetry breaking parameters in Fig.~\ref{fig:LOSPdecayLength}, and make lifetime predictions in Figs.~\ref{fig:ctauHighDFSZ0}, \ref{fig:ctauHighDFSZ+}, \ref{fig:ctauLowDFSZ0}, and \ref{fig:ctauLowDFSZPlus}.  We supplement our work by Appendices with useful results employed in our analysis.

\section{Saxion Cosmology}
\label{sec:SaxiCosmo}

Inflationary dynamics sets the initial conditions for the saxion condensate, typically displacing it by an amount $s_I$ from the minimum today. After inflation ends, Hubble friction keeps the saxion field fixed until the universe expands and cools sufficiently that $3 H \sim m_s$, with $m_s$ the saxion mass, when the saxion condensate oscillates around its minimum. The energy density stored in such oscillations red-shifts like non-relativistic matter at a rate slower than radiation, and eventually dominates the energy budget. 

The specific temperature where saxion oscillations begin is crucial for the evolution of the saxion condensate. In particular, we identify two main regimes according to the size of the reheat temperature after inflation 
\be 
T_R {> \atop <} \sqrt{m_s M_{Pl}}  \, \sim \, \mathcal{O} (10^{10}) \GeV 
\ee
and describe the cosmologies separately.  In the case where $T_R \gsim 10^{10}$ GeV, saxion oscillations start during the radiation-dominated era after inflationary reheating has ended, while in the case where $T_R \lsim 10^{10}$ GeV, they start during inflationary reheating.

In addition to the saxion condensate that arises from inflationary dynamics, the saxion can also be generated from the thermal processes such as the scattering with gluinos via the dimension-5 operators generated by the heavy colored fermions with PQ charges in the DFSZ$_+$ theories. This contribution is sub-dominant when $s_I > 10^{13-14}$ GeV. For $s_I < 10^{13-14}$ GeV, we find that the extra thermal saxions still fail to provide the dilution necessary for the overproduced axinos from UV scattering. Therefore, this thermal contribution of saxions is never relevant in the parameter space of interest.

\subsection{High reheat temperature after inflation: $T_R \gsim 10^{10}$ GeV}
\label{subsec:highTR}

Saxion oscillations begin when the radiation bath has a temperature
\be
\label{eq:SaxTosc}
T_{osc} = \left(  \frac{10}{\pi^2 g_*(T_{osc})}  \right)^{ \scalebox{1.01}{$\frac{1}{4}$} } \sqrt{m_s M_{\rm Pl}} \ ,
\ee
where we introduce the effective number of relativistic degrees of freedom in the thermal bath $g_*(T)$ and the reduced Planck mass $M_{\rm Pl} = 2.4 \times 10^{18}$ GeV. The onset of the saxion oscillation happens during an early Radiation Dominated era (RD$'$). The energy stored in saxion oscillations red-shifts with the scale factor $a$ as $a^{-3}$, namely with the same behavior of non-relativistic matter. If the condensate is long-lived enough, it eventually takes over the radiation energy and the universe enters an early matter-dominated (MD) era at a temperature $T_M$. This characteristic temperature is found by imposing that the saxion and the radiation bath equally contribute to the energy density, or in other words by solving the equation
\be
m_s^2 s_I^2 \, \frac{g_*(T_M)}{g_*(T_{osc})} \left( \frac{T_M}{T_{osc}} \right)^3 = \frac{ \pi^2}{30} g_*(T_M) \ T_M^4 \ ,
\ee
where we use the conservation of the total entropy to properly red-shift the saxion energy. Only for the purpose of an analytical estimate, we set $g_*(T_{osc}) = g_*(T_M) = 228.75$, corresponding to the full MSSM field content, giving
\be
T_M  \simeq 10 \, \text{TeV} \left( \frac{m_s}{\text{ TeV}} \right)^{ \scalebox{1.01}{$\frac{1}{2}$} }  \left( \frac{s_I}{ 10^{15} \text{ GeV} } \right)^2 \ .
\label{eq:SaxTM}
\ee

This early MD era consists of two distinct phases, as detailed in Refs.~\cite{Co:2015pka,Co:2016vsi}. In the beginning, the radiation energy density is dominated by the red-shifted initial component. We dub this initial part an adiabatic matter-dominated (MD$_A$) era. However, radiation constantly produced from saxion visible decays red-shifts slower and in the end accounts for most of the radiation present in the universe. Once the contribution from saxion decays dominates, at a temperature $T_{NA}$, we enter a non-adiabatic matter-dominated (MD$_{NA}$) era, where the total entropy is not conserved. At temperatures below $T_{NA}$, saxion decays reheat the universe and a large amount of entropy is released. Finally, most of the saxions decay when the Hubble parameter is of the order the saxion decay width, $H \sim \Gamma_s$. This is the beginning of the last phase of a radiation-dominated universe (RD), which starts when the radiation bath has the reheat temperature $T_{Rs}$.

We require that saxion decays to R-odd neutralinos and charginos are kinematically forbidden, and that decays to axions are sub-dominant. The saxion width is thus dominated by decays to MSSM Higgs bosons and longitudinal electroweak gauge bosons. We use the expression in \Eq{eq:Gammastovisible} for the saxion visible width, valid in the decoupling limit and for large $\tan\beta$. The transition from MD$_A$ to MD$_{NA}$ occurs at temperature
\be
T_{NA} \simeq 0.2 \, \text{GeV} \, q_\mu^{\scalebox{1.01}{$\frac{4}{5}$}}  \left(\frac{\mathcal{D}}{4} \right)^{\scalebox{1.01}{$\frac{2}{5}$}} \left( \frac{\mu}{\text{ TeV}} \right)^{\scalebox{1.01}{$\frac{13}{10}$}} \left( \frac{\mu}{m_s} \right)^{\scalebox{1.01}{$\frac{3}{10}$}}  \left( \frac{s_I}{V_{PQ}} \right)^{\scalebox{1.01}{$\frac{2}{5}$}} \left( \frac{10^{15} \ \text{GeV}}{V_{PQ}} \right)^{\scalebox{1.01}{$\frac{2}{5}$}}   \ ,
\label{eq:SaxTNA}
\ee
where we used $g_*(T_{Rs}) = 10.75$ and $\mathcal{D}$ denotes the number of final states kinematically accessible ($\mathcal{D} = 4$ for SM and $\mathcal{D} = 8$ for the full MSSM). The non-adiabatic era ends once the saxion condensate decays, reheating the universe at the temperature
\be
T_{Rs}  \simeq 10 \, \text{MeV} \ q_\mu  \left(\frac{10.75}{g_*(T_{Rs})} \right)^{ \scalebox{1.01}{$\frac{1}{4}$} } \left(\frac{\mathcal{D}}{4} \right)^{ \scalebox{1.01}{$\frac{1}{2}$} }  \left( \frac{\mu}{\text{ TeV}} \right)^{ \scalebox{1.01}{$\frac{3}{2}$} } \left( \frac{\mu}{m_s} \right)^{ \scalebox{1.01}{$\frac{1}{2}$} }   \left( \frac{10^{15} \ \text{GeV}}{V_{PQ}} \right)  \ .
\label{eq:SaxTR}
\ee

\begin{figure}
\begin{center}
\includegraphics[width=0.495\linewidth]{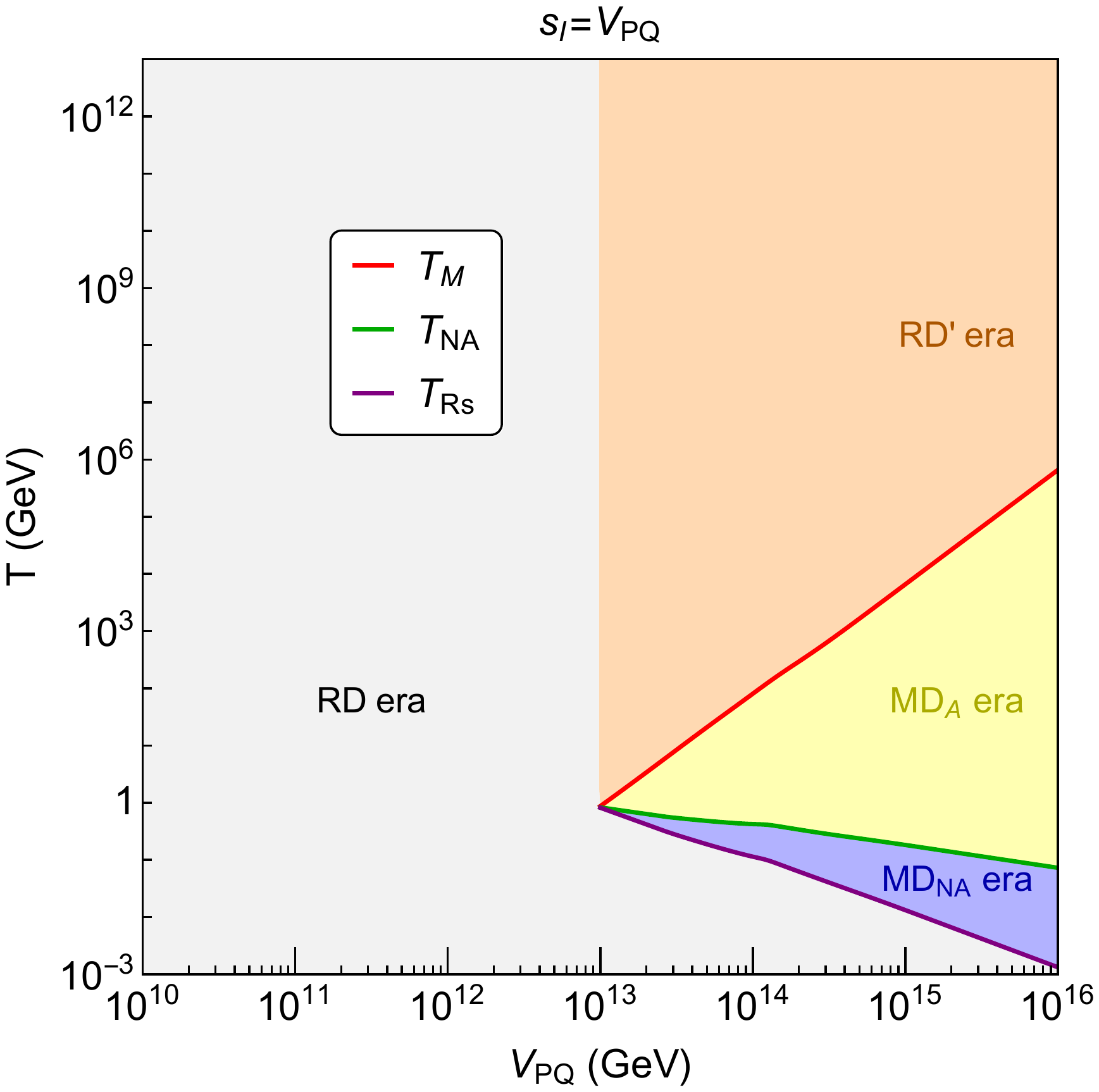} \includegraphics[width=0.495\linewidth]{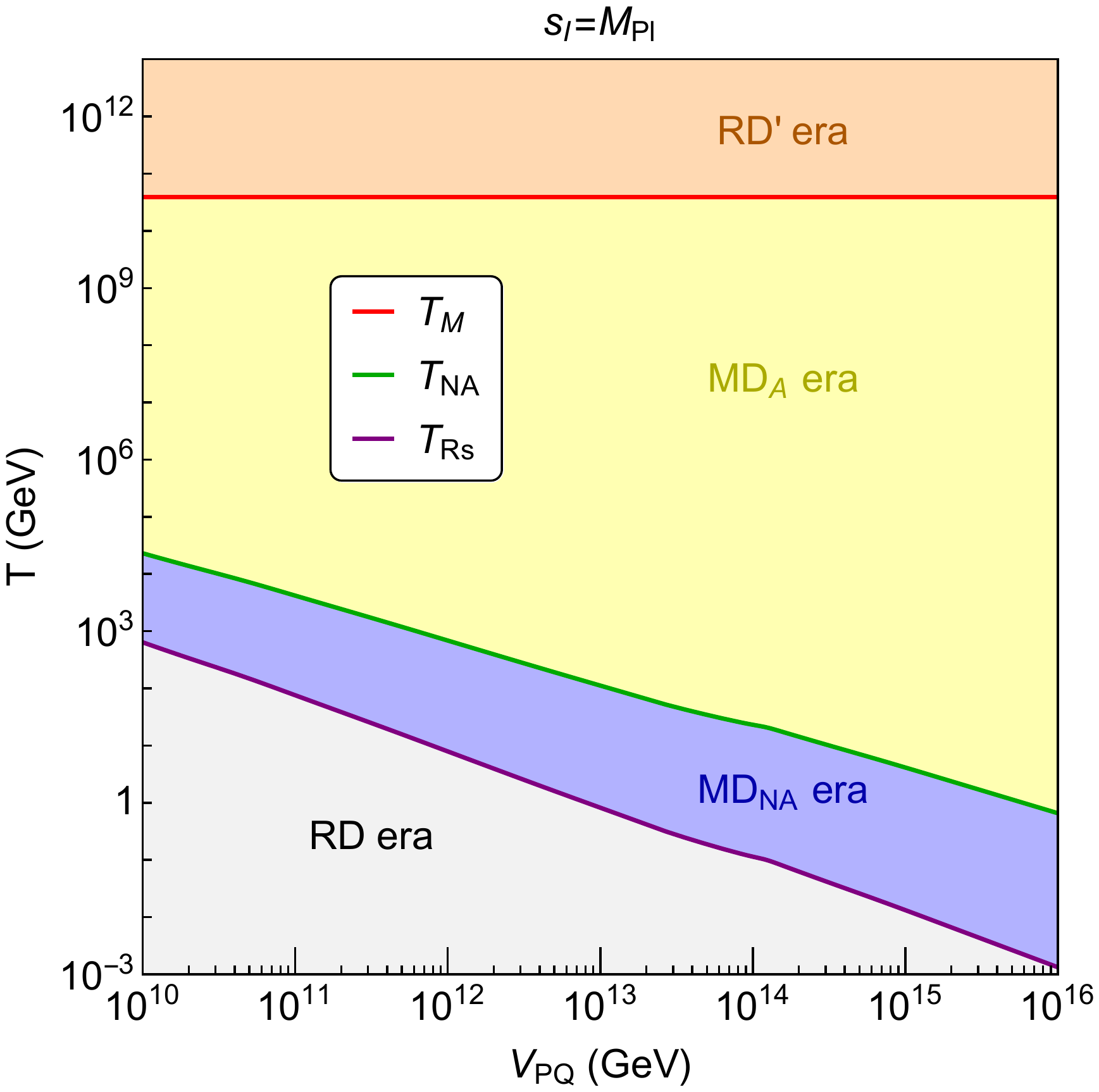}
\end{center}
\caption{Cosmological eras for $T_R \ge 10^{10}$ GeV.  $T_M$, $T_{NA}$ and $T_{Rs}$ as functions of $V_{PQ}$ with $s_I = V_{PQ}$ ($M_{Pl}$) in the left (right) panel, $\mu=m_s=1$ TeV, $q_\mu=1$, and $\mathcal{D}=4$. The RD$'$, MD$_A$, MD$_{NA}$, RD eras are individually shaded.}
\label{fig:TMvsTRa}
\end{figure}

The three characteristic temperatures $T_M$, $T_{NA}$ and $T_{Rs}$ are shown in \Fig{fig:TMvsTRa}. We plot their values as a function of $V_{PQ}$ for two different initial conditions: $s_I = V_{\rm PQ}$ (left panel) and $s_I = M_{\rm Pl}$ (right panel). In the first case, we notice that for $V_{PQ} \lesssim 10^{13} \, {\rm GeV}$ there is no matter dominated era. For such low values of $V_{\rm PQ}$, the initial saxion energy density is not enough to overtake radiation before it decays. Thus, $T_M$ and $T_{NA}$ are not defined and there is no significant entropy production at $T_{Rs}$. We define $V_{\rm PQ}^{(c)}$ this critical value of $V_{\rm PQ}$, corresponding to the intersection of the three lines in the left panel of \Fig{fig:TMvsTRa}.

A physically meaningful and useful quantity is the amount of entropy released by saxion decays during the reheating process. We quantify this by introducing the dilution factor $D(T_i)$, defined as the ratio of the entropy after saxion decays to that at an initial temperature $T_i$
\be
\label{eq:Ddef}
D (T_i) \equiv \frac{S_{Rs}}{S_i} = \frac{g_*(T_{Rs}) T_{Rs}^3}{g_*(T_i) T_i^3} \left(\frac{a_{Rs}}{a_i} \right)^3 .
\ee
We interpret $T_i$ as the temperature when dark matter is produced, e.g. the  freeze-in temperature. In the MD$_{NA}$ era, we make use of the scaling $a^3 \propto T^{-8}$, whereas in the MD$_A$ and RD eras, $a^3 \propto T^{-3}$. With $T_{NA}^5 \simeq T_M T_{Rs}^4$, we obtain the following dilution factor
\be
D (T_i) \approx \begin{cases}
\frac{T_M}{T_{Rs}} & T_i \ge T_{NA}  \\
\left( \frac{T_i}{T_{Rs}} \right)^5  & T_{NA} \ge T_i \ge T_{Rs} .
\end{cases}
\ee
We can thus find $D$ from Eqs.~(\ref{eq:SaxTM}) and (\ref{eq:SaxTR})
\be
\label{eq:SaxD}
D (T_i)
\simeq \begin{cases}
10^6 \, \left(\frac{2}{q_\mu}\right) \left(\frac{4}{\mathcal{D}}\right)^{\scalebox{1.01}{$\frac{1}{2}$}} \left( \frac{\text{ TeV}}{ \mu } \right) \left( \frac{m_s}{\mu} \right) \left( \frac{s_I}{10^{15} \ \text{GeV}} \right)^2 \left( \frac{ V_{PQ} }{ 10^{15} \ \text{GeV} } \right)   \ 
 & T_i \ge T_{NA}  \\
10^5 \, q_\mu^{-5} \left( \frac{T_i}{100 \rm{MeV}} \right)^5 \left(\frac{4}{\mathcal{D}} \right)^{\scalebox{1.01}{$\frac{5}{2}$}}  \left( \frac{\text{ TeV}}{\mu} \right)^{\scalebox{1.01}{$\frac{15}{2}$}} \left( \frac{m_s}{\mu} \right)^{\scalebox{1.01}{$\frac{5}{2}$}}   \left( \frac{V_{PQ}}{10^{15} \ \text{GeV}} \right)^5 
& T_{NA} \ge T_i \ge T_{Rs} .
\end{cases}
\ee
The dilution factor is extremely large for the initial condition $s_I = M_{Pl}$, ranging from $10^9$ to $10^{15}$ as $V_{PQ}$ increases from $10^{10}$ GeV to $10^{16}$ GeV.  On the other hand, the $s_I = V_{PQ}$ case has a much small dilution, varying from 1 to $10^9$ as $V_{PQ}$ increases from $10^{13}$ GeV to $10^{16}$ GeV. Each comoving number density frozen-in or -out before the entropy release, namely at temperatures $T_i > T_{NA}$, gets maximally depleted by $D$. 

The quantity $D$ also allows us to quantify the critical value $V_{\rm PQ}^{(c)}$, defined as the point where the three lines in the left panel of \Fig{fig:TMvsTRa} intersect. Its value is obtained by imposing $D =1$
\begin{align}
\label{eq:VPQMD}
V_{PQ}^{(c)} = \ & \left( \frac{ q_\mu \sqrt{ \mathcal{D} }    }{4 \sqrt{3 \pi } } \frac{\mu^2 M_{Pl}^2}{m_s} \right)^{ \scalebox{1.01}{$\frac{1}{3}$} } 
\simeq \ 10^{13} \, \text{GeV} \ q_\mu^{ \scalebox{1.01}{$\frac{1}{3}$} } \left( \frac{\mathcal{D}}{4} \right)^{ \scalebox{1.01}{$\frac{1}{6}$} }  \left( \frac{\mu}{\text{TeV} } \right)^{ \scalebox{1.01}{$\frac{2}{3}$} } \left( \frac{\text{TeV}}{m_s} \right)^{ \scalebox{1.01}{$\frac{1}{3}$} }.
\end{align}

\subsection{Low reheat temperature after inflation: $T_R \lsim 10^{10}$ GeV}
\label{subsec:lowTR}

So far we have assumed that saxion oscillations begin during a radiation dominated era. However, this need not be the case as we do not know the temperature of reheating after inflation, which is set by the decay width of the inflaton. We now extend our analysis to low $T_R$.

During inflationary reheating, the temperature dependence of the Hubble parameter is different than during a radiation dominated era. An approximate solution to the coupled Boltzmann equations describing the evolution of inflaton and radiation energy densities, at early times before the inflaton decays, gives the expression for the Hubble parameter~\cite{Chung:1998rq,Giudice:2000ex}
\be
H(T) = \frac{\sqrt{5} \, \pi}{6 \sqrt{2}} \frac{g_*(T)}{g_*(T_R)^{1/2}} \frac{T^4}{M_{\rm Pl} \, T_R^2} \ , \qquad \qquad 
\qquad T > T_R \ .
\label{eq:Hinf}
\ee
Oscillations for the saxion condensate begin at $T_{\rm osc}'$, when $3 H \simeq m_s$.  Assuming that $T_{\rm osc}'$ is less than the maximum temperature reached during this reheating era, so that $H$ is now given by \Eq{eq:Hinf}, we obtain
\be
\begin{split}
T_{\rm osc}' = & \, \left( \frac{8 \, g_*(T_R)}{5 \, \pi^2 \, g_*(T_{\rm osc}')^2} \right)^{ \scalebox{1.01}{$\frac{1}{8}$} } \left(m_s  M_{\rm Pl} T_R^2 \right)^{ \scalebox{1.01}{$\frac{1}{4}$} } \\
= \, &  10^{10} \, {\rm GeV} \,
 \left(\frac{g_*(T_R)}{228.75}\right)^{ \scalebox{1.01}{$\frac{1}{8}$} } \left(\frac{228.75}{g_*(T_{\rm osc}')}\right)^{ \scalebox{1.01}{$\frac{1}{4}$} } 
 \left( \frac{m_s}{{\rm TeV}} \right)^{ \scalebox{1.01}{$\frac{1}{4}$} } \left( \frac{T_R}{10^{10} \, {\rm GeV}} \right)^{ \scalebox{1.01}{$\frac{1}{2}$} } \ .
\end{split}
\label{eq:ToscI}
\ee
This equation makes sense only if $T_{\rm osc}' > T_R$, and this condition is satisfied as long as $T_R < T_{\rm osc} \simeq 10^{10}$ GeV, namely the oscillation temperature given in \Eq{eq:SaxTosc}.  

The saxion oscillations for $T < T_{\rm osc}'$ still red-shift as non-relativistic matter, and once the inflaton finally decays the energy density stored in them results in
\be
\rho_s(T_R) = m_s^2 s_I^2 \left( \frac{g_*(T_R)}{g_*(T'_{\rm osc})} \right)^2  
\left( \frac{T_R}{T'_{\rm osc}} \right)^8 \ .
\label{eq:rhosaRI}
\ee
Afterwards, we have the conventional thermal history described above. The only difference is the initial condition for the saxion energy density as in \Eq{eq:rhosaRI}. This in turn gives a different expression for $T_M$, which is now found by imposing the condition
\be
\rho_s(T_M) = \rho_s(T_R) \frac{g_*(T_M) T_M^3}{g_*(T_R) T_R^3}  = \rho_R(T_M) =
\frac{\pi^2}{30} g_*(T_M) T_M^4 \ . 
\ee
The solution of this equation simply gives
\be
T'_M = \frac{30}{\pi^2} \frac{g_*(T_R)}{g_*(T'_{\rm osc})^2} \frac{m_s^2 \, s_I^2 \, T_R^5}{(T'_{\rm osc})^8} = \frac{75}{4}  \left(\frac{s_I}{M_{\rm Pl}}\right)^2 \, T_R \ ,
\label{eq:TMI}
\ee
where in the last step we have used the expression in \Eq{eq:ToscI}. This expression breaks down when $T_M'$ becomes larger than $T_R$, which happens if $s_I \geq M_{Pl} / \sqrt{3}$. We avoid this situation because the saxion condensate dominates the energy density before it starts to oscillate and consequently the Universe enters inflation.

The expression for $T_{Rs}$ is of course unchanged, and still given by \Eq{eq:SaxTR}. However, the temperature for the transition to the non-adiabatic phase is changed 
\be
\begin{split}
T'_{NA} = & \, \left( T'_M T_{Rs}^4 \right)^{ \scalebox{1.01}{$\frac{1}{5}$} } \simeq 2.5 \times 10^4 \, {\rm GeV} \, \left( \frac{T_R s_I^2}{m_s^2 V_{PQ}^4} \right)^{ \scalebox{1.01}{$\frac{1}{5}$} } \\ \simeq & \, 632 \, {\rm GeV} \, \left( \frac{T_R}{10^{10} \, {\rm GeV}} \right)^{ \scalebox{1.01}{$\frac{1}{5}$} } \left( \frac{s_I}{10^{16} \, {\rm GeV}} \right)^{ \scalebox{1.01}{$\frac{2}{5}$} } \left( \frac{ {\rm TeV}}{m_s} \right)^{ \scalebox{1.01}{$\frac{2}{5}$} } \left( \frac{10^{11} \, {\rm GeV}}{V_{PQ}} \right)^{ \scalebox{1.01}{$\frac{4}{5}$} } \ . 
\end{split}
\ee 

\begin{figure}[t]
\begin{center}
\includegraphics[width=0.495\linewidth]{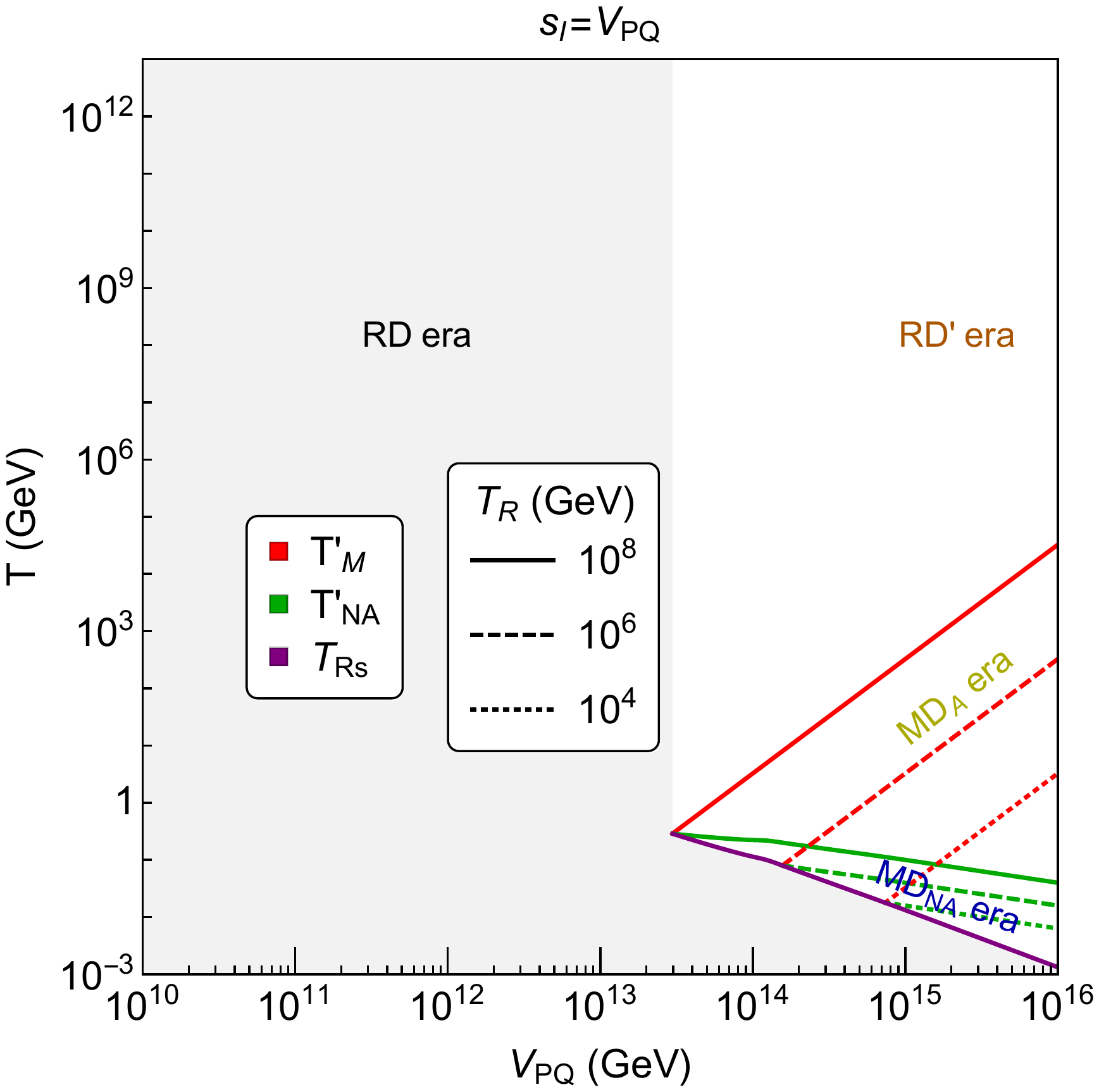} \includegraphics[width=0.495\linewidth]{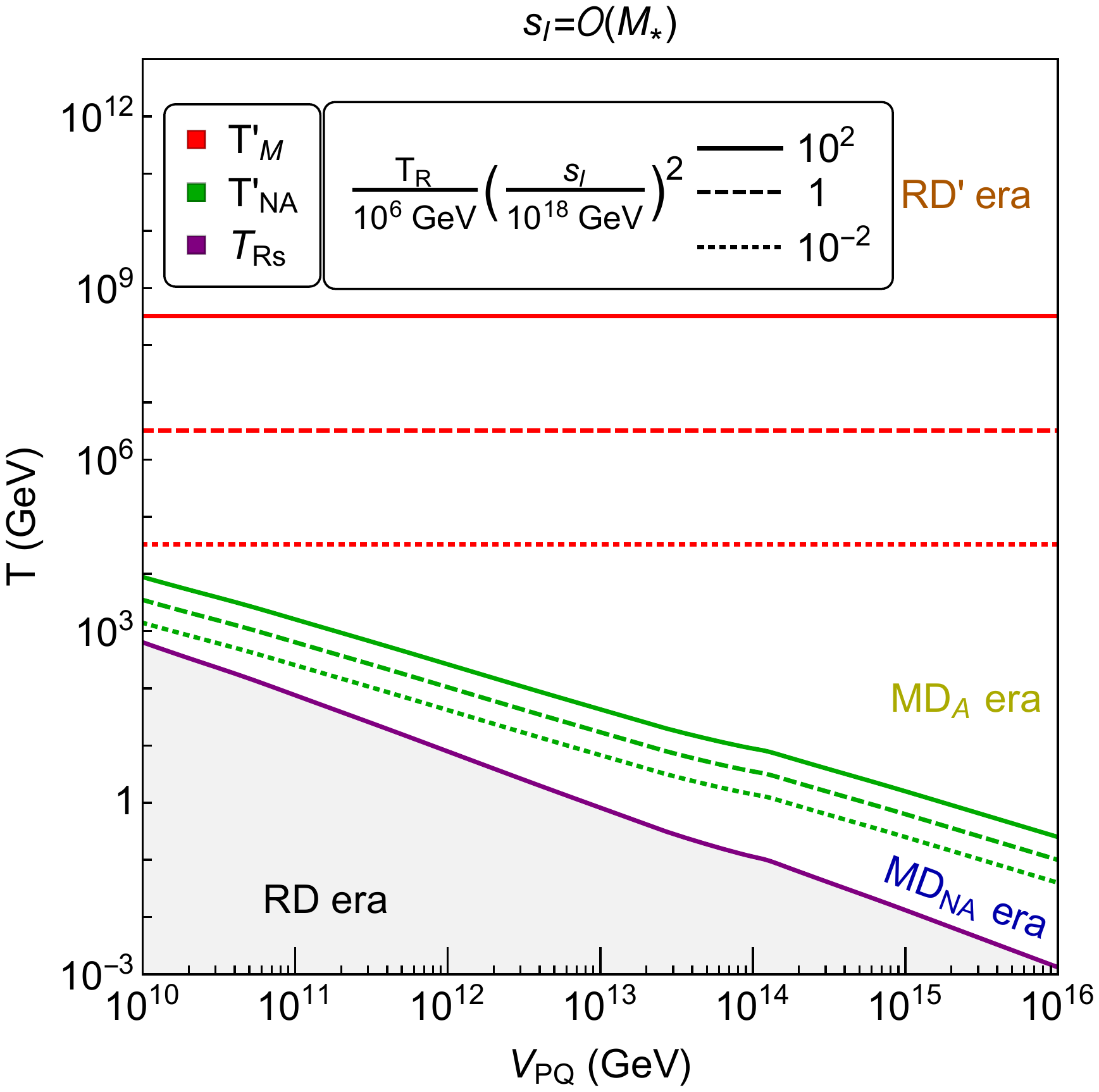}
\end{center}
\caption{Cosmological eras for $T_R \le 10^{10}$ GeV.  $T'_M$, $T'_{NA}$ and $T_{Rs}$ as functions of $V_{PQ}$ with $s_I = V_{PQ}$ ($M_*$) in the left (right) panel, $\mu=m_s=1$ TeV, $q_\mu=1$, and $\mathcal{D}=4$. The RD$'$, MD$_A$, MD$_{NA}$, RD eras are labeled similar to \Fig{fig:TMvsTRa}, whereas some of the shadings are removed for clarity. In the right panel, both $T_R$ and $s_I$ are needed to specify $T'_M$.}
\label{fig:TMvsTRb}
\end{figure}

Moreover, also the expression for the dilution factor changes
\be
D' (T_i) = \begin{cases}
4 \times 10^5 \, \left( \frac{s_I}{10^{16} \, {\rm GeV}} \right)^{2} \left( \frac{T_R}{10^{10} \, {\rm GeV}} \right)
\left( \frac{{\rm TeV}}{\mu} \right)^{2} \left( \frac{m_s}{{\rm TeV}} \right)^{ \scalebox{1.01}{$\frac{1}{2}$} } \left( \frac{V_{PQ}}{10^{11} \, {\rm GeV}} \right) \ 
 & T_i \ge T'_{NA}  \\
10^5 \ q_\mu^{-5} \left( \frac{T_i}{\TeV} \right)^5   \left(\frac{4}{\mathcal{D}} \right)^{\scalebox{1.01}{$\frac{5}{2}$}}  \left( \frac{\text{ TeV}}{\mu} \right)^{\scalebox{1.01}{$\frac{15}{2}$}} \left( \frac{m_s}{\mu} \right)^{\scalebox{1.01}{$\frac{5}{2}$}}   \left( \frac{V_{PQ}}{10^{11} \ \text{GeV}} \right)^5
& T'_{NA} \ge T_i \ge T_{Rs} .
\end{cases}
\label{eq:DI}
\ee
We observe the remarkable fact that the dilution factor is proportional to $T_R$ when $T_i \ge T'_{NA}$. The three characteristic temperatures $T'_M$, $T'_{NA}$ and $T_{Rs}$ are shown in \Fig{fig:TMvsTRb}.

\subsection{Field equations}

The analytical expressions for the characteristic temperatures and the dilution factor are very useful for an order
of magnitude estimate of the effect. However, they are not suited for precision calculation of relic densities. In this work we solve numerically the Boltzmann equation system describing the evolution of the saxion condensate coupled to the radiation bath.

We begin with the case where the saxion starts to oscillate after inflationary reheating, presented in \Sec{subsec:highTR}. The energy density of the saxion condensate, after the onset of oscillations at $T_{\rm osc}$ of \Eq{eq:SaxTosc} or at $T'_{\rm osc}$ of \Eq{eq:ToscI}, evolves according to
\be 
\label{eq:Hubble_s}
\frac{d \rho_s}{d t} + 3 H \rho_s  = - \Gamma_s \rho_s \ .
\ee
The redshift due to the Hubble expansion is accompanied by the term proportional to the saxion total decay width. 

The radiation bath temperature evolves according to
\be
\label{eq:TempPDE}
\frac{\pi^2}{30} g_*(T) \left(1 + \frac{1}{3} \frac{d \ln g_*}{d \ln T} \right) \frac{d T^4}{dt} +
4 H  \, \frac{\pi^2}{30} g_*(T)  T^4 = \Gamma_s \rho_s  \ ,
\ee
where $g_{*}$ is the effective number of degrees of freedom contributing to the entropy density. In the limiting case where $g_*$ is a constant, the equation takes a more familiar form
\be
\frac{d \rho_{\rm rad}}{dt} + 4 H \rho_{\rm rad} = \Gamma_s \rho_s \ ,
\ee
where the radiation energy density results in
\be
\rho_{\rm rad} =  \frac{\pi^2}{30} g_*(T) T^4 \ .
\label{eq:BEradApprox}
\ee
This approximation is certainly valid at very high temperature, where the full spectrum is relativistic. At lower temperature the approximation breaks down and the error one makes by using \Eq{eq:BEradApprox} is at most of few percent. In our work we use \Eq{eq:TempPDE}, and  $g_*(T)$ is computed using the masses of the SM particles and of the SUSY particles, assumed degenerate at 1 TeV.

Finally, the time-temperature relation can be found by solving the Friedmann equation
\be
\label{eq:Hubble}
H  = \frac{1}{ \sqrt{3}M_{Pl}} \sqrt{ \rho_s +  \frac{\pi^2}{30} g_*(T) T^4} .
\ee
The initial condition for this case is set at some high temperature $T_0$ by
\be
\label{eq:SaxIniCon}
\rho_{Mi} =  \frac{\pi^2}{30} g_*(T_M) \, T_M \, T_0^3 ,
\ee
where $T_M$ is defined in \Eq{eq:SaxTM} and our numerical studies are completely insensitive to $T_0$.

For the case where the saxion starts to oscillate during inflationary reheating, the numerical setup needs to be extended as follows. Firstly, we cannot ignore the inflaton anymore and we couple \Eqs{eq:Hubble_s}{eq:TempPDE} to the one describing the evolution of the inflaton energy density $\rho_I$, which takes the same form as \Eq{eq:Hubble_s} with the modifications $\rho_s \rightarrow \rho_I$ and $\Gamma_s \rightarrow \Gamma_I$. Furthermore, we need to add the inflaton decay contribution $\Gamma_I \rho_I$ to the right-hand side of \Eq{eq:TempPDE}, and the inflaton energy density on the right-hand side of \Eq{eq:Hubble}. Secondly, we set the initial conditions for the saxion oscillation at the time $3 H(t_{\rm osc}) = m_s$, with $H$ in this case dominated by the inflaton energy density. Since inflationary reheating is in an MD era, we can identify $3 H(t_{\rm osc}) = 2 / t_{\rm osc}$, and thus the initial condition $\rho_s(t_{\rm osc}) = m_s^2 s_I^2$ is set at the time $t_{\rm osc} = 2 / m_s$.

\section{Axino and Gravitino Production}
\label{sec:axinoprod}

In this Section we quantify axino and gravitino production by accounting for the saxion condensate effects. We consider three different mechanisms: axino production from freeze-in, axino UV production (present only in DFSZ$_+$ theories) and gravitino UV production. For each case, we show results for the comoving number density as a function of $V_{\rm PQ}$, and we consider both $s_I = V_{\rm PQ}$ and $s_I = M_*$. We also comment on gravitino freeze-in production, a sub-dominant source. The results presented here are completely general and independent of where the axino and gravitino sit in the superpartner spectrum. We apply the framework to two specific spectra corresponding to High Scale and Low Scale mediation in Secs.~\ref{sec:HighScale} and \ref{sec:LowScale} respectively.

\subsection{Freeze-In Production of Axinos}
\label{sec:IRAxinoProd}

The freeze-in production of axinos is controlled by neutralino and chargino decays and inverse decays, $\tilde{\chi} \rightarrow \tilde{a}$. Explicit decay widths relevant for this case are given in App.~\ref{app:Decay}. The evolution of the axino number density $n_{\tilde{a}}$ is governed by the Boltzmann equation
\be
\frac{d n_{\tilde{a}}}{dt} + 3 H n_{\tilde{a}} = \mathcal{C}_{\rm FI} \ . 
\label{eq:BoltzFI}
\ee
Our goal here is to provide the expression for the collision operators $\mathcal{C}_{\rm FI}$.

For a light axino, lighter than all the MSSM superpartners, freeze-in comes from neutralinos and charginos decays to axinos
\begin{align}
\widetilde{N}_i \; \rightarrow \; & \tilde{a} h , \; \tilde{a} Z \ , \\
\widetilde{C}^\pm_i  \; \rightarrow \; & \tilde{a} W^\pm \ .
\end{align}
The partial widths for these channels are given in \Eqs{eq:Gammaitoa}{eq:Gammaitoacharged}, respectively. On the contrary, for a heavy axino, heavier than all the neutralinos and charginos, freeze-in production is dominated by the inverse decays
\begin{align}
\widetilde{N}_i h , \widetilde{N}_i Z \; \rightarrow \; & \tilde{a} \ , \\
\widetilde{C}^\pm_i \, W^\mp \; \rightarrow \; & \tilde{a}  \ .
\end{align}
We use the detailed-balance principle to write the collision operator in the Boltzmann equation by using the inverse reaction, namely the axino decay, with partial decay widths given in \Eqs{eq:Gammaatoi}{eq:Gammaatoicharged}. In the intermediate case, with the axino mass within the neutralinos and charginos masses, we have both types of reactions, but only the ones allowed by kinematics.

The freeze-in collision operator is a sum of the possible sources
\be
\mathcal{C}_{\rm FI} = 
\mathcal{C}_{\rm FI}^{\rm n-decay} + \mathcal{C}_{\rm FI}^{\rm n-inverse} + \mathcal{C}_{\rm FI}^{\rm c-decay} + \mathcal{C}_{\rm FI}^{\rm c-inverse} \ . 
\ee
The first two terms account for the processes involving the neutralinos
\begin{align}
\label{eq:nDecay}
\mathcal{C}_{\rm FI}^{\rm n-decay} = & \,
\frac{T^3}{\pi^2}\sum_{i = 1}^4 \theta( m_{\widetilde{N}_i} - m_{\tilde{a}} ) \;
\Gamma_{\widetilde{N}_i \; \rightarrow \; \tilde{a} H} \; \left( \frac{m_{\widetilde{N}_i}}{T}\right)^2 \; 
K_1[m_{\widetilde{N}_i}/T]  \ , \\
\label{eq:nInvDecay}
\mathcal{C}_{\rm FI}^{\rm n-inverse} = & \,
\frac{T^3}{\pi^2}\sum_{i = 1}^4 \theta( m_{\tilde{a}}  - m_{\widetilde{N}_i}) \;
\Gamma_{\tilde{a} \; \rightarrow \; \widetilde{N}_i H} \; \left( \frac{m_{\tilde{a}}}{T}\right)^2 \; 
K_1[m_{\tilde{a}}/T]  \ ,
\end{align}
where $K_1$ is the first modified Bessel function of the second kind.
The Heaviside step function $\theta$ makes sure that only the kinematically allowed channels are accounted for. The correspondent collision operators for the charginos are
\begin{align}
\label{eq:cDecay}
\mathcal{C}_{\rm FI}^{\rm c-decay} = & \,
\frac{2 \, T^3}{\pi^2}\sum_{i = 1}^2 \theta( m_{\widetilde{C}_i} - m_{\tilde{a}} ) \;
\Gamma_{\widetilde{C}_i \; \rightarrow \; \tilde{a} H} \; \left( \frac{m_{\widetilde{C}_i}}{T}\right)^2 \; 
K_1[m_{\widetilde{C}_i}/T]  \ , \\
\label{eq:cInvDecay}
\mathcal{C}_{\rm FI}^{\rm c-inverse} = & \,
\frac{T^3}{\pi^2}\sum_{i = 1}^2 \theta( m_{\tilde{a}}  - m_{\widetilde{C}_i}) \;
\Gamma_{\tilde{a} \; \rightarrow \; \widetilde{C}_i H} \; \left( \frac{m_{\tilde{a}}}{T}\right)^2 \; 
K_1[m_{\tilde{a}}/T]  \ .
\end{align}

\begin{figure}[t]
\begin{center}
\includegraphics[width=0.495\linewidth]{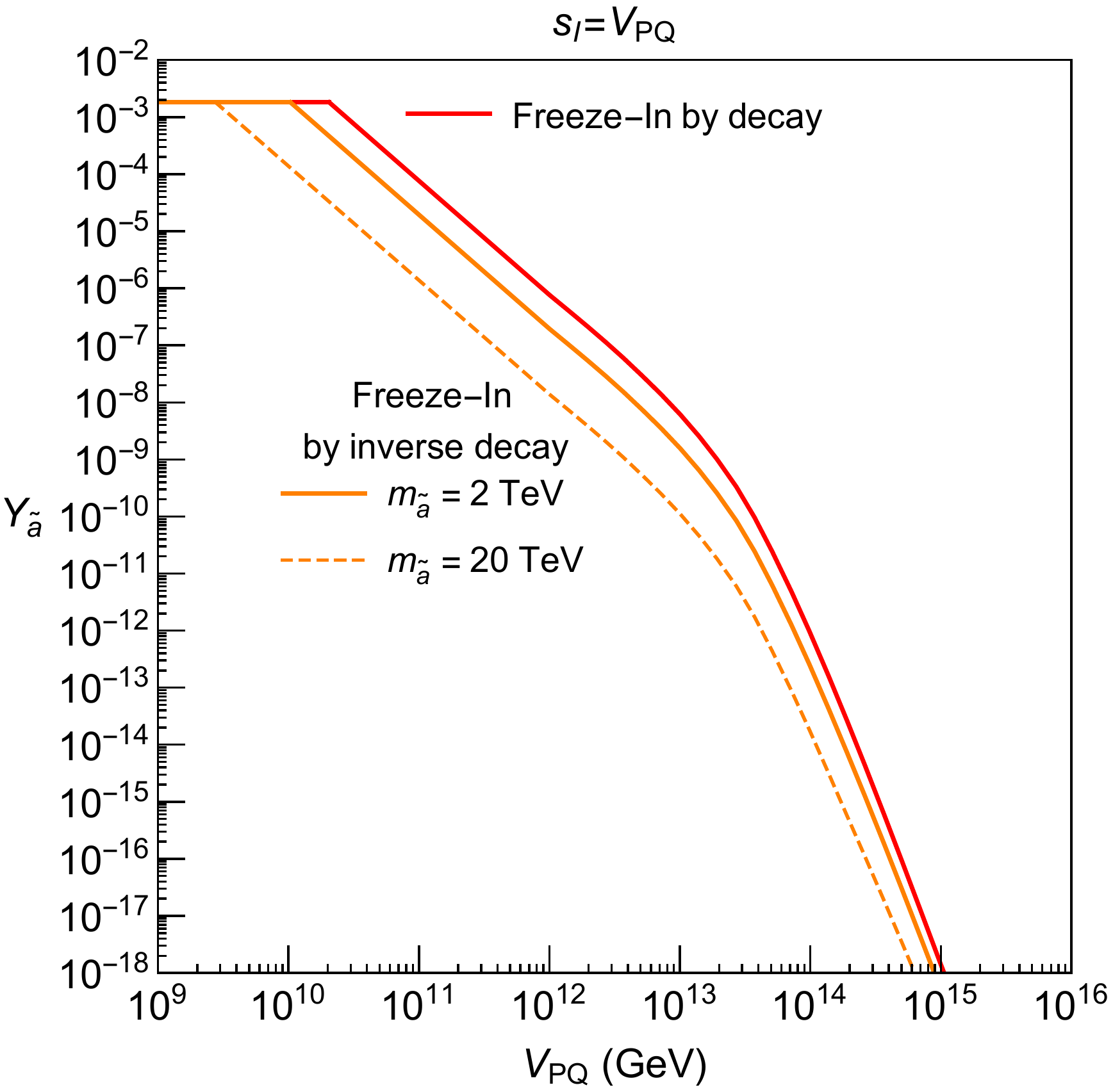} \includegraphics[width=0.495\linewidth]{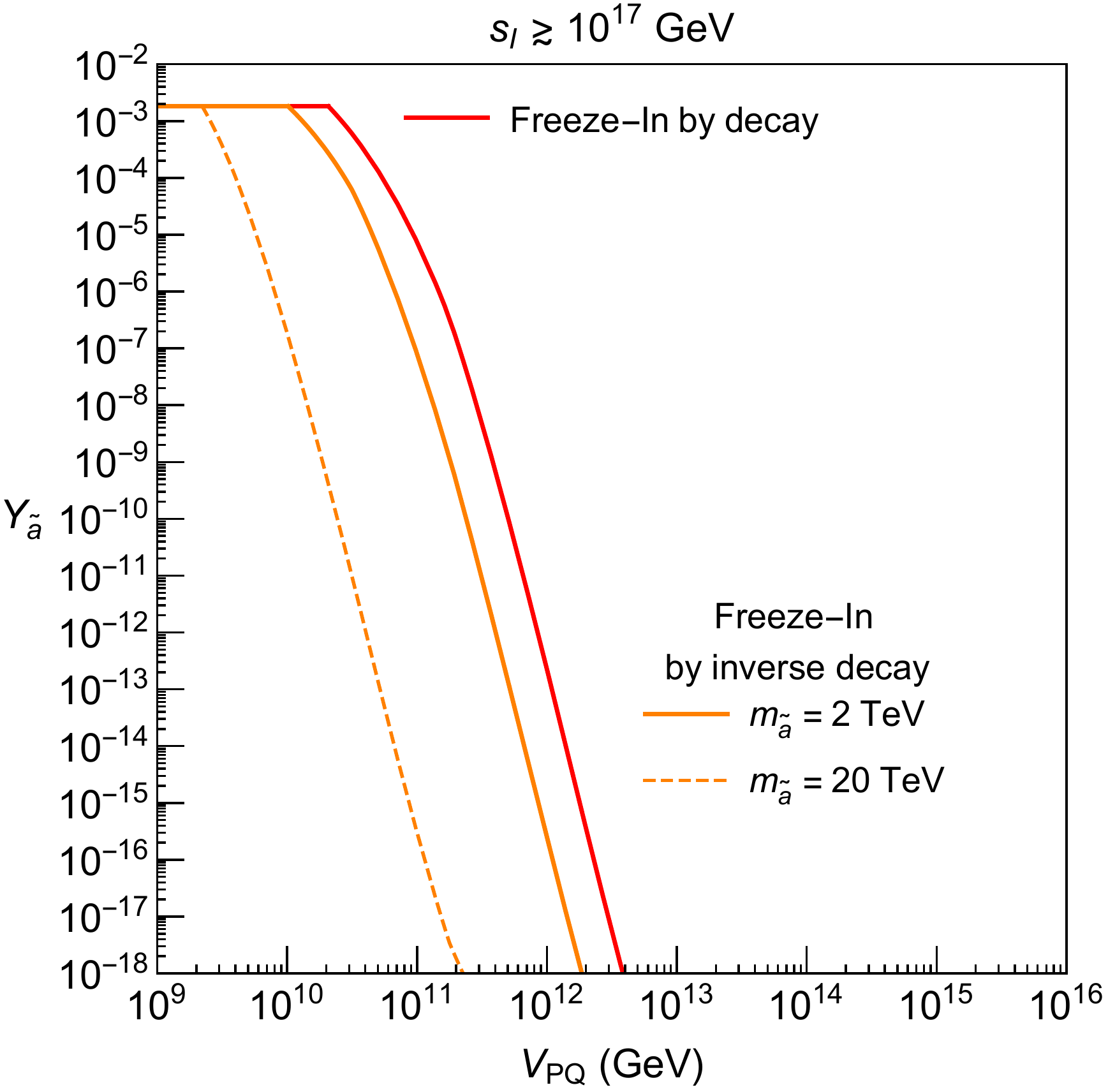}
\end{center}
\caption{The axino yield from neutralino decays (red) for $m_{\tilde{a}} \ll 1$ TeV, and neutralino inverse decays (orange) for $m_{\tilde{a}} = 2$ and 20 TeV. In both panels, $2 M_1 = M_2= \mu =1$ TeV, $m_s=500$ GeV, $\tan \beta = 2$, $q_\mu = 2$, and $\mathcal{D}=4$; while $s_I = V_{PQ} \ (M_* \gsim 10^{17} \GeV)$ for the left (right) panel.}
\label{fig:AxinoYield}
\end{figure}

The results for the axino comoving density are shown in \Fig{fig:AxinoYield}, where the saxion dilution is computed using the cosmology of section \ref{subsec:highTR} with $T_R \gsim 10^{10}$ GeV.  In each of the two panels, we fix the MSSM mass parameters as in the caption, and compute the axino comoving density as a function of $V_{\rm PQ}$.


\subsection{UV Production of Axinos}
\label{sec:UVAxinoProd}

This gravitino problem is greatly exacerbated in PQ theories with heavy matter since then UV production of axinos also generally occurs.  The combination of the two UV production mechanisms provides a particularly powerful upper bound on $T_R$ \cite{Cheung:2011mg}.   
A complication in computing the UV contribution to axino production is that if the heaviest matter carrying both PQ charges and gauge charges, $\Phi$, has a mass $M_\Phi < T_R$ then the UV production is cutoff at $M_\Phi$~\cite{Bae:2011jb}; thus $Y_{\tilde{a}}^{UV}$ is model-dependent.   In the DFSZ$_+$ theory we take $M_\Phi \gsim T_R$, so that the UV axino production is cutoff at $T_R$.  The axino mass is expected to be of order the gravitino mass or larger, in which case, in DFSZ$_+$ with $V_{PQ} \sim 10^{12}$ GeV and TeV scale supersymmetry, $T_R \lsim 10^6$ GeV \cite{Cheung:2011mg}.  Here we show that the limits from UV production of axinos is greatly ameliorated by the decay of the saxion condensate, with resulting limits depending on $(V_{PQ}, m_{\tilde{a}})$.

With UV production of axinos cut off at $T_R$, scattering leads to an axino undiluted yield \cite{Strumia:2010aa} 
\be
\label{eq:Yaxino}
Y_{\tilde{a}}^{UV}|_+ \, \sim \, 2 \times 10^{-6} \, g_3^6 \, \ln\left(\frac{3}{g_3}\right) \, \left( \frac{ N_{DW}}{6} \right)^2  \left( \frac{T_R}{10^{10} \GeV} \right) \left( \frac{10^{14} \GeV}{V_{PQ}} \right)^2 \ ,
\ee
where $g_3$ is the strong gauge coupling.
As elsewhere, we take $N_{DW} =6$.  The yields of axinos and gravitinos should not exceed the thermal equilibrium value
\be
\label{eq:Yeq}
Y_{eq} = \frac{135 \zeta(3) g}{8 \pi^4 g_*} \simeq 1.8\times10^{-3} \left( \frac{g}{2} \right) \left( \frac{228.75}{g_*} \right) ,
\ee
where $\zeta$ is the zeta function and the internal degrees of freedom $g$ is 2 (4) for the axino (gravitino). 

\begin{figure}[t]
\begin{center}
\includegraphics[width=0.495\linewidth]{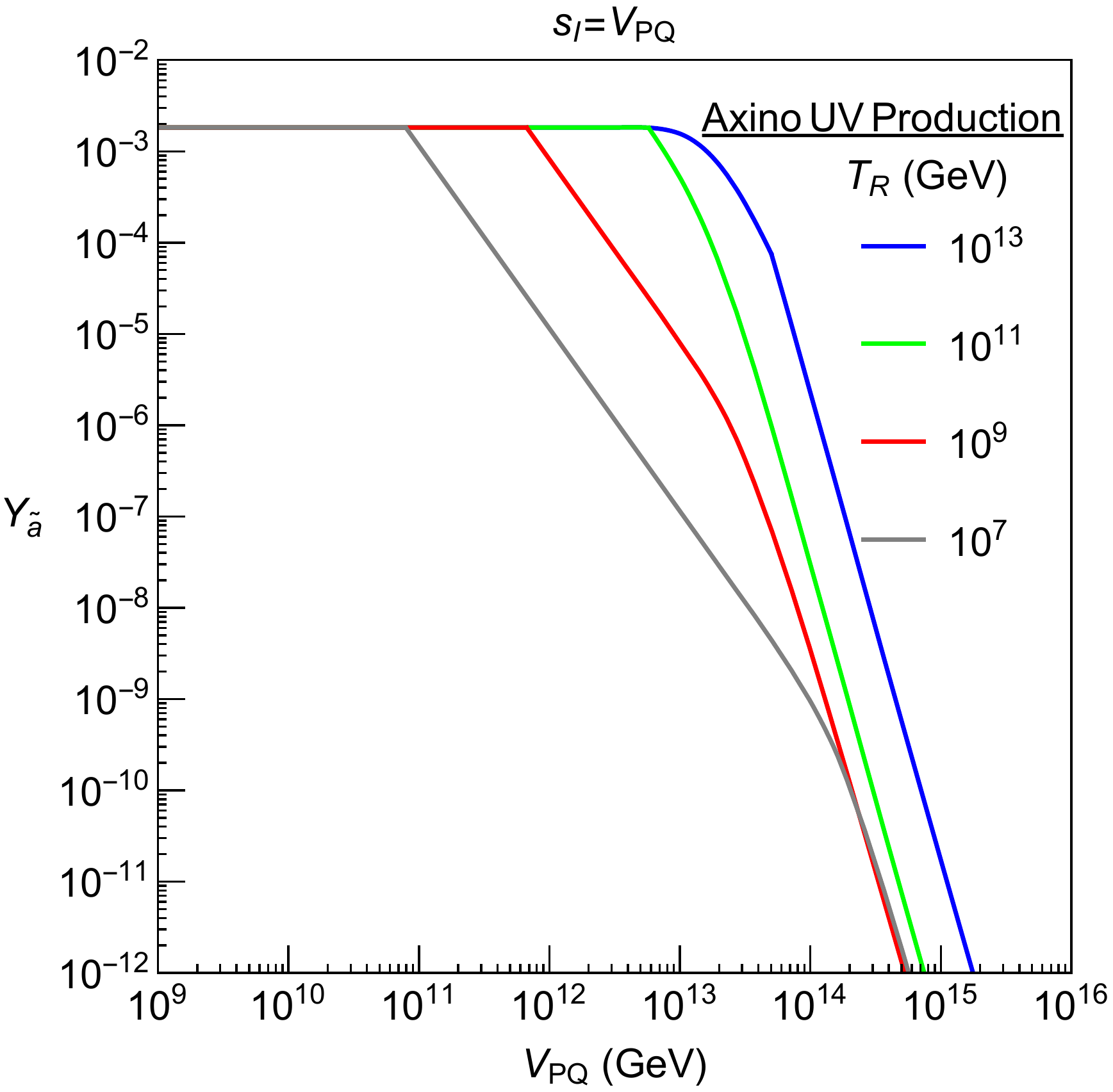} \includegraphics[width=0.495\linewidth]{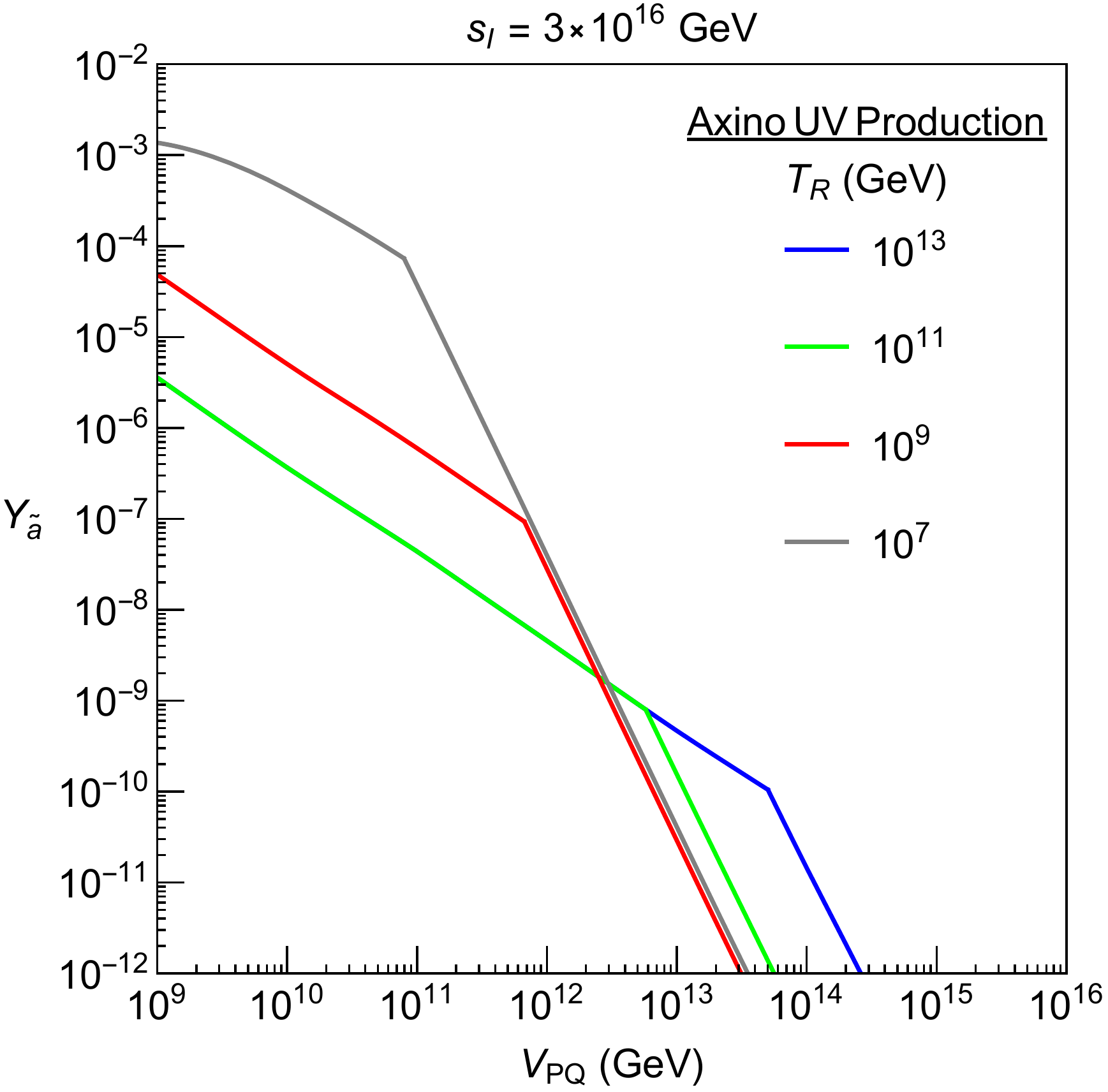}
\end{center}
\caption{The axino yield from UV scattering. In both panels, $m_s = \mu =1$ TeV, $q_\mu = 2$, and $\mathcal{D}=4$, while $s_I = V_{PQ} \ (s_I = 3\times10^{16} \GeV)$ for the left (right) panel. We take the axion domain wall number $N_{DW}=6$ for the axion decay constant $f_a = \sqrt{2} V_{PQ}/N_{DW}$.}
\label{fig:AxinoYieldUV}
\end{figure}

To include the effect of saxion dilution, we divide the axino abundance by the dilution factor computed numerically. Since axino UV production occurs before the saxion injects the entropy, the dilution factor is simply the ratio of the total entropy before and after the saxion MD era. The analytic formulas for the dilution factors are also given in \Eqs{eq:SaxD}{eq:DI}.

The results for the axino abundance are shown in \Fig{fig:AxinoYieldUV}, for some values of $T_R \gsim 10^{10}$~GeV that have saxion dilution of \Sec{subsec:highTR} and others of \Sec{subsec:lowTR} with $T_R \lsim 10^{10}$ GeV. We take $s_I = V_{PQ} \ (M_*=3\times10^{16} \GeV)$ in the left (right) panel. Note that for a sufficiently low $V_{PQ}$, the axino reaches thermal equilibrium, in which case we will use the equilibrium value of the yield in \Eq{eq:Yeq}. In particular, the sharp kink in each of the curve is due to the transition from the thermal value $Y_{eq}$ to the yield from UV scattering. In the left panel, the smooth change of the slope at $V_{PQ} \sim 10^{13-14}$ GeV is due to the emergence of the saxion MD era demonstrated in \Figs{fig:TMvsTRa}{fig:TMvsTRb}. In the right panel, the dilution effect is present for the entire range of $V_{PQ}$. In the case where $T_R \lsim 10^{10}$ GeV, one interesting feature is that the diluted abundance is (almost) independent of $T_R$ because both dilution in \Eq{eq:DI} and production in \Eq{eq:Yaxino} are proportional to $T_R$ (other than the mild dependence on $T_R$ in $g_3$). 

\subsection{UV Production of Gravitinos}
\label{sec:UVGraviProd}

\begin{figure}[t]
\begin{center}
\includegraphics[width=0.495\linewidth]{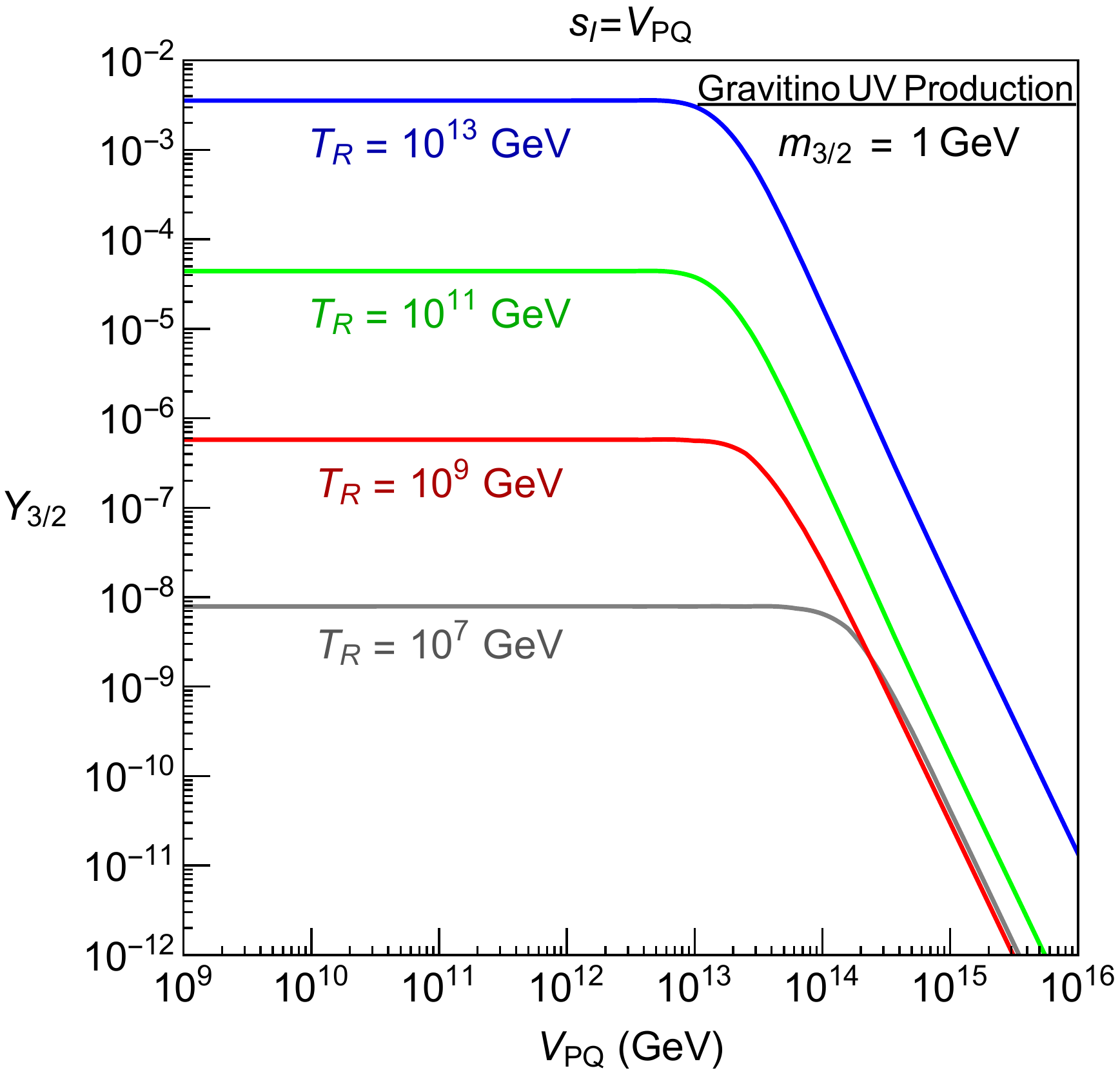} \includegraphics[width=0.495\linewidth]{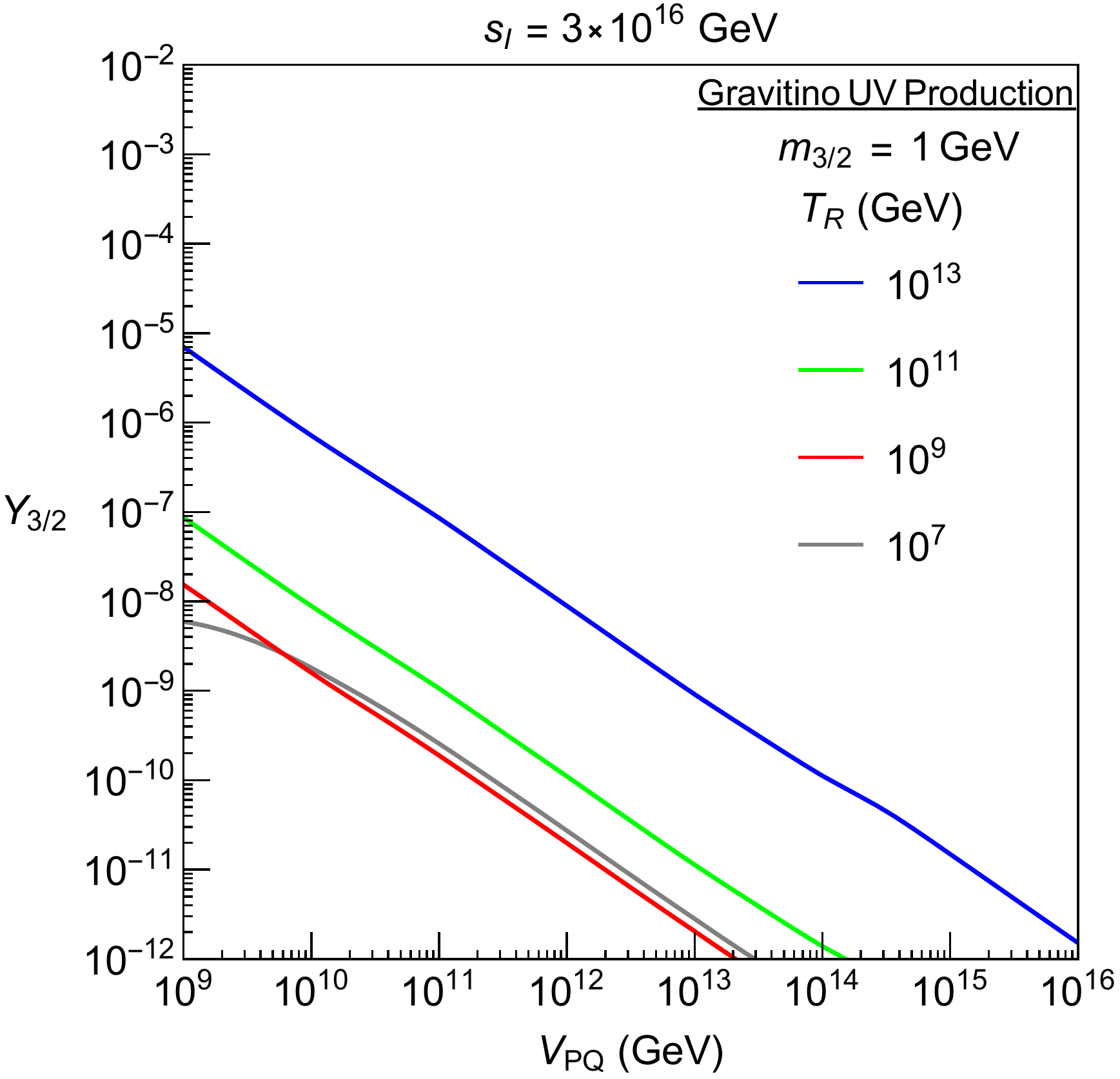}
\end{center}
\caption{The gravitino yield from UV scattering and saxion dilution for $s_I = V_{PQ} \ (s_I = 3\times10^{16} \GeV)$ in the left (right) panel.  In both panels, $m_s = \mu =1$ TeV, $q_\mu = 2$, $\mathcal{D}=4$, and the unified gaugino mass is $2$ TeV and $m_{3/2}=1$ GeV.  The yield scales as $Y_{3/2} \propto 1/m_{3/2}^2$, as long as $Y_{3/2}$ is below its equilibrium value. 
}
\label{fig:GravYield}
\end{figure}

In supersymmetric theories UV production of gravitinos generally limits the reheat temperature after inflation, $T_R$ \cite{Nanopoulos:1983up}.  The undiluted abundance of gravitinos from scattering at $T_R$ is \cite{Moroi:1993mb}
\be
\label{eq:Ygravitino}
Y_{3/2}^{UV} \,\sim \,  6 \times 10^{-12} \frac{T_R}{10^{10} \GeV} \sum_i \gamma_i(T_R) \left( 1+  \frac{m_i^2}{3m_{3/2}^2} \right),
\ee
where $\gamma_i (T_R) \sim (0.02, 0.08, 0.25-0.4, 0.02)$ and $m_i = ( m_{(1,2,3)}, \, A_t )$ is the gaugino mass of $\left( U(1), SU(2), SU(3)\right)$ and the $A$-term of the top Yukawa coupling.

Similar to \Sec{sec:UVAxinoProd}, we calculate the final gravitino abundance by dividing the yield by the dilution factor computed numerically. The analytic estimate of the dilution factors are also provided in \Eqs{eq:SaxD}{eq:DI}.

The numerical results for the gravitino abundance are shown in \Fig{fig:GravYield} for $T_R \ge 10^{10}$ GeV ($T_R \le 10^{10}$ GeV) with saxion dilution of \Sec{subsec:highTR} (\Sec{subsec:lowTR}).   In the left panel, with $s_I = V_{PQ}$, there is no dilution for $V_{PQ} < V_{PQ}^{(c)}$ because the condensate is too small for the saxions to dominate before they decay.   The key feature is the rapid dilution for $V_{PQ} > V_{PQ}^{(c)}$, with $V_{PQ}^{(c)} \sim 10^{13}$ GeV for $T_R \gsim 10^{10}$ GeV and growing for smaller $T_R$.  In the right panel $s_I = M_* = 3\times10^{16} \GeV$ is large everywhere, so a saxion dominated MD era occurs at much lower values of $V_{PQ}^{(c)}$, leading to dilution at low $V_{PQ}$. For $T_R < 10^{10}$~GeV, the final diluted yield is nearly independent of $T_R$ (other than $\gamma_i(T_R)$ in \Eq{eq:Ygravitino}). 

\subsection{Freeze-In Production of Gravitinos}
\label{sec:FIGraviProd}

Gravitinos are also produced via decays of thermal charginos and neutralinos, $\widetilde{N}_i/\widetilde{C}_i \rightarrow \tilde{G}$~\cite{Cheung:2011nn}. These freeze-in processes are IR dominated and independent of $T_R$. The resulting yield, which is proportional to the decay width to the gravitino, is enhanced for low values of the gravitino mass. This is just a consequence of the production of longitudinal gravitinos (goldstinos), as manifestly shown in Eqs.~(\ref{eq:axinodecaygrav2})--(\ref{eq:graviToh}). For the same reason, gravitino UV production is also enhanced for small $m_{3/2}$, as shown in \Eq{eq:Ygravitino}. We checked that the gravitino FI contribution is always sub-dominant compared to those from gravitino UV and axino FI in the parameter space of interest, and therefore we do not consider it in this work.

\section{Axino and Gravitino as the Lightest Superpartners}
\label{sec:AxGravLSP}

We classify the superpartner spectra according to the size of the mediation scale for supersymmetry breaking, $M_{\rm mess}$.  Superpartner masses arise from the effective operators
\begin{align}
\mathcal{L}_{\rm mess} \, =  \, \frac{c_{a}}{M_{\rm mess}} \int d^2 \theta \, X \, W_a^\alpha W_{a \alpha}  \, +  \, \frac{c_{Q}^2}{M_{\rm mess}^2}  \int d^2 \theta \, X^\dag X \, Q^\dag Q
\end{align}
where the superfield $X$, defined in \Eq{eq:Xdef}, has a SUSY breaking F-component.
We take the model-dependent coefficients $c_{a}$ and $c_{Q}$ to be comparable $c_{Q} \simeq c_{a}$; the spectrum is not split. 

In \Sec{sec:HighScale} we consider $M_{\rm mess}$ of order $M_{\rm Pl}$, which can be broadly identified with ``gravity mediation.'' The gravitino for this case cannot be much lighter than the other superpartners. This has to contrasted with the low mediation scale case elaborated in \Sec{sec:LowScale}, $M_{\rm mess}  \ll M_{\rm Pl}$, where the gravitino is much lighter than other superpartners. We focus our attention on the spectra where the axino and the gravitino are both lighter than all the other superpartners. However, for gravity mediation we do not commit to any relative hierarchy between them, and consider both axino and gravitino LSP cases.

For both high and low mediation scales, axinos and gravitinos are produced in the early Universe through the various mechanisms discussed in the previous Section. These processes produce both the NLSP and the LSP, and they all contribute to the present dark matter abundance, which today is made of LSP particles only. Highly relativistic axions are produced in NLSP decays, which make a negligible contribution to dark radiation. Furthermore, since the final products are the LSP and the axion, these decays are not subject to BBN limits~ \cite{Jedamzik:2004er,Kawasaki:2004yh,Ellis:2005ii}. As we will see in this Section, the origin of the current dark matter abundance is typically due to either NLSP or LSP production, unless we consider peculiar parameter space regions. 

Before we discuss the high and low scale cases in detail, we highlight the main features of our framework.

\subsection{Warm Dark Matter from NLSP Decays}
\label{sec:WDM}
 Dark matter from LSP production is always cold. On the contrary, dark matter from NLSP production and decay  could be hot, warm or cold, depending on the ratio of NLSP and LSP masses and the NLSP decay lifetime. The LSP is a gauge singlet extremely weakly coupled to the radiation bath, and thus DM particles coming from NLSP decays just lose their momenta by pure free streaming. This has the effect of potentially washing out cosmological perturbations at small scales through free streaming, and this scenario is severely constrained by Lyman-$\alpha$ forest observations~\cite{Boyarsky:2008xj,Harada:2014lma,Kamada:2016vsc}, bounding the free streaming length to be less than $\simeq 1 \, {\rm Mpc}$. Interestingly, if the free streaming length is consistent with large scale structure and not too small it can address some large scale structure (LSS) issues that are indicated by simulations of collisionless cold dark matter~\cite{Weinberg:2013aya}.  Baryonic feedback effects can explain some discrepancies \cite{El-Badry15,Wetzel:2016wro}, although there is much debate on this~\cite{Oman:2015xda,Papastergis:2015}.

\begin{figure}[t]
\begin{center}
\includegraphics[width=0.495\linewidth]{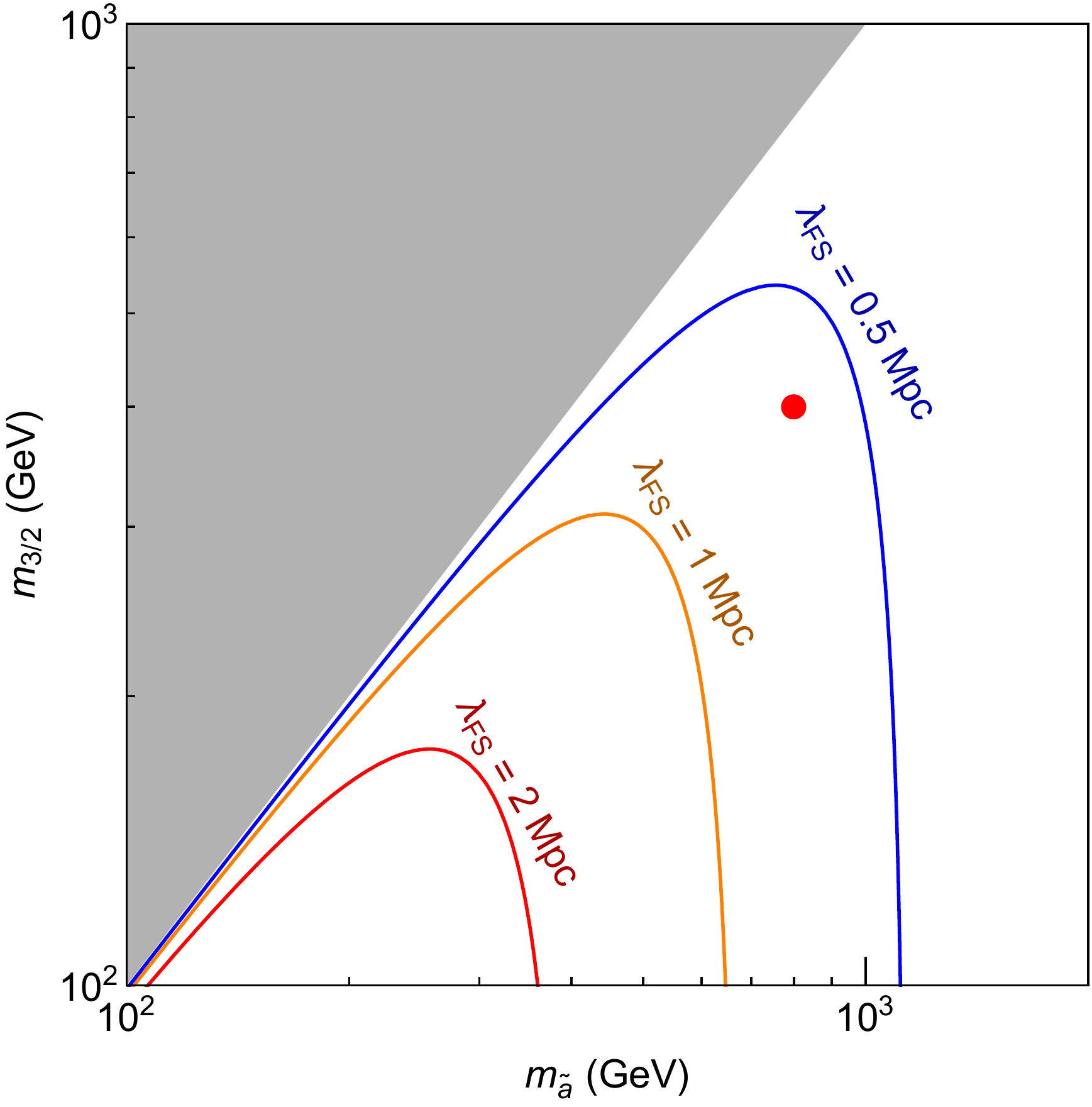}
\includegraphics[width=0.495\linewidth]{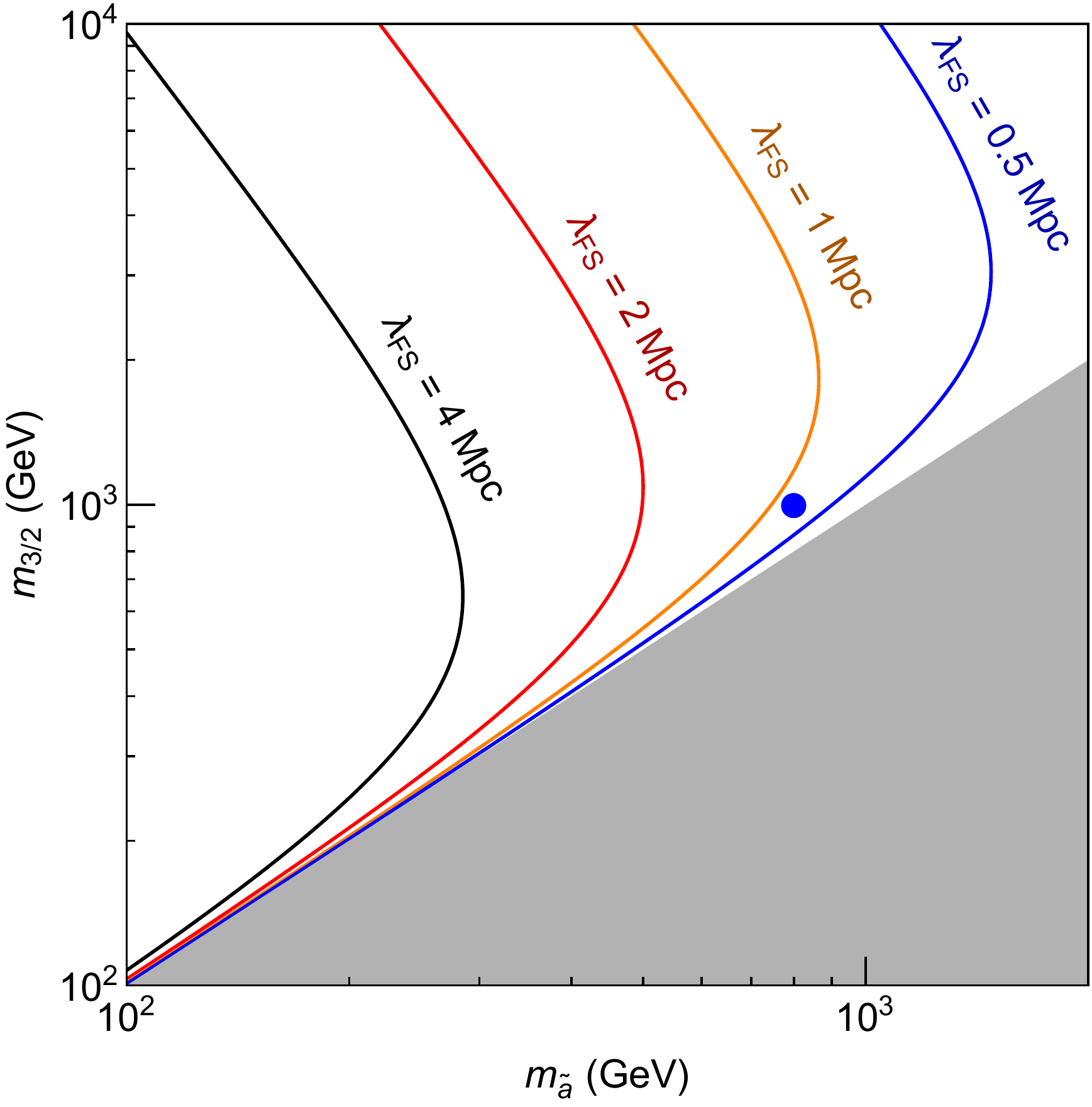}
\end{center}
\vspace{-0.2cm}
\caption{Free streaming length $\lambda_{FS}$ for the warm dark matter component. In the left (right) panel we show the result for gravitino (axino) LSP produced through the axino (gravitino) NLSP decay process $\tilde{a} \rightarrow \tilde{G} a$ ($\tilde{G} \rightarrow \tilde{a} a$). The red and blue dots label the spectra we choose for High Scale mediation in \Sec{sec:HighScale}. For Low Scale mediation in \Sec{sec:LowScale}, we only need to require $m_{\tilde{a}} \gsim 650$ GeV since the free streaming length becomes independent of $m_{3/2}$.}
\label{fig:FSLength}
\end{figure}

If the axino is the NLSP, it decays to longitudinal gravitinos (i.e. goldstino) via the effective operators given in \Eq{eq:gravitinodecayop2}, giving a lifetime from \Eq{eq:axinodecaygrav2} of
\be
\tau_{\tilde a} = \Gamma^{-1}_{\tilde a \rightarrow  a \, \tilde G} \simeq 1.2 \times 10^4 \, {\rm sec} \, 
\left( \frac{m_{3/2}}{100 \, {\rm GeV}} \right)^2 \left( \frac{1 \, {\rm TeV}}{m_{\tilde a}} \right)^5 . 
\label{eq:axinolifetime}
\ee
The associated free streaming length for free streaming gravitinos is approximately given by the expression in \Eq{eq:lambdaFSpprox}, which for the lifetime above reads
\be
\lambda^{\tilde G}_{\rm FS} \simeq 0.6 \, {\rm Mpc} \, \left( \frac{1 \, {\rm TeV}}{m_{\tilde a}} \right)^{3/2}
 \left[ 1 +  0.15 \log\left( \frac{m_{\tilde a}}{1 \, {\rm TeV}} \right)\right] \ .
\label{eq:lambdaFSpproxGRAVITINO}
\ee
This expression holds as long as $m_{\tilde a} \gg m_{3/2}$. Remarkably, the gravitino free streaming length depends only on the axino mass, and for TeV scale axinos is not in conflict with Lyman-$\alpha$ forest observations and in the correct ballpark to address LSS anomalies. Numerical results are shown in the left panel of Fig.~\ref{fig:FSLength}, where we draw $\lambda_{\rm FS}$ isocontours in the $(m_{\tilde a}, m_{3/2})$ plane. These results are derived by using the full derivation of $\lambda_{\rm FS}$ in App.~\ref{app:FreeStreaming}. In the $m_{\tilde a} \gg m_{3/2}$ regime, where the free streaming length is approximated by \Eq{eq:lambdaFSpproxGRAVITINO}, the isocontours are vertical lines in the  $(m_{\tilde a}, m_{3/2})$ plane.

For gravitino NLSP, the decays to axinos are mediated by the operator in \Eq{eq:gravitinodecayop}. We assume that the decay to saxion and axino final state is kinematically forbidden, and thus  the decay width is half of the expression in \Eq{eq:gravdecchiral}, to account for decays to axion and axino only. The resulting lifetime is
\be
\tau_{3/2} = \Gamma^{-1}_{\tilde{G} \rightarrow  a \, \tilde a} \simeq 8.5 \times 10^6 \, {\rm sec} \, \left( \frac{1 \, {\rm TeV}}{m_{3/2}} \right)^3 \ .
\label{eq:gravitinolifetime}
\ee
To simplify the discussion we take the axino mass to be of order 1 TeV -- i.e. not far below the other superpartner masses. Thus, while \Eq{eq:axinolifetime} can be used for both High and Low Scale mediation, \Eq{eq:gravitinolifetime} is used only in High Scale mediation. In both equations we have ignored phase space factors that become relevant when the NLSP and LSP masses are comparable. The axino free streaming length, in the limit where $m_{\tilde a} \ll m_{3/2}$, is approximately given by
\be
\lambda^{\tilde a}_{\rm FS} \simeq 1.2 \, {\rm Mpc} \, \left( \frac{2 \, {\rm TeV}}{m_{3/2}} \right)^{1/2} 
\left( \frac{1 \, {\rm TeV}}{m_{\tilde a}} \right)
\left[ 1 +  0.1 \log\left(\left( \frac{m_{3/2}}{2 \, {\rm TeV}} \right)^{1/2} 
\left( \frac{m_{\tilde a}}{1 \, {\rm TeV}} \right) \right)\right] \ .
\label{eq:lambdaFSpproxAXINO}
\ee
The full result is shown in the right panel of Fig.~\ref{fig:FSLength}. In the $m_{\tilde a} \ll m_{3/2}$ region, the free streaming isocontours are along the lines $m_{3/2} m^2_{\tilde a} = {\rm const}$, consistently with the approximate expression in \Eq{eq:lambdaFSpproxAXINO}. Lyman-$\alpha$ bounds are evaded for sufficiently heavy gravitino and/or axino, and issues with LSS can be addressed again by TeV scale superpartners. 

A related noteworthy example is the case of gravitinos coming from LOSP freeze-out and decays~\cite{Kaplinghat:2005sy,Cembranos:2005us}, which in our case is made irrelevant by the saxion condensate dilution. 

\subsection{Displaced Signals at Colliders}
\label{sec:displaced}

\begin{figure}[t]
\begin{center}
\includegraphics[width=0.495\linewidth]{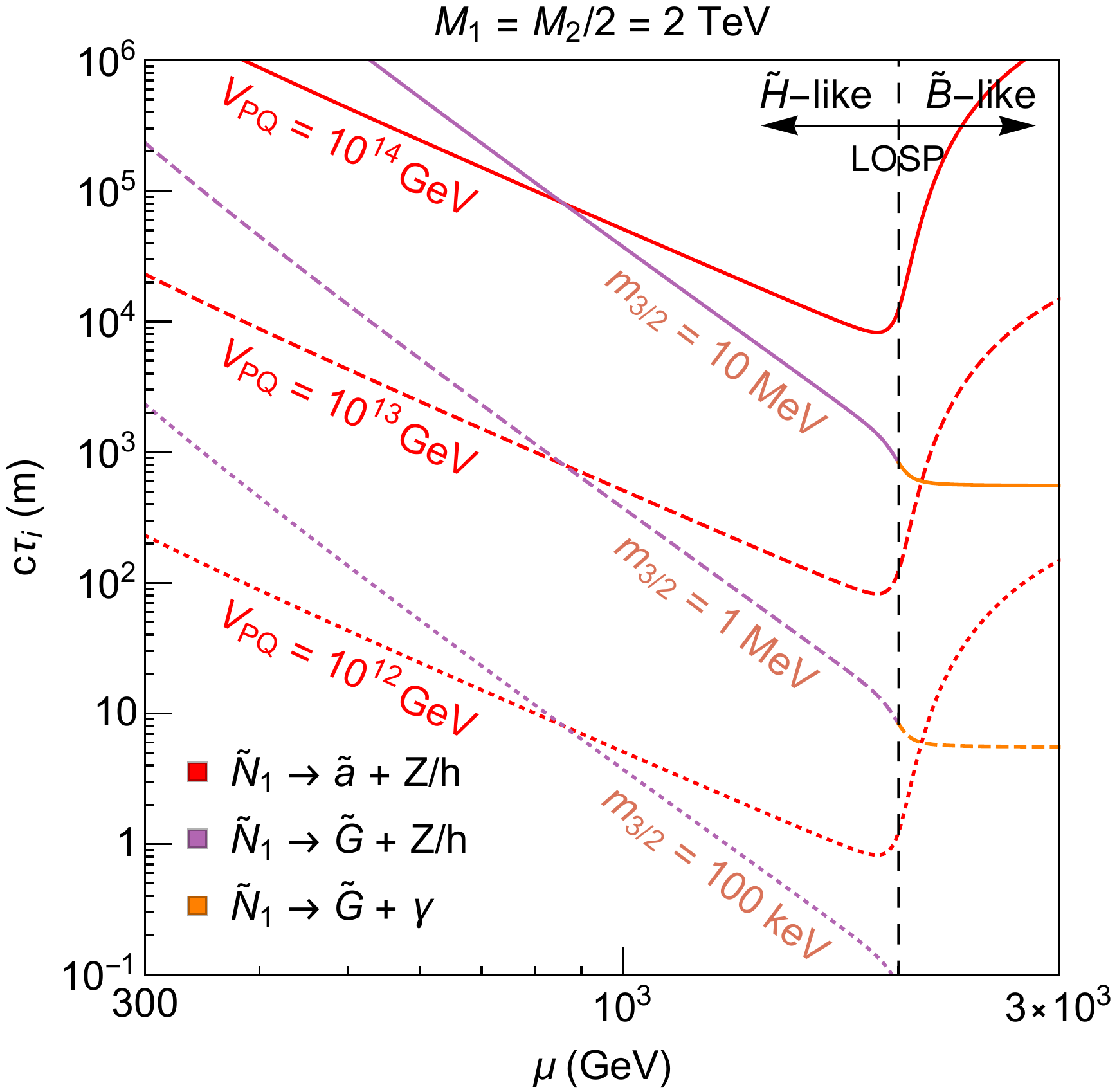}
\includegraphics[width=0.495\linewidth]{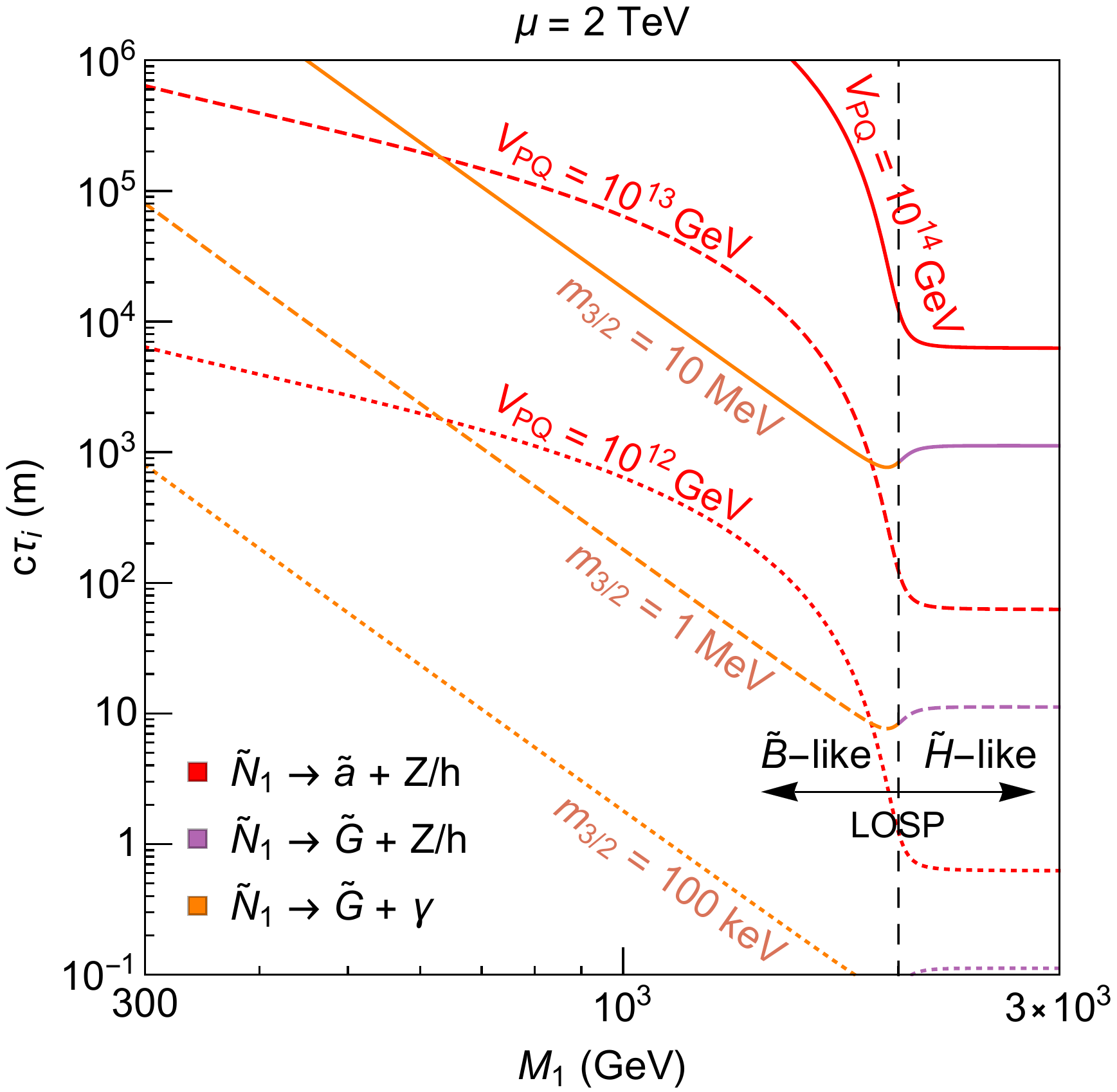}
\end{center}
\vspace{-0.2cm}
\caption{
The neutralino LOSP inverse partial decay width $c\tau_i = c\Gamma_{\tilde{N}_1 \rightarrow i}^{-1}$ ($i = \tilde{a}, \tilde{G}$) as a function of $\mu \ (M_1)$ for fixed $M_1 \ (\mu)$ in the left (right) panel, with $q_\mu=2$, $\tan\beta=2$, and $M_2 \simeq 2 M_1$.  
Curves for $c\tau_{\tilde{G}}$ are shown in red and are labelled by values of $V_{PQ}$.  
Curves for $c\tau_{\tilde{a}}$ are labelled by values of $m_{3/2}$ and are shown in purple when the LOSP is Higgsino-like and in orange when it is bino-like. 
The LOSP lifetime is given approximately by the smallest $c\tau_i$. 
}
\label{fig:LOSPdecayLength}
\end{figure}

 If the lightest observable-sector supersymmetric partner (LOSP) is a neutralino ${\widetilde{N}}_1$, its decay to the axino and Higgs/longitudinal Z bosons will leave missing energy and a displaced vertex. The ``lifetime" for this channel can be obtained from the associated decay width given in \Eq{eq:Gammaitoa} of \App{app:Decay},
\be
c\tau_{\tilde{a}} \equiv \frac{c}{\Gamma_{\tilde{N}_1 \rightarrow \tilde{a}}} \simeq  2.5 \, \text{m} \; \frac{1}{k_{\tilde{a}}} \left( \frac{2}{q_\mu} \right)^2 \left( \frac{ \mu }{ m_{{\widetilde{N}}_1} }  \right)  \left( \frac{ 10^3 \ \text{GeV} }{\mu}  \right)^3  \left( \frac{ V_{PQ} }{10^{12} \ \text{GeV}}  \right)^2  \ .
\label{eq:LOSPctauAxi}
\ee
Here, we define the mixing factor $k_{\tilde{a}} \equiv \squared{s_\beta R_{31} } + \squared{c_\beta R_{41} } $, with $R_{ij}$ the neutralino mixing matrix (for details see Eqs.~\eqref{eq:NvsChi} and \eqref{eq:Gammaitoa}). In the pure Higgsino LOSP limit, with decoupled bino and wino, $k_{\tilde{a}}$ becomes $1/2$ and is thus independent of $\beta$. On the other hand, $k_{\tilde{a}}$ will be suppressed when the LOSP is bino- or wino-like.

For Low Scale mediation with $m_{3/2} \lsim 1$ MeV, the decay channel of neutralinos to gravitinos becomes sufficiently enhanced to dominate over the axino final state, giving a ``lifetime"
\be
c\tau_{\tilde{G}} \equiv \frac{c}{\Gamma_{\tilde{N}_1 \rightarrow \tilde{G}}} \simeq  2 \, \text{m} \; \frac{1}{k_{\tilde{G}}} \left( \frac{ 1 \TeV }{ m_{{\widetilde{N}}_1} }  \right)^5  \left( \frac{ m_{3/2} }{ 100 \ \text{keV} }  \right)^2 ,
\label{eq:LOSPctauGravi}
\ee
where $k_{\tilde{G}}$ contains the analogous mixing factors given in the decay widths Eqs.~(\ref{eq:graviToGamma})--(\ref{eq:graviToh}).


The lifetime that can be probed at the LHC and future colliders depends on the total production cross section of supersymmetric particles, which we quote for the case of degenerate squark and gluino masses, $\tilde{m}$ \cite{Borschensky:2014cia}.  For $\tilde{m} = (1.5,2.5)$ TeV, at $\sqrt{s} = 14$ TeV this cross section is of order (100,1) fb, so that planned runs of the LHC will allow ATLAS and CMS to reach $c \tau$ of order (100, 10)m.  
Recently, a surface detector called MATHUSLA \cite{Chou:2016lxi} has been proposed to search for (ultra) long-lived particles at the LHC and future colliders.  At the LHC, with 30 ab$^{-1}$ at $\sqrt{s} = 14$ TeV, Fig. 4 of \cite{Chou:2016lxi} implies a reach in $c \tau$ of order $(10^5, 10^3)$m for $\tilde{m} = (1.5,2.5)$ TeV.   At a future 100 TeV collider \cite{Tang:2015qga}, with a susy production cross section of $(10^3,10)$ fb for $\tilde{m} = (3,8)$ TeV,  Fig. 5 of \cite{Chou:2016lxi} implies a reach in $c \tau$ of order $(10^7, 10^5)$m. 

In \Secs{sec:HighScale}{sec:LowScale} we show predictions for $c \tau_i$,  ($i = \tilde{a}, \tilde{G}$), following from the constraint $\Omega h^2 = 0.11$, in theories with High and Low Scale mediation for the particular point in supersymmtric parameter space of $(\mu, M_1, \tan \beta)$ = (1 TeV, 1 TeV, 2).  Here we illustrate the variation in the $c \tau_i$ as the parameter space changes, always keeping the unified gaugino mass relation, $M_2 \simeq 2 M_1$.  In \Fig{fig:LOSPdecayLength}, we display curves for $c\tau_{\tilde{G}}$ in red, labelled by values of $V_{PQ}$. The  
curves for $c\tau_{\tilde{a}}$ are in purple when the LOSP is Higgsino-like and in orange when it is bino-like, labelled by values of $m_{3/2}$. 
The LOSP lifetime $c\tau_{\tilde{N}_1}$ is given approximately by the smallest $c\tau_i$. In the left (right) panel, we vary $\mu$ ($M_1$) while fixing $M_1$ ($\mu$). The change of behavior in the lifetime curves at $\mu = 2$ TeV ($M_1 = 2$ TeV) in the left (right) panel reflects the fact that the mixing factors in \Eq{eq:LOSPctauAxi} and \Eq{eq:LOSPctauGravi} drastically change as $\mu$ becomes larger or smaller than $M_1$. 
The neutralino LOSP decay can lead to observable displaced signals at the LHC and future colliders over a remarkably wide range of parameter space.

\subsection{Axion Dark Radiation}
\label{sec:DR}

Saxions can decay to axions with a rate given by \Eq{eq:stoaa} if $\kappa$ does not vanish due to symmetry. Using the branching ratio of the saxion to the visible sector and to axions, we predict the amount of dark radiation to be \cite{Co:2016vsi}
\be
\Delta N_{eff} = \ 3 \ \frac{\rho_{a}}{\rho_\nu} \simeq  \frac{43}{112}  \left( \frac{4}{\mathcal{D}} \right) \left( \frac{g_*(1\text{ MeV})}{10.75}  \right) \left( \frac{\kappa}{q_\mu} \right)^2 \,  \left(\frac{m_s}{\mu} \right)^{4} .
\ee
The Planck experimental bound \cite{Ade:2015xua} is $\Delta N_{eff} < 0.6$. The proposed CMB Stage-IV experiment \cite{Abazajian:2016yjj} can be sensitive to $\Delta N_{eff} = 0.03$.

\section{Results for High Scale or ``Gravity" Mediation}
\label{sec:HighScale}
\subsection{The DFSZ$_0$ Theory}
\label{sec:DFSZ0}

\begin{figure}
\begin{center}
\caption*{\bf High Scale mediation: DFSZ$_0$}
\includegraphics[width=0.495\linewidth]{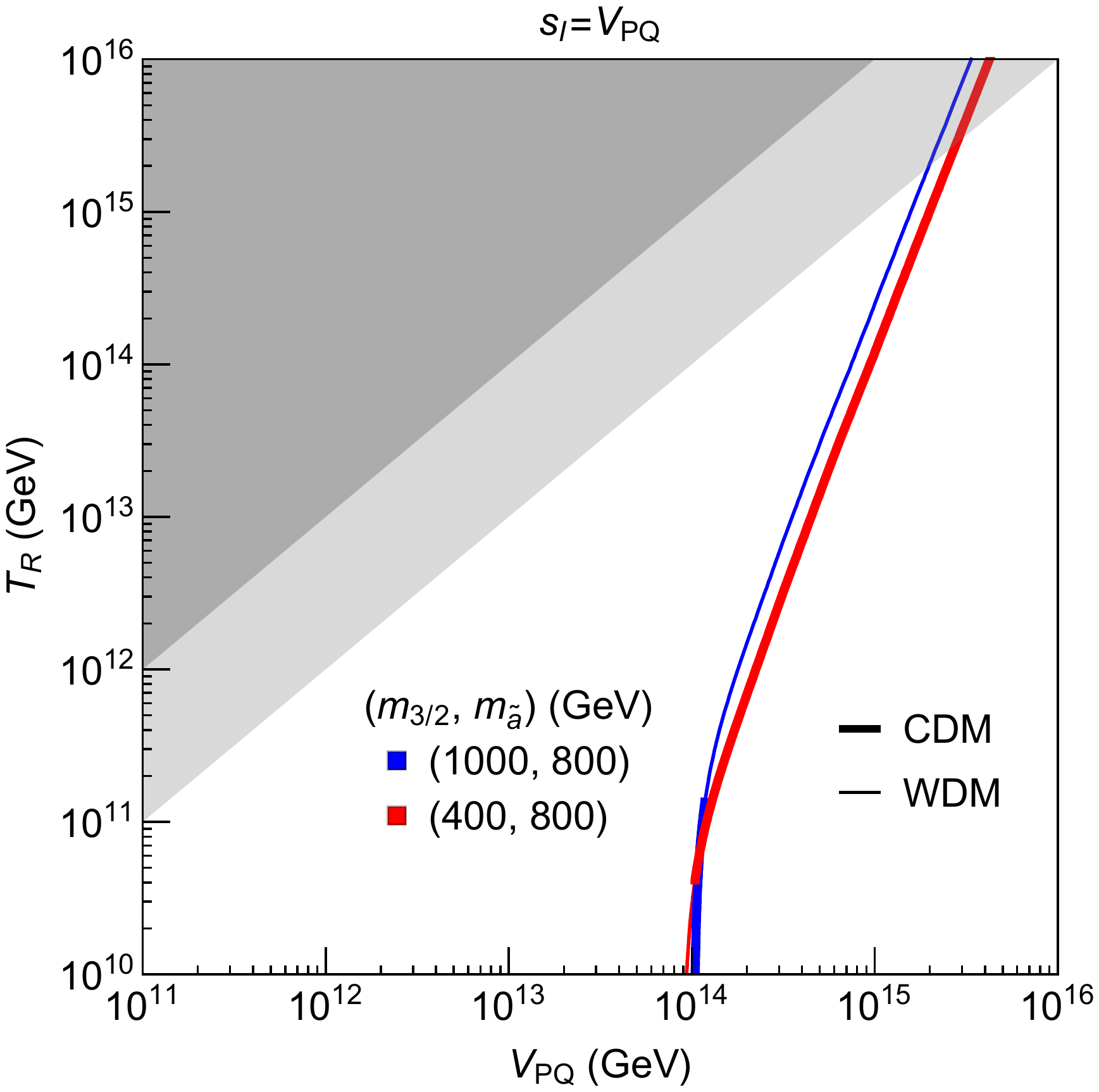} \includegraphics[width=0.495\linewidth]{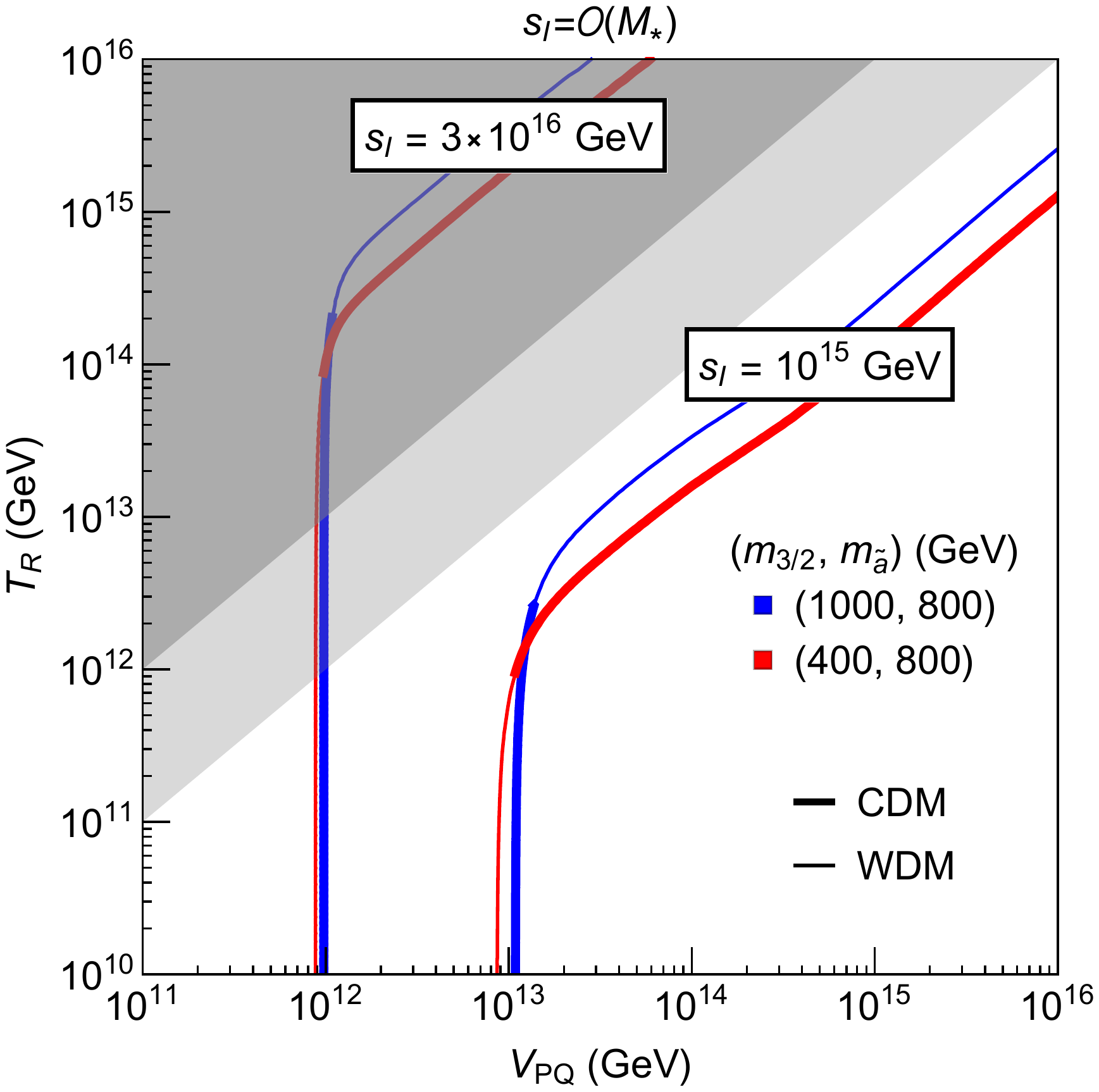}
\includegraphics[width=0.495\linewidth]{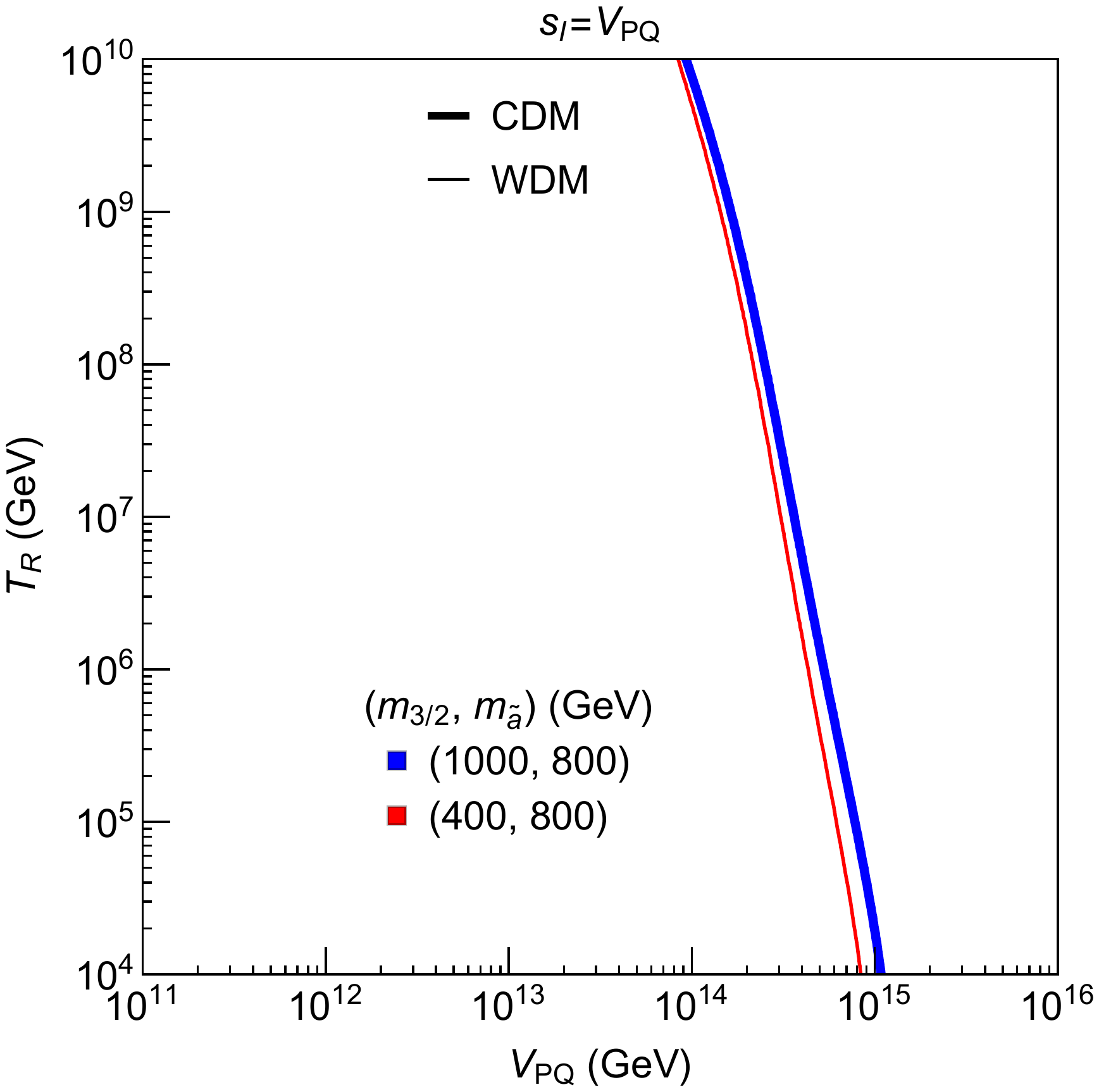} \includegraphics[width=0.495\linewidth]{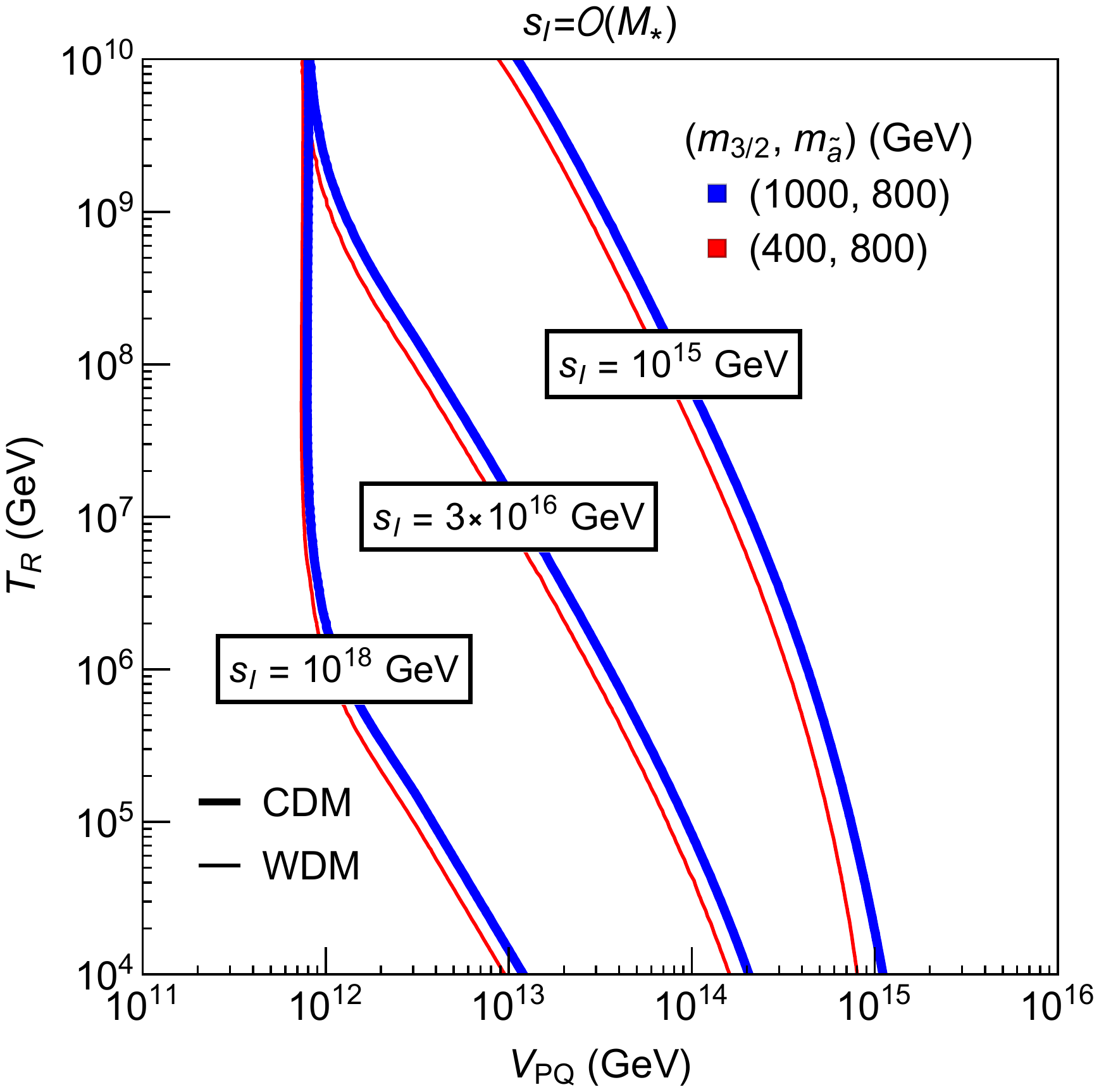}
\caption{Contours of $\Omega h^2 = 0.11$ from axino freeze-in and gravitino UV production. We fix $q_\mu=2$, $\mathcal{D}=4$, $\tan\beta=2$, and $M_2/2 = M_1 = \mu = 1$ TeV and $m_s = 600$ GeV. The top (bottom) row is for the cosmology with $T_R \gsim (\lsim) 10^{10}$ GeV discussed in \Sec{subsec:highTR} (\Sec{subsec:lowTR}). }
\label{fig:HighDFSZ0}
\end{center}
\end{figure}

\begin{figure}[t]
\begin{center}
\caption*{\bf High Scale mediation: DFSZ$_0$ + Neutralino LOSP}
\includegraphics[width=0.495\linewidth]{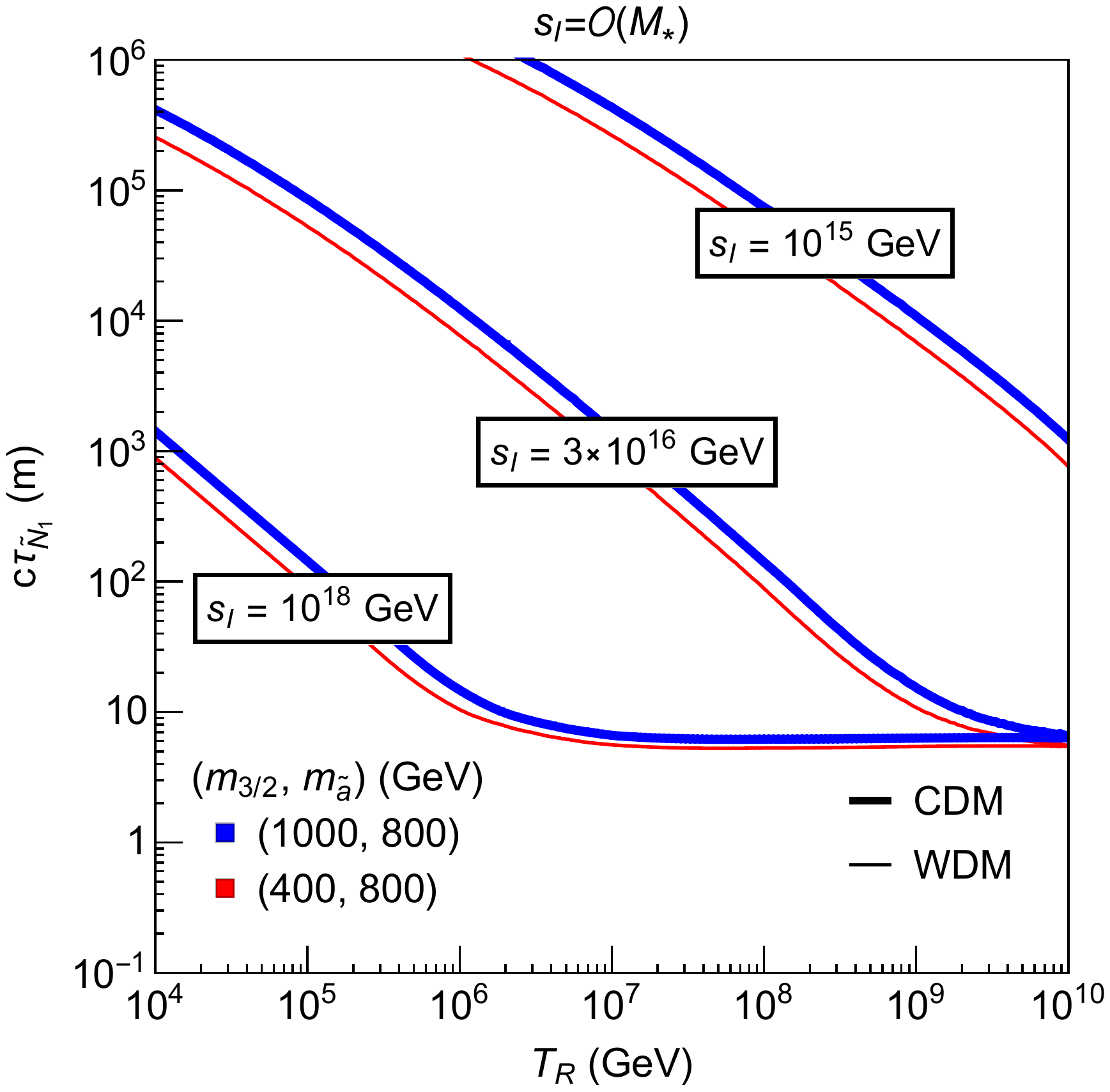} \includegraphics[width=0.495\linewidth]{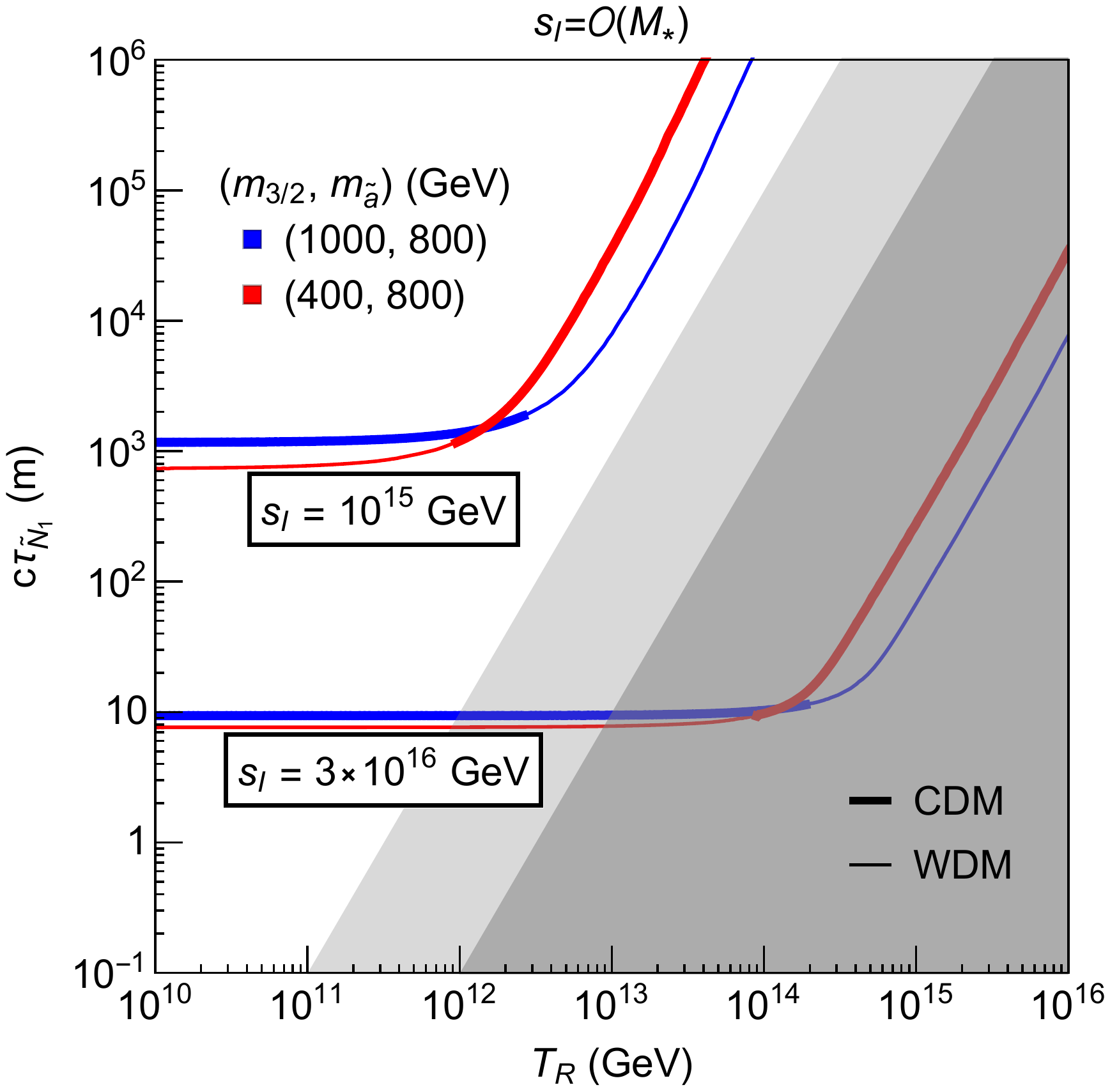}
\vspace{-0.2cm}
\caption{The lifetime of the neutralino LOSP, which decays dominantly to $\tilde{a} + h/Z$, predicted by determining $V_{PQ}$ from  $\Omega h^2 = 0.11$ using \Fig{fig:HighDFSZ0}. We fix $q_\mu=2$, $\mathcal{D}=4$, $\tan\beta=2$, and $M_2/2 = M_1 = \mu = 1$ TeV and $m_s = 600$ GeV. The left (right) panel is for the cosmology with $T_R \lsim (\gsim) 10^{10}$ GeV discussed in \Sec{subsec:lowTR} (\Sec{subsec:highTR}). }
\label{fig:ctauHighDFSZ0}
\end{center}
\end{figure}

In the DFSZ$_0$ theory, the axino is dominantly produced by IR freeze-in, as discussed in \Sec{sec:IRAxinoProd}. Since the axino is lighter than the superpartners in the thermal bath, FI occurs via decays of neutralinos and charginos with rates proportional to $1/V_{PQ}^2$ given by \Eqs{eq:nDecay}{eq:cDecay}. On the other hand, gravitinos are populated by the UV scattering of quarks, gluons, and their superpartners with an abundance given by \Eq{eq:Ygravitino}, which is proportional to $T_R$.   Both production sources are heavily diluted by the decay of the saxion condensate, which also makes the LOSP freeze-out and decay contribution negligible.

We compute the total DM abundance from these LSP and NLSP production sources in terms of $V_{PQ}$ and $T_R$ and draw contours of $\Omega h^2 = 0.11$ in the  $(V_{PQ}, T_R)$ plane in \Fig{fig:HighDFSZ0}.  After decaying to the LSP, the NLSP number density is transferred to that of the LSP, and therefore we only need the LSP mass to compute the final DM abundance.  In the left two panels, we take $s_I = V_{PQ}$, and in the right two panels we take $s_I = M_*$ and give contours for $M_* = (10^{15}, 3 \times 10^{16}, 10^{18} )$ GeV. The top two panels have $T_R \ge 10^{10}$ GeV and are therefore described by the cosmology in \Sec{subsec:highTR}, while the bottom two panels have $T_R \le 10^{10}$ GeV and are described by the cosmology in \Sec{subsec:lowTR}.   

In the upper two panels, for each value of $s_I$ two contours are shown. The blue one is an example of an axino LSP, while the red one is an example of gravitino LSP.   The vertical parts of the contours have axino FI as the dominant production mechanism and hence are independent of $T_R$, while the parts of the contour with positive constant slope have UV gravitino production dominate.   When freeze-in occurs above $T_{NA}$, the dilution factor from the saxion condensate is proportional to $s_I^2 \, V_{PQ}$, which is much larger in the right panel than in the left panel.  This means that the production needs to be much larger in the right panel than the left, and hence the contours in the right panel are at much lower $V_{PQ}$. This also explains why in the right panel, as $s_I $ is increased from $10^{15}$ GeV to $3 \times 10^{16}$ GeV, the contours move to lower $V_{PQ}$.  However, for $s_I > 10^{16}$ GeV a new regime is entered where freeze-in occurs during the MD$_{NA}$ era and dilution becomes independent of $T_M$ and therefore of $s_I$.  Hence, the vertical contours for $s_I = 3 \times 10^{16}$~GeV and $10^{18}$~GeV differ slightly only because of a different $m_{LSP}$.  

When UV gravitino production dominates, the difference in the slopes of the contours results from the dilution factor, scaling as $V_{PQ}^3$ in the left panel and $V_{PQ}$ in the right panel.   At large $T_R$ where UV gravitino production dominates, the blue contours are well above the red ones; this is because $Y_{3/2} \propto 1/m_{3/2}^2$ and the blue contours have larger $m_{3/2}$ and hence need larger $T_R$ to compensate.  On the other hand, at lower $T_R$ where axino FI dominates, $Y_{\tilde{a}}$ is independent of $m_{\tilde{a}}$ and $m_{3/2}$, so that the red and blue $\Omega h^2$ contours differ only because of $m_{LSP}$.  

Each contour is divided into thick and thin parts.  The thick parts indicate cold dark matter (CDM) where NLSP production is sub-dominant. The thin parts label the warm dark matter (WDM) case where the component of dark matter from the NLSP decay constitutes more than 50\% of the total abundance. The thin lines may be relevant for understanding possible difficulties with pure CDM, as in core-cusp, too big to fail and missing satellite problems.

A key point emerges from comparing the contours in the two upper panels of \Fig{fig:HighDFSZ0} with the contour for High Scale mediation ($m_{3/2} = 100$ GeV) in \Fig{fig:AxinoProblem}, where the saxion condensate is absent.  {\it The saxion condensate increases the maximum value of $T_R$ from $10^8$ GeV to $10^{16}$ GeV.}  This allows very high reheat temperatures after inflation\footnote{The largest possible values of $T_R$ and $V_{PQ}$, of order $(10^{15}-10^{16})$ GeV, are excluded by isocurvature density perturbations if there is a sufficient contribution of misalignment axions to dark matter.  In our figures we do not show any such excluded region, as the bound from such perturbations can be avoided if the axion misalignment angle $\theta_{mis}$ is sufficiently small, typically 0.1 - 0.3 suffices,  and/or $N_{DW}$ is sufficiently large. }, so that baryogenesis may occur at very high temperatures.  The upper bound on $T_R$ now arises because inflation only gives a saxion condensate if PQ symmetry is broken before inflation.  The precise constraint on $T_R$ is dependent on the model for the PQ phase transition and on the model for reheating after inflation.  For instantaneous reheating we expect the condition to typically be $T_R < V_{PQ}$ corresponding to the unshaded region of \Fig{fig:HighDFSZ0}.  However, certain theories may have PQ breaking before inflation even for $T_R$ somewhat higher than $V_{PQ}$, so that for these theories the lightly shaded region also becomes physical.  We expect the dark shaded region to be unphysical in all models.  On the other hand, if the inflaton decay rate is slow, $T_R$ could be much below the energy scale of inflation, lowering the shaded bands and reducing the maximal allowed value of $T_R$.  A reduction by a few orders of magnitude would still allow $T_R$ to be sufficiently large for leptogenesis. 

Another key point emerges from comparing the contours in the two upper panels of \Fig{fig:HighDFSZ0} with the contour for High Scale mediation ($m_{3/2} = 100$ GeV) in \Fig{fig:AxinoProblem}. {\it  For large $s_I = M_*$, the saxion condensate lowers the minimal value of $V_{PQ}$ by several orders of magnitude.}  An important part of the axino problem is that the minimal value of $V_{PQ}$ is so large -- $4 \times 10^{14}$ GeV for $m_{3/2} = 100$ GeV -- that misalignment axions overclose the universe unless the misalignment angle is very small.  This difficulty is removed with a saxion condensate because the decay of the condensate also dilutes misalignment axions, so that they typically give sub-dominant contributions to dark matter for $f_a \lsim 10^{15}$ GeV and misalignment angles order unity \cite{Hashimoto:1998ua, Co:2016vsi}.  Since $V_{PQ}$ is larger than $f_a$ by $N_{DW} / \sqrt{2}$, which is often an order of magnitude, axions are typically sub-dominant on all the contours of \Fig{fig:HighDFSZ0}, although they could give a comparable contribution when $V_{PQ}$ is of order $(10^{15}-10^{16})$ GeV.

In the lower two panels, for $s_I = 10^{15}$ GeV and $3 \times 10^{16}$ GeV, freeze-in occurs above $T'_{NA}$, so that the dilution factor is proportional to $T_R$, as seen in \Eq{eq:TMI}, and the axino FI abundance depends on $T_R$. In particular, a decrease in $T_R$ is compensated by an increase in $V_{PQ}$ at a rate that depends on the cosmological era during FI.  However, for $s_I = 10^{18}$ GeV much of the contour is still vertical as freeze-in occurs during MD$_{NA}$.  For very high $s_I$ there is a very robust prediction for $V_{PQ}$ from axino freeze-in dark matter.

If the LOSP is a neutralino, its lifetime, Eq.~(\ref{eq:LOSPctauAxi}), can be predicted by determining $V_{PQ}$ from the dark matter abundance. The prediction shown in \Fig{fig:ctauHighDFSZ0} is obtained by inverting the two axes in the right panels of \Fig{fig:HighDFSZ0} and converting the $V_{PQ}$ axis to the lifetime using \Eq{eq:LOSPctauAxi}. The left (right) panel corresponds to the cosmology of $T_R$ less (greater) than $10^{10}$ GeV discussed in \Sec{subsec:lowTR} (\Sec{subsec:highTR}). As explained for \Fig{fig:HighDFSZ0}, for axino FI a larger saxion condensate (higher $s_I$) leads to a lower $V_{PQ}$, which in turn gives a shorter LOSP lifetime. However, this behavior does not continue for an arbitrarily high $s_I$: once $s_I$ is sufficiently high, the axino freeze-in occurs during the MD$_{NA}$ era and the abundance becomes insensitive to $s_I$ \cite{Co:2015pka}. Consequently, for $T_R > 3 \times 10^9$ GeV and any  $s_I > 10^{16}$ GeV, there is a very robust prediction of $c \tau_{LOSP} \simeq 10$m. In the left panel, as $T_R$ drops FI transitions to occurring in MD$_A$, so that the dilution factor also drops and $V_{PQ}$ and $c \tau_{LOSP}$ increase.   As in \Fig{fig:HighDFSZ0}, the thin parts of contours give warm dark matter from NLSP decay.
 
We do not show the LOSP lifetime for $s_I = V_{PQ}$.  The large values of $V_{PQ}$ indicated by the left panels of \Fig{fig:HighDFSZ0} lead to $c \tau \sim 10^{(5-7)}$m.
 
\subsection{The DFSZ$_+$ Theory}
\label{sec:DFSZ+}

\begin{figure}
\begin{center}
\caption*{\bf High Scale mediation: DFSZ$_+$}
\includegraphics[width=0.495\linewidth]{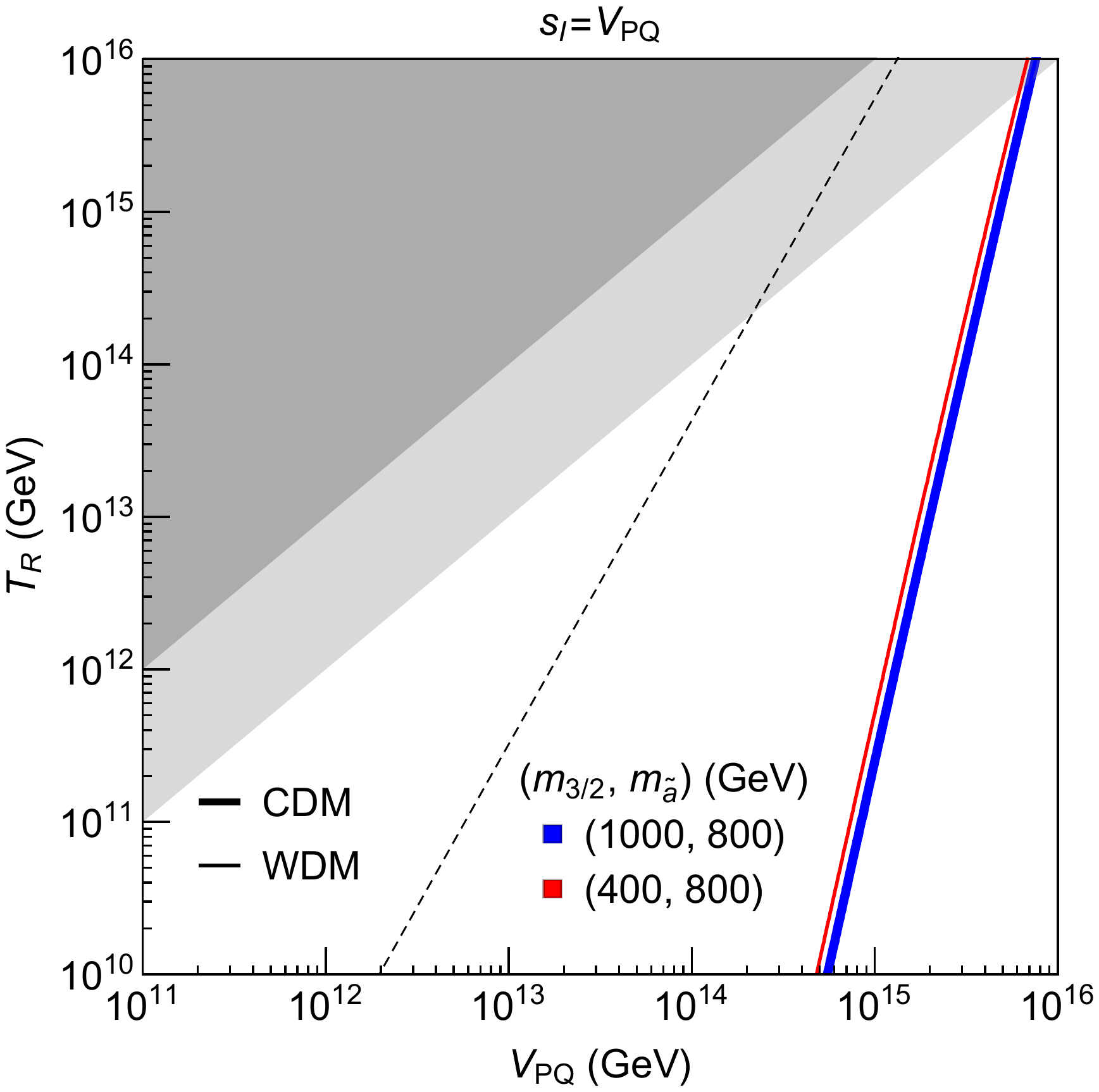} \includegraphics[width=0.495\linewidth]{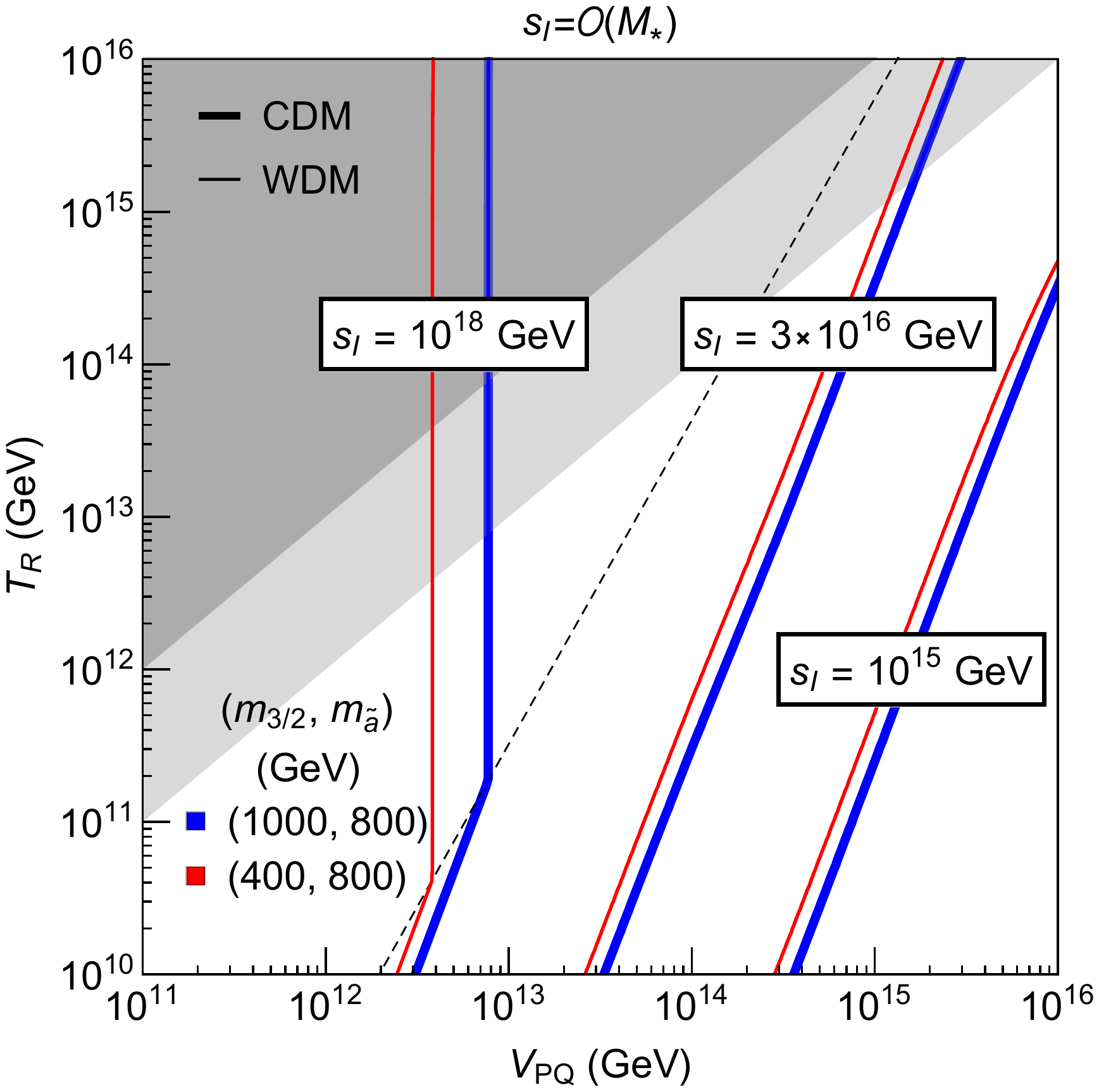}
\includegraphics[width=0.495\linewidth]{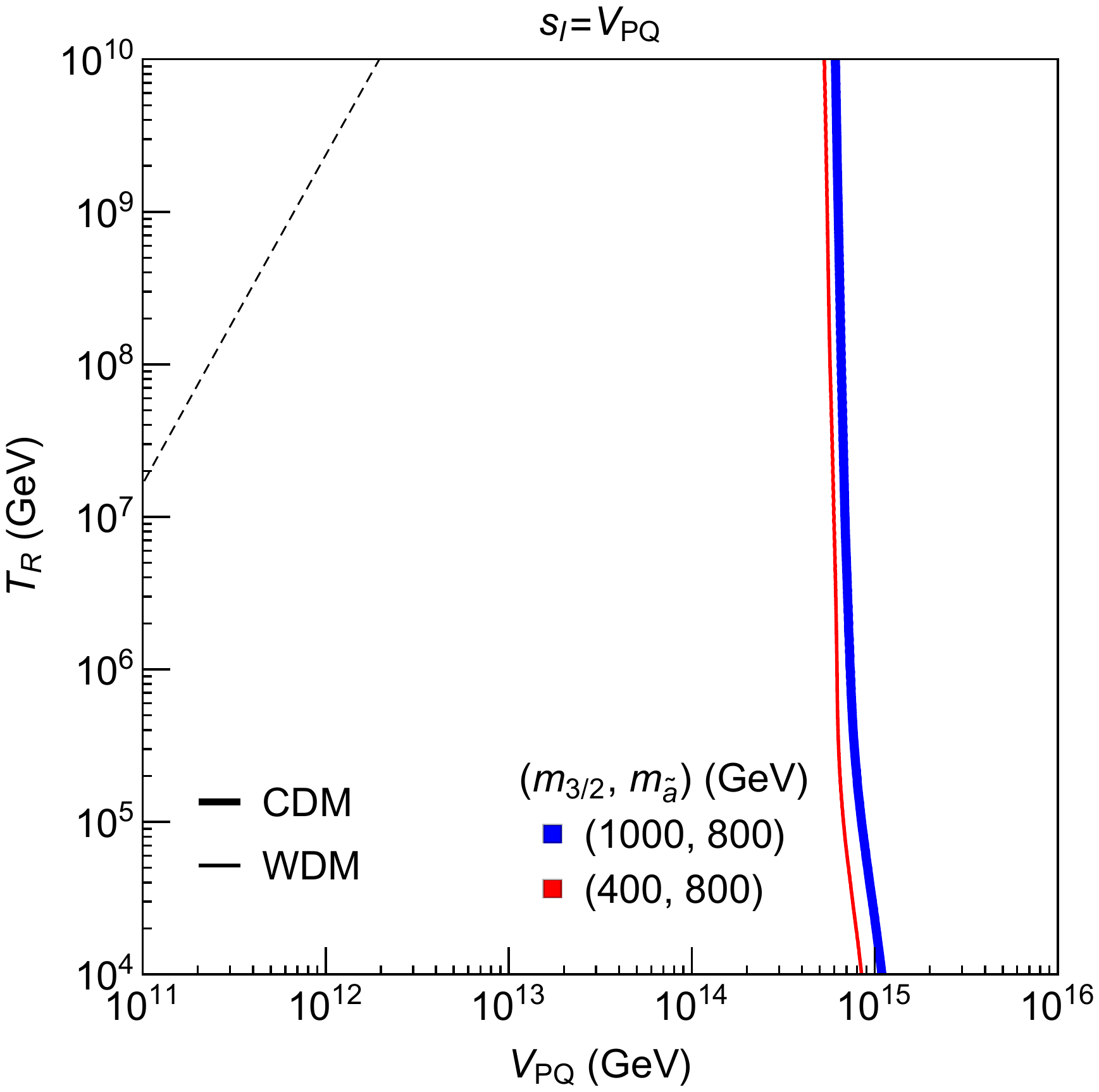} \includegraphics[width=0.495\linewidth]{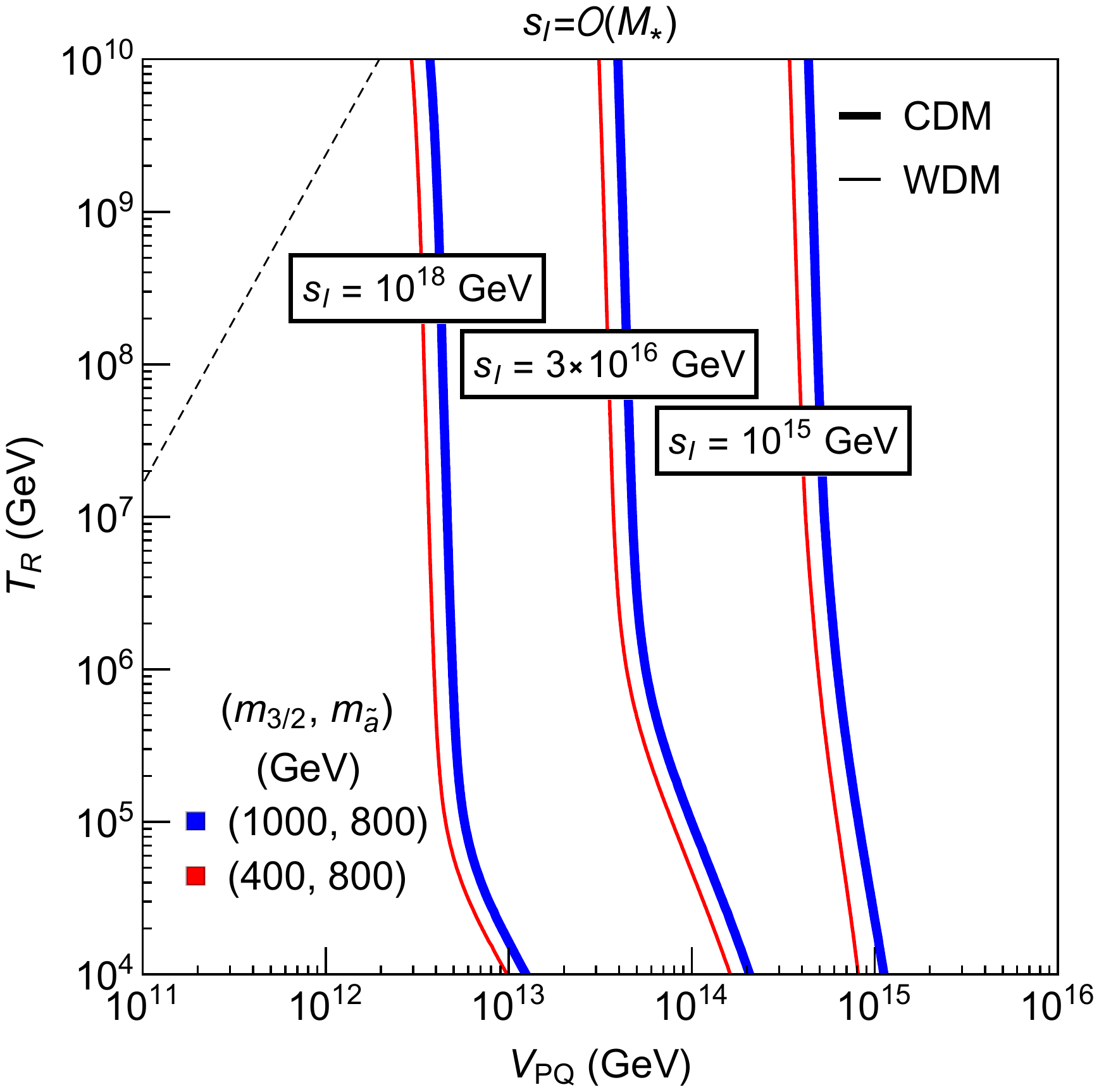}
\end{center}
\caption{Contours of $\Omega h^2 = 0.11$ from axino freeze-in, axino UV, and gravitino UV production. We fix $q_\mu=2$, $\mathcal{D}=4$, $\tan\beta=2$, and $M_2/2 = M_1 = \mu = 1$ TeV and $m_s = 600$ GeV. The top (bottom) row is for the cosmology with $T_R \gsim (\lsim) 10^{10}$ GeV discussed in \Sec{subsec:highTR} (\Sec{subsec:lowTR}).}
\label{fig:HighDFSZ+}
\end{figure}

\begin{figure}[t]
\begin{center}
\caption*{\bf High Scale mediation: DFSZ$_+$  + Neutralino LOSP}
\includegraphics[width=0.495\linewidth]{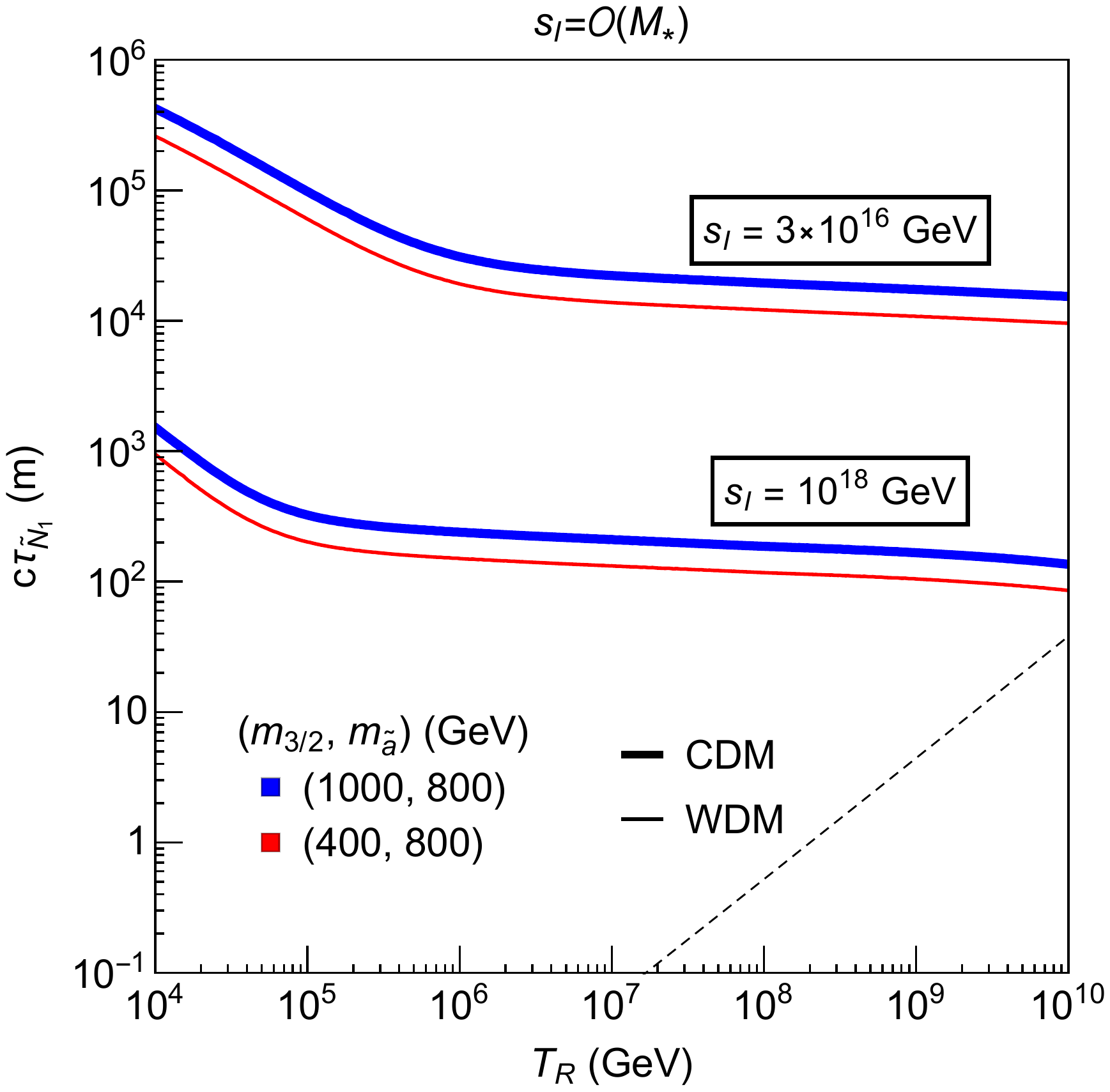} \includegraphics[width=0.495\linewidth]{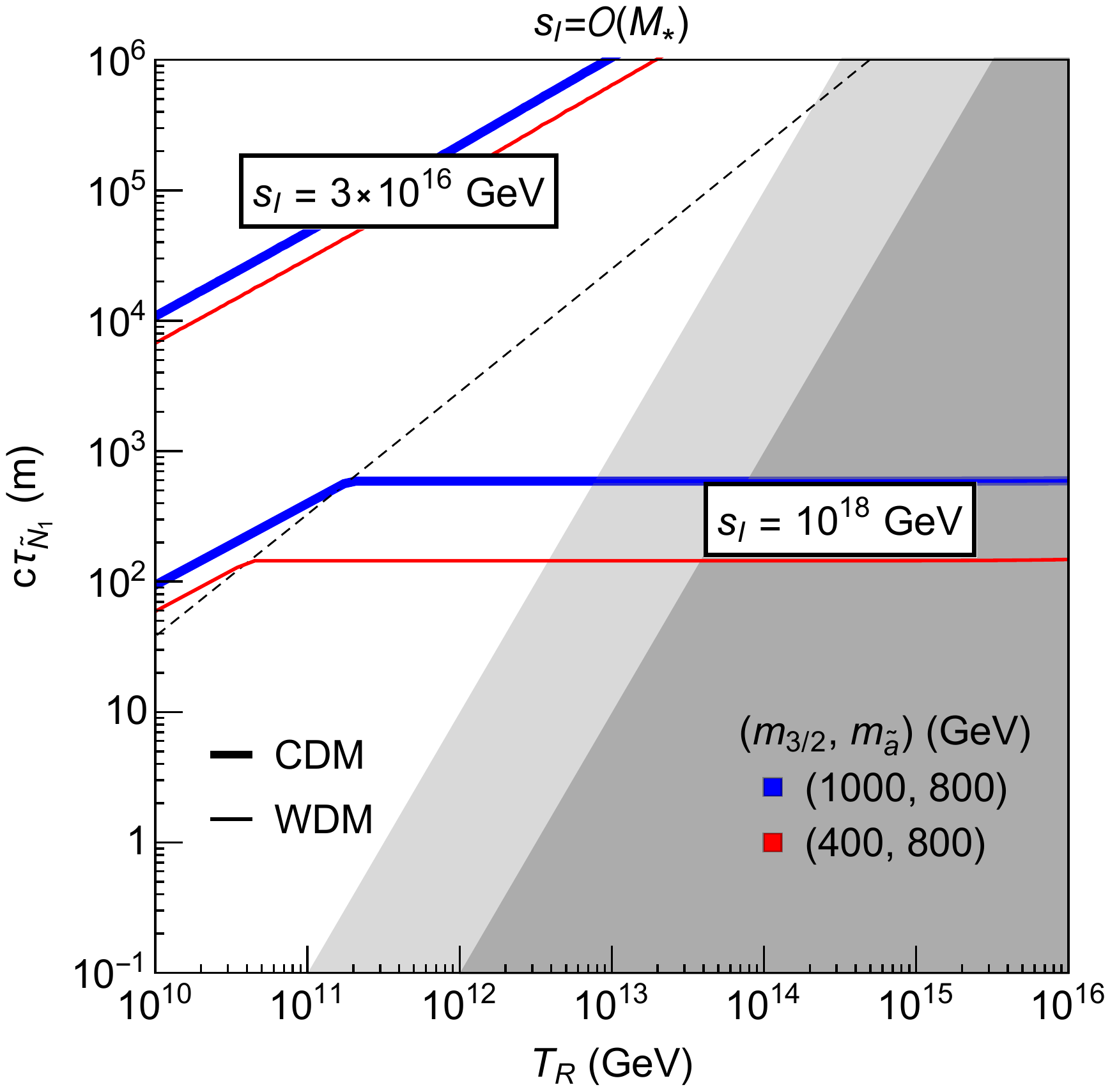}
\end{center}
\caption{The lifetime of the neutralino LOSP, which decays dominantly to $\tilde{a} + h/Z$, predicted from the $\Omega h^2 = 0.11$ contours of \Fig{fig:HighDFSZ+}. We fix $q_\mu=2$, $\mathcal{D}=4$, $\tan\beta=2$, and $M_2/2 = M_1 = \mu = 1$ TeV and $m_s = 600$ GeV. The left (right) panel is for the cosmology with $T_R \lsim (\gsim) 10^{10}$ GeV discussed in \Sec{subsec:lowTR} (\Sec{subsec:highTR}).}
\label{fig:ctauHighDFSZ+}
\end{figure}

In addition to the axino FI and gravitino UV contributions existing in DFSZ$_0$ theories, there is also axino production from UV scattering discussed in \Sec{sec:UVAxinoProd} for DFSZ$_+$ theories. In a setting similar to \Fig{fig:HighDFSZ0}, we show the results for the total abundance in \Fig{fig:HighDFSZ+} as a function of $T_R$ and $V_{PQ}$.  Gravitino production is everywhere sub-dominant.  In fact UV axino production dominates everywhere, except for $T_R$ less than about $10^6$ GeV where axino freeze-in becomes important.  The blue and red curves, corresponding to axino and gravitino LSP, nearly coincide; they differ only because the LSP mass differs by a factor of two between the curves.


The parametric dependence of the contours can be understood from $\Omega h^2 \propto Y_{\tilde{a}} m_{LSP}/D$, where, for UV production, the dilution factor $D \propto s_I^2 V_{PQ} (1,T_R)$ for ($T_R \gsim 10^{10}$ GeV, $T_R \lsim 10^{10}$ GeV).  For UV axino production, $Y_{\tilde{a}} \propto (T_R/V_{PQ}^2, 1)$, with the constant value applying only if the equilibrium abundance is reached, which happens above and to the left of the dashed line.  For $T_R > 10^{11}$ GeV, the contour for $s_I = 10^{18}$ GeV is vertical, corresponding to an equilibrium axino abundance, while the contours with $s_I = 10^{15}, 3 \times 10^{16}$ GeV have $T_R \propto V_{PQ}^3$, and those with $s_I = V_{PQ}$ are steeper with $T_R \propto V_{PQ}^5$ .  For $10^6 \, \mbox{GeV} <T_R < 10^{10}$ GeV, both $Y_{\tilde{a}}$ and $D$ are linear in $T_R$, so that the contours are vertical.
At very low $T_R$ the contours become sloped as the axino yield becomes dominated by freeze-in; in this low $T_R$ region they are identical to the curves in \Fig{fig:HighDFSZ0}. 


In \Fig{fig:ctauHighDFSZ+}, we predict the neutralino LOSP lifetime, \Eq{eq:LOSPctauAxi}, from the values of $V_{PQ}$ fixed by the observed dark matter abundance, in a way completely analogous to \Fig{fig:ctauHighDFSZ0}, with the left (right) panel for $T_R$ less (greater) than $10^{10}$ GeV. The key feature in DFSZ$_+$ is that, for all $T_R > 10^6$ GeV, UV axino production dominates dark matter production and hence, compared to DFSZ$_0$, more dilution (i.e. higher $s_I$) is required for a given $V_{PQ}$.   \Fig{fig:ctauHighDFSZ+} shows that for $s_I$ of order, or larger than, the scale of supersymmetric grand unification, displaced vertices should be observable at LHC.  
For fixed $T_R$ and $s_I$, the lifetime scales as $m_{LSP}^{2/3}$, unless the axino abundance from UV scattering reaches equilibrium, when it scales as $m_{LSP}^2$.

\section{Results for Low Scale or ``Gauge" Mediation}
\label{sec:LowScale}

\subsection{The DFSZ$_0$ Theory}
\label{sec:DFSZ0low}

\begin{figure}
\begin{center}
\caption*{\bf Low Scale mediation: DFSZ$_0$}
\includegraphics[width=0.495\linewidth]{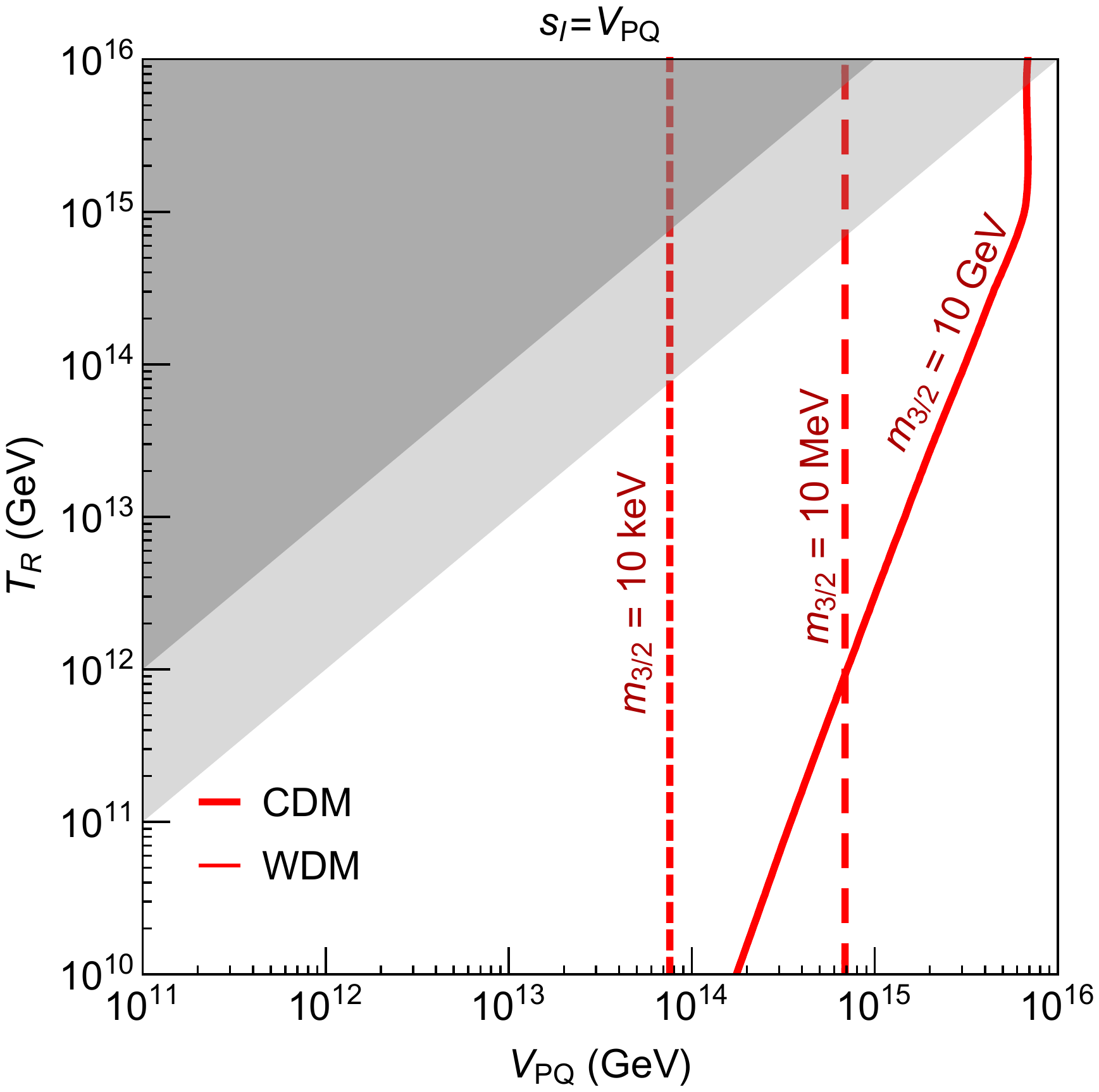} \includegraphics[width=0.495\linewidth]{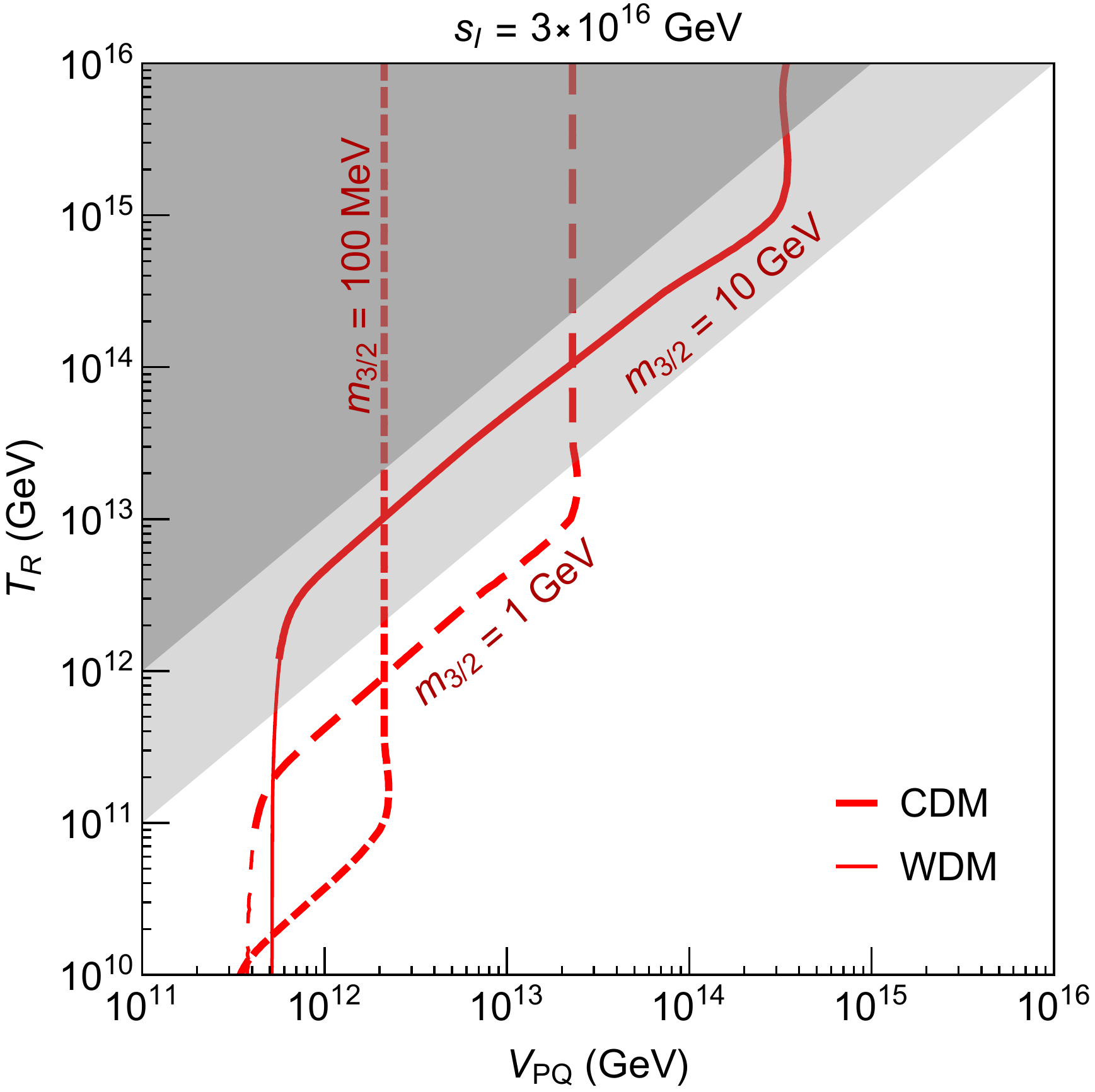}
\includegraphics[width=0.495\linewidth]{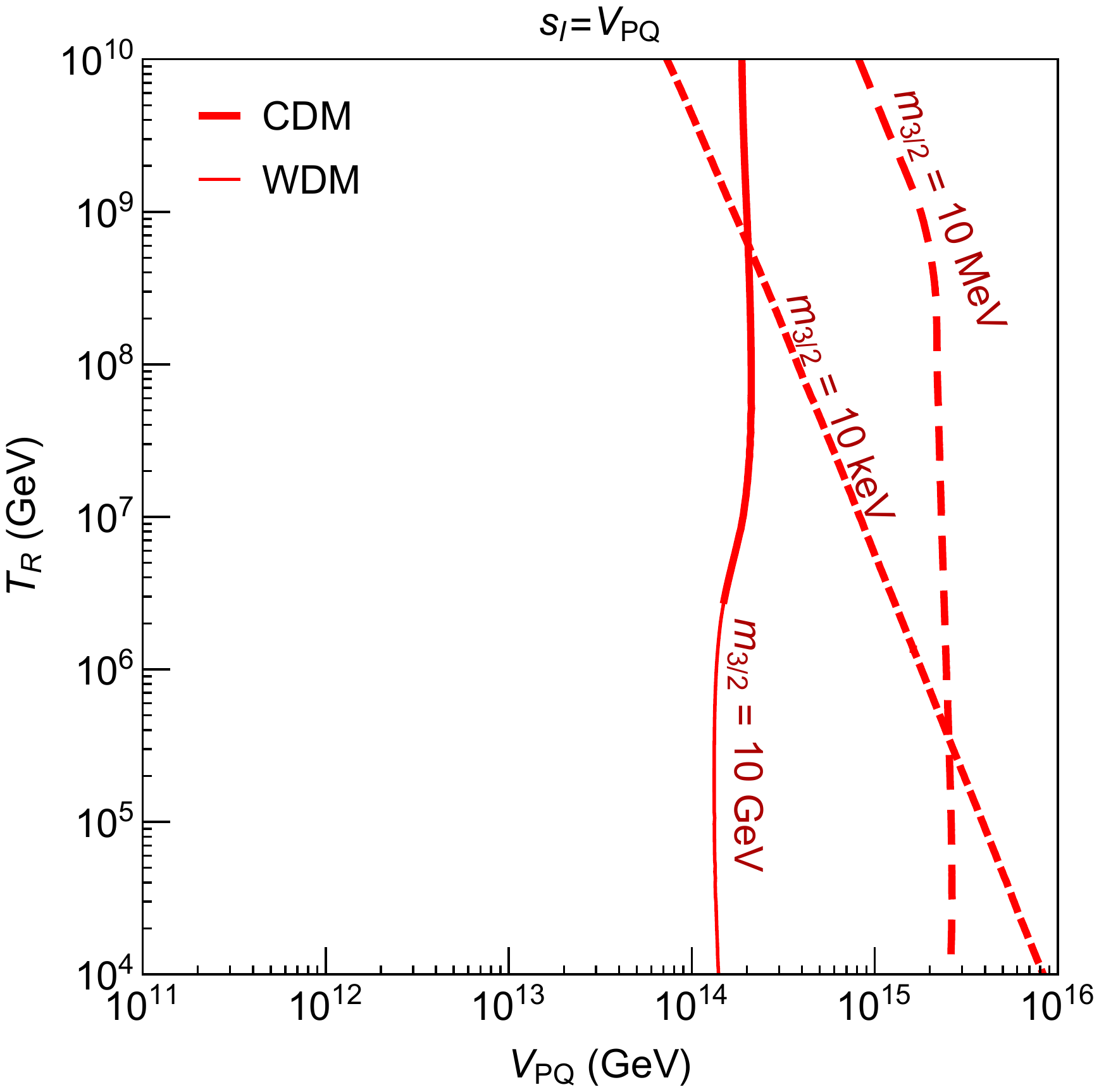} \includegraphics[width=0.495\linewidth]{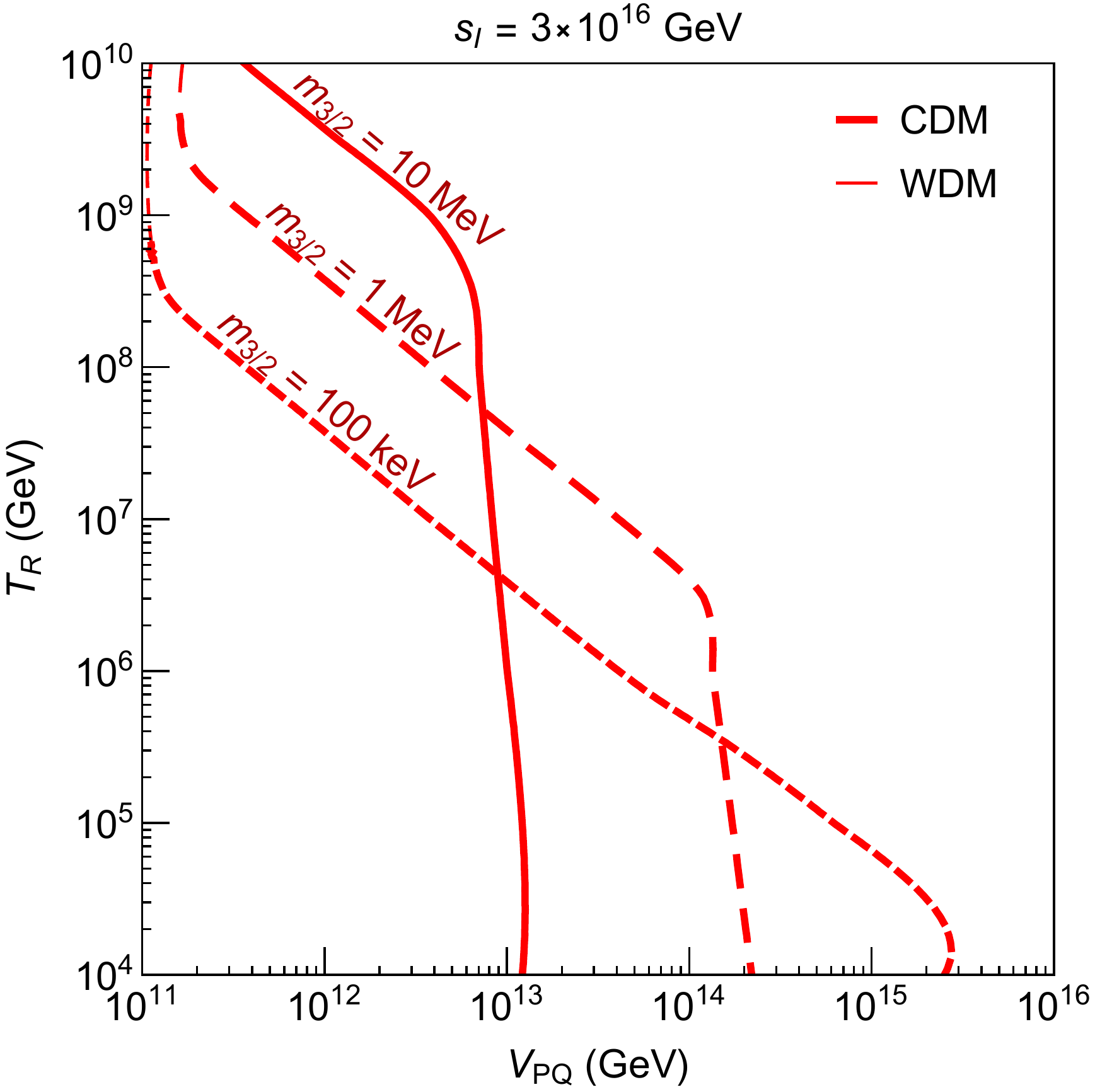}
\end{center}
\caption{Contours of $\Omega h^2 = 0.11$ from axino freeze-in and gravitino UV production. We fix $q_\mu=2$, $\mathcal{D}=4$, $\tan\beta=2$, $M_2/2 = M_1 = \mu = 1$ TeV, and $m_s = 600$ GeV. The top (bottom) row is for the cosmology with $T_R \gsim (\lsim) 10^{10}$ GeV discussed in \Sec{subsec:highTR} (\Sec{subsec:lowTR}).}
\label{fig:LowDFSZ0}
\end{figure}

\begin{figure}[t]
\begin{center}
\caption*{\bf Low Scale mediation: DFSZ$_0$ + Neutralino LOSP}
\includegraphics[width=0.495\linewidth]{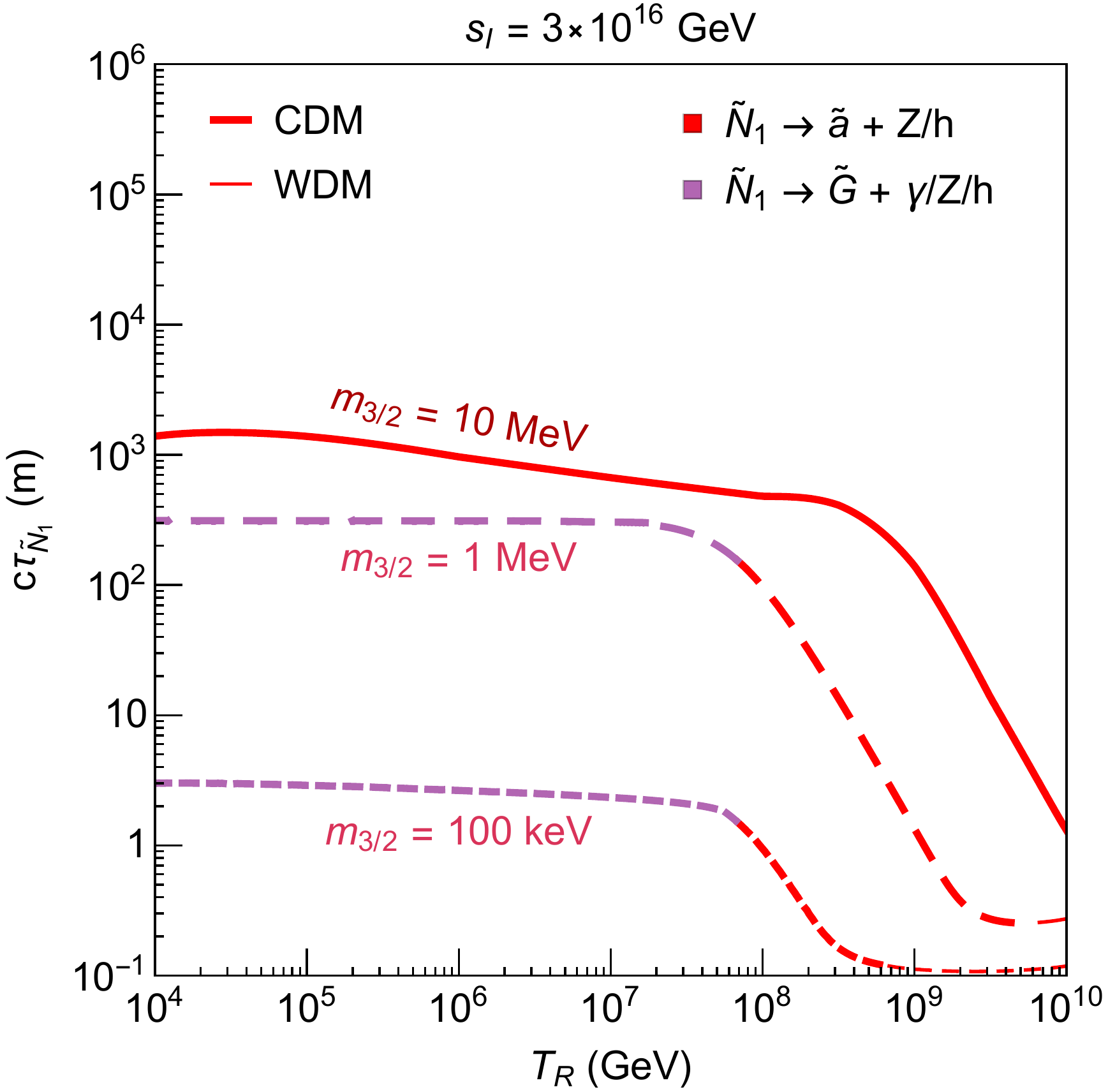} \includegraphics[width=0.495\linewidth]{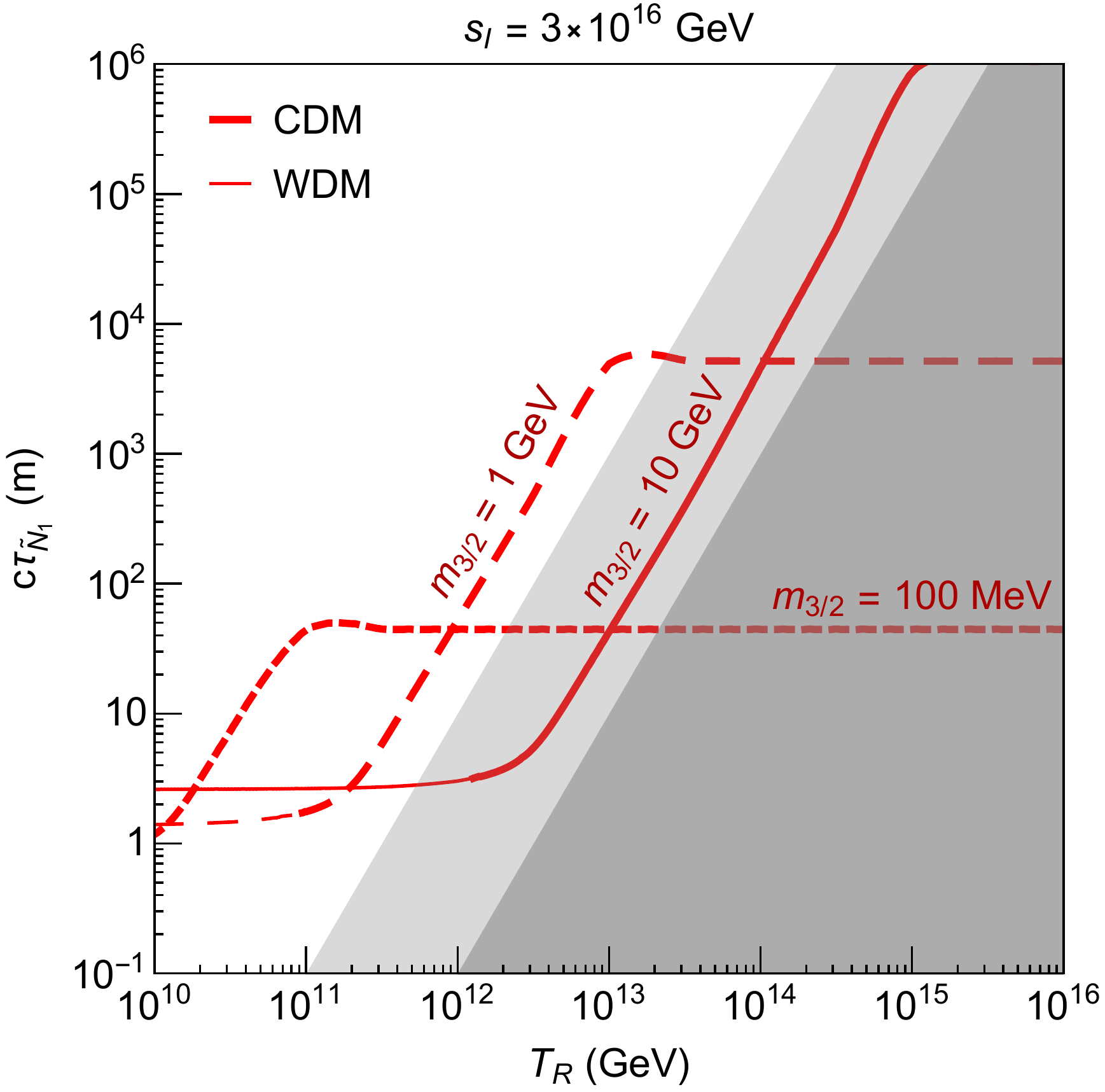}
\end{center}
\caption{The lifetime of the neutralino LOSP, decaying dominantly to $\tilde{a} + h/Z$ (red) or to $ \tilde{G}+\gamma/Z$(pink), predicted from the $\Omega h^2 = 0.11$ contours of \Fig{fig:LowDFSZ0}. We fix $q_\mu=2$, $\mathcal{D}=4$, $\tan\beta=2$, and $M_2/2 = M_1 = \mu = 1$ TeV and $m_s = 600$ GeV. The left (right) panel is for the cosmology with $T_R \lsim (\gsim) 10^{10}$ GeV discussed in \Sec{subsec:lowTR} (\Sec{subsec:highTR}).}
\label{fig:ctauLowDFSZ0}
\end{figure}

Dark matter production in DFSZ$_0$ theories is dominantly from axino freeze-in and Gravitino UV production as explained in \Sec{sec:DFSZ0}. Here we are concerned with Low Scale mediation of supersymmetry breaking, as in gauge mediation, and hence take the gravitino to be the LSP.  Axinos produced from freeze-in subsequently decay and produce a warm component of gravitino dark matter. 

In \Fig{fig:LowDFSZ0}, we show contours of $\Omega_{3/2} h^2 = 0.11$. Following the same setup as in \Figs{fig:HighDFSZ0}{fig:HighDFSZ+}, the thick (thin) parts of the curves indicate that dark matter arises dominantly from LSP (NLSP) production. The gray region is excluded because PQ breaks after inflation. The upper (lower) panels are for the cosmology of $T_R$ greater (lower) than $10^{10}$ GeV discussed in \Sec{subsec:highTR} (\Sec{subsec:lowTR}). The left (right) panels assume $s_I = V_{PQ}$ ($M_* = 3\times 10^{16}$ GeV). It is important to remember that the dilution factor is larger for larger $V_{PQ}$.
 
Gravitino UV production given by \Eq{eq:Ygravitino} is much enhanced for low $m_{3/2}$. For low enough $m_{3/2}$ and high enough $T_R$, gravitino UV production is so efficient that gravitinos thermalize, with an abundance independent of $m_{3/2}$ and $T_R$ given in \Eq{eq:Yeq}.  In the upper panels, the vertical thick lines correspond to this thermal gravitino production. As $T_R$ decreases, the gravitino may become non-thermal with an abundance decreasing with $T_R$, as in \Eq{eq:Ygravitino}, corresponding to the sloped parts of the thick curves in the upper panels. Lastly, the thin vertical curves label the axino FI domination. Since axino FI production in \Eqs{eq:nDecay}{eq:cDecay} is independent of $m_{\tilde{a}}$, the result generically applies for any $m_{\tilde{a}}$ between $650 \GeV - 1 \TeV$ as required by the free-streaming length for cold dark matter and neutralino decay kinematics respectively.

In the lower panels, $T_R < 10^{10}$ GeV and the dilution factor in the cosmology described in \Sec{subsec:lowTR} scales linearly with $T_R$. Therefore, the curves for thermal gravitino production (with a constant undiluted yield) are now at a slope. On the other hand, non-thermal gravitino UV production is close to being proportional to $T_R$ as well, and with dilution, the final yield becomes independent of $T_R$, corresponding to the vertical thick parts. Lastly, the axino FI production dominates at the vertical thin parts of the curves. 

In \Fig{fig:ctauLowDFSZ0}, the LOSP lifetimes can again be predicted from \Eq{eq:LOSPctauAxi} with the values of $V_{PQ}$ that give the observed dark matter abundance. The left (right) panel corresponds to the cosmology of $T_R$ less (greater) than $10^{10}$ GeV discussed in \Sec{subsec:lowTR} (\Sec{subsec:highTR}). It is worth noting that we are choosing different values of the gravitino mass in these two panels to demonstrate interesting LOSP decay signals for the relevant regions. In particular, we open up a new decay channel when the gravitino mass is less than $\mathcal{O}(1)$ MeV-- the neutralino LOSP decays to the gravitino and $\gamma/h/Z$. This decay channel, with the rate given in \Eq{eq:LOSPctauGravi}, dominates in the purple portions of the curves and gives the very well known displaced vertex signal of gauge mediation.  Our framework yields a cosmology for this scenario with $T_R$ far above the TeV scale.  As in \Fig{fig:LowDFSZ0}, the thin parts of the lines are dominated by the axino FI contribution and lead to warm dark matter because of the late axino decay to axions and gravitinos. For clarity, only one value of $s_I$ is shown but there exists a set of correlated $m_{3/2}$ and $s_I$ that can lead to LOSP displaced signals. In particular, a low $m_{3/2}$ enhances the production and thus requires a larger $s_I$ for dilution. However, once $m_{3/2}$ is sufficiently low for the gravitino to be thermalized, any lower $m_{3/2}$ results in a decrease in its energy density and a smaller $s_I$ is needed. It is remarkable that, due to the interplay between the two decay channels, the LOSP lifetimes for low $m_{3/2}$ are always within the reach of current and future colliders.

\subsection{The DFSZ$_+$ Theory}
\label{sec:DFSZ+low}

\begin{figure}
\begin{center}
\caption*{\bf Low Scale mediation: DFSZ$_+$}
\includegraphics[width=0.495\linewidth]{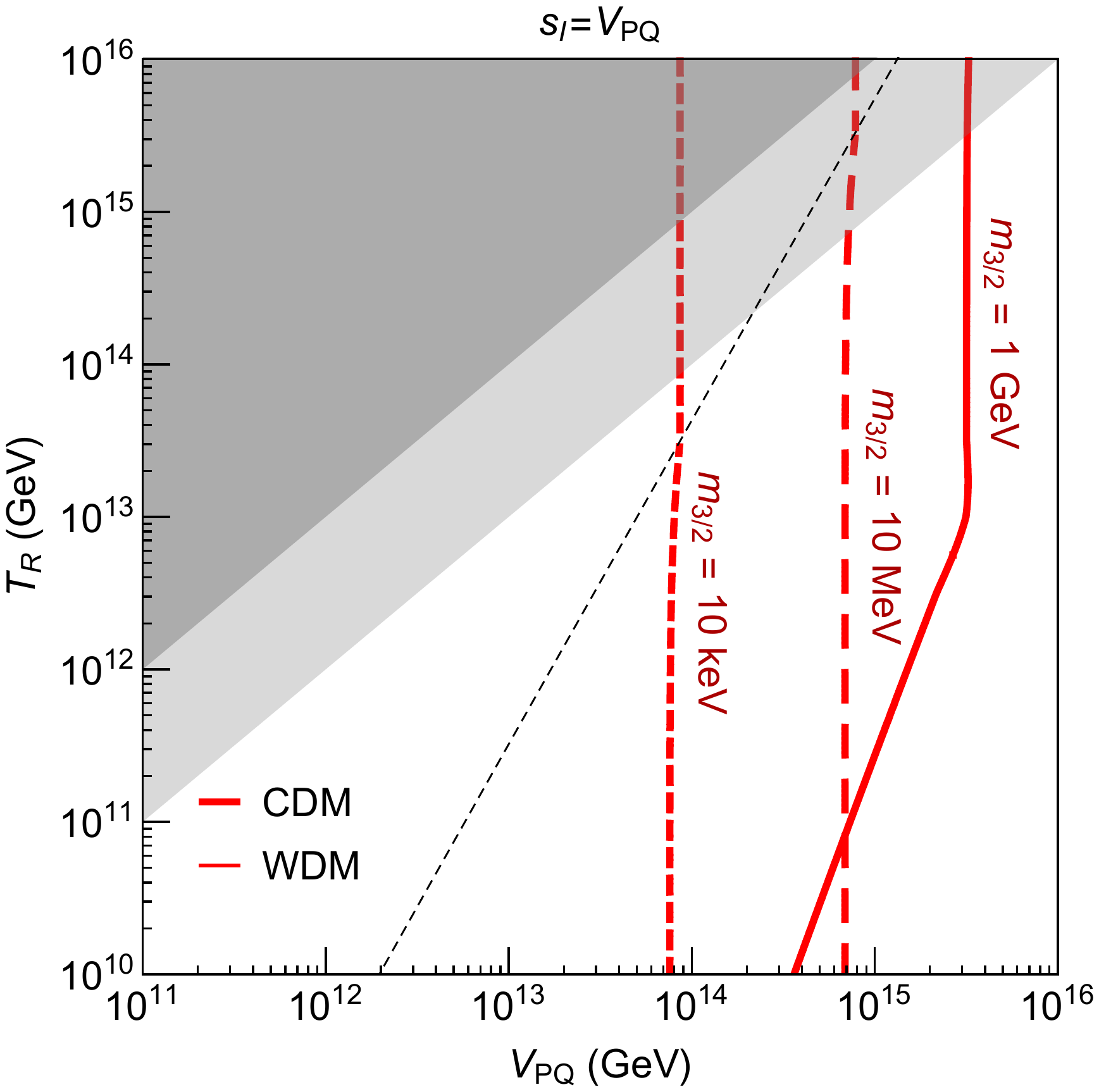} \includegraphics[width=0.495\linewidth]{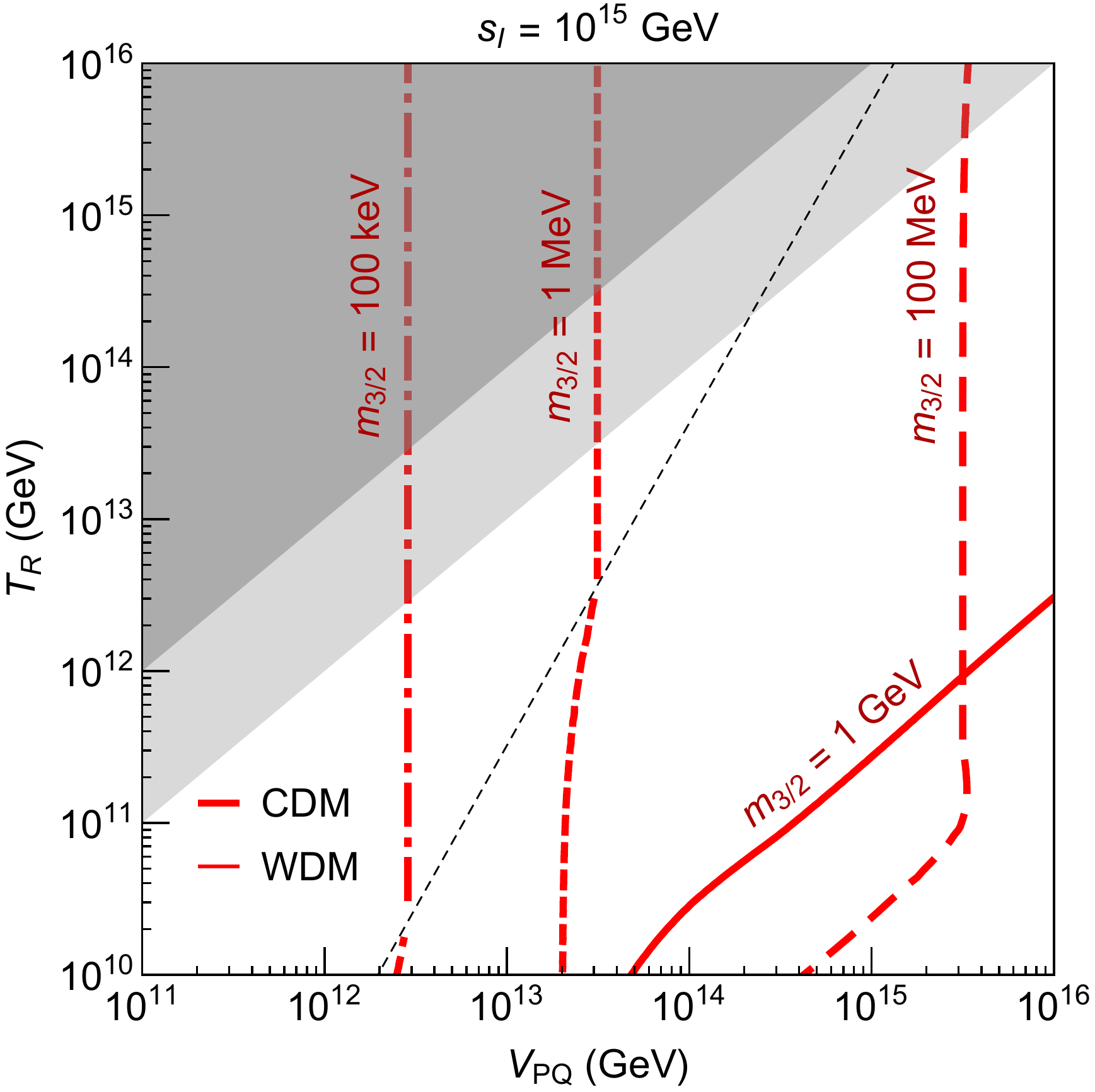}
\includegraphics[width=0.495\linewidth]{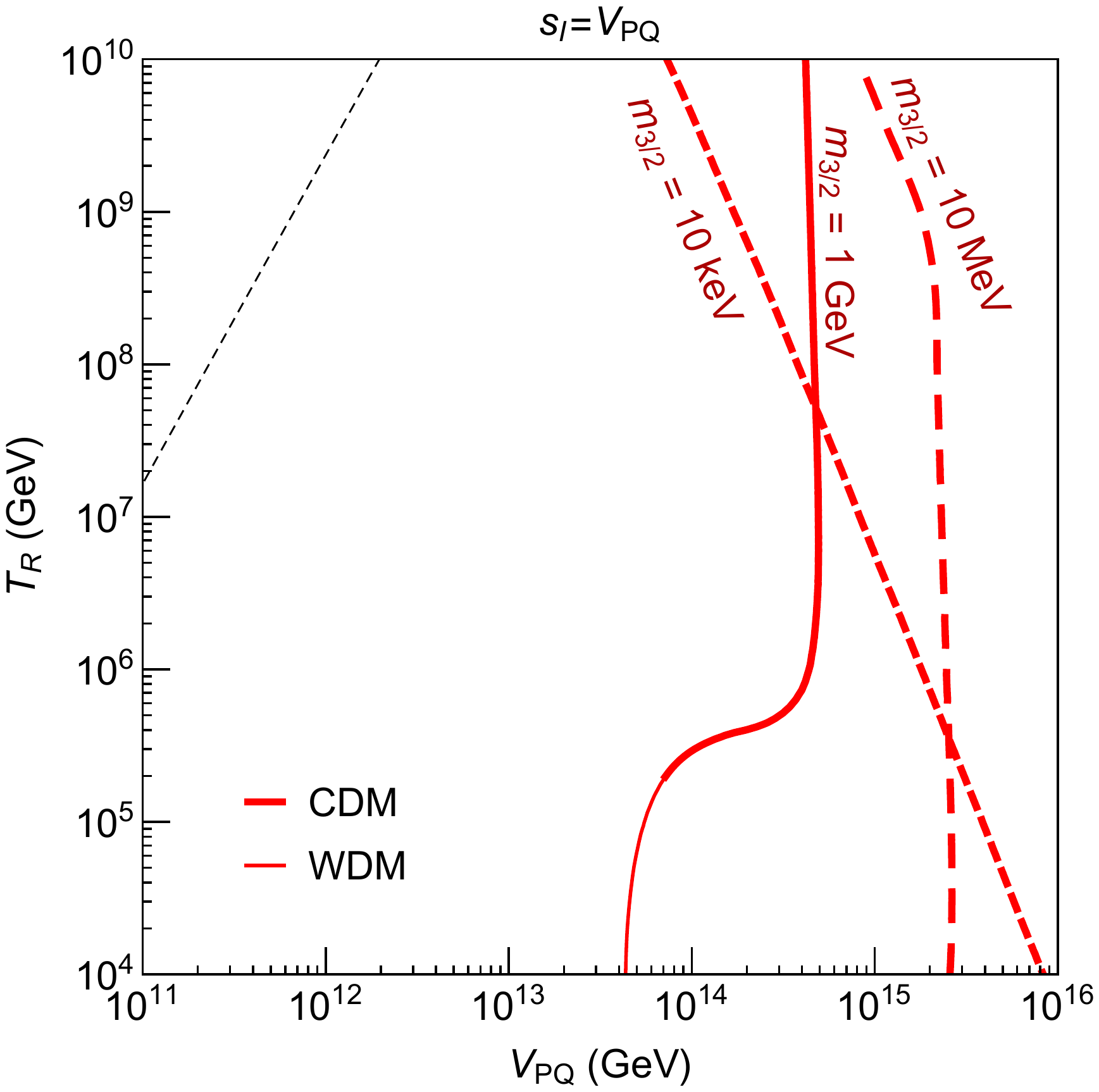} \includegraphics[width=0.495\linewidth]{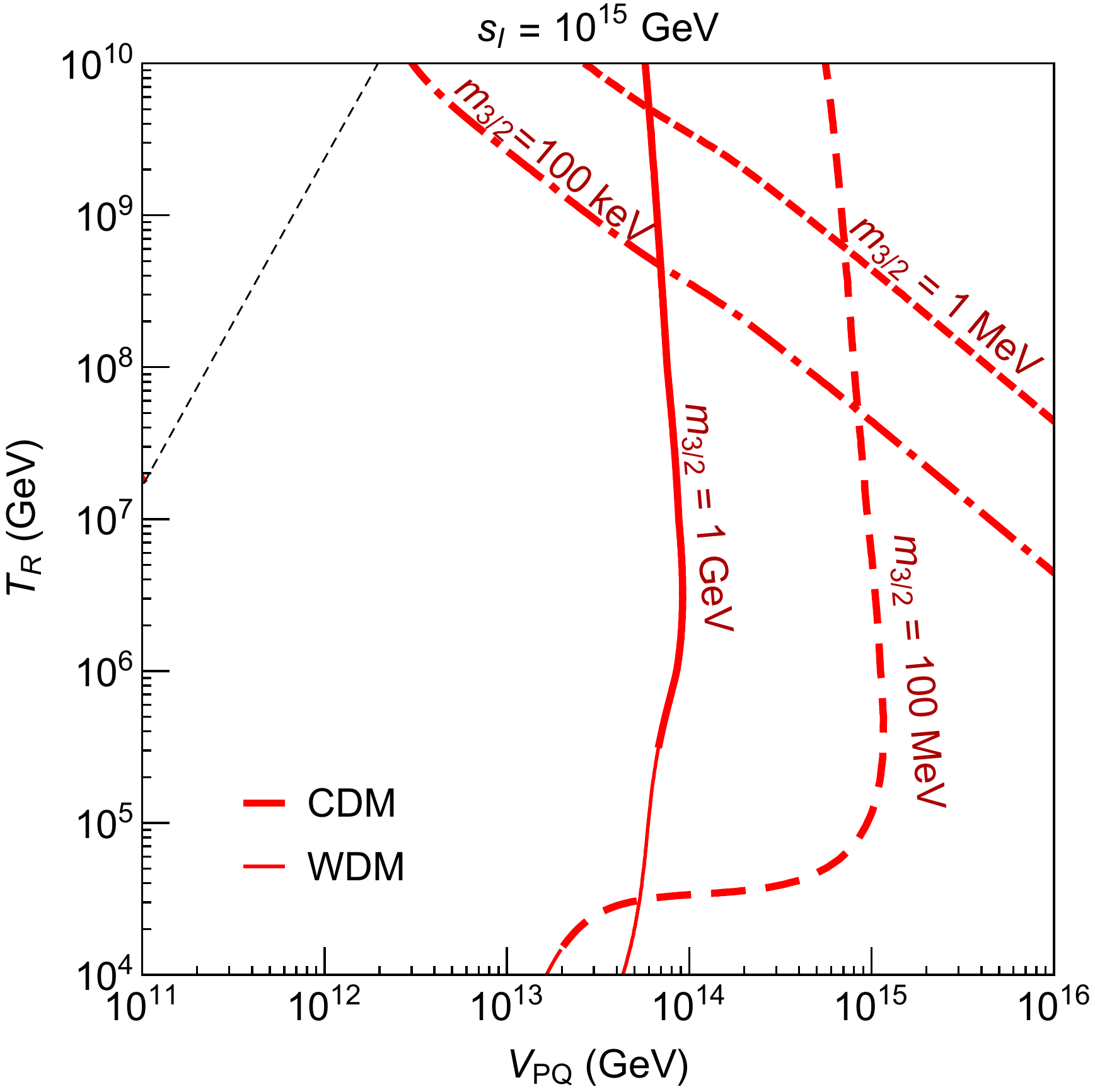}
\end{center}
\caption{Contours of $\Omega h^2 = 0.11$ from axino UV and gravitino UV production. We fix $q_\mu=2$, $\mathcal{D}=4$, $\tan\beta=2$, and $M_2/2 = M_1 = \mu = 1$ TeV and $m_s = 600$ GeV. The top (bottom) row is for the cosmology with $T_R \gsim (\lsim) 10^{10}$ GeV discussed in \Sec{subsec:highTR} (\Sec{subsec:lowTR}). }
\label{fig:LowDFSZ+}
\end{figure}

\begin{figure}[t]
\begin{center}
\caption*{\bf Low Scale mediation: DFSZ$_+$ + Neutralino LOSP}
\includegraphics[width=0.495\linewidth]{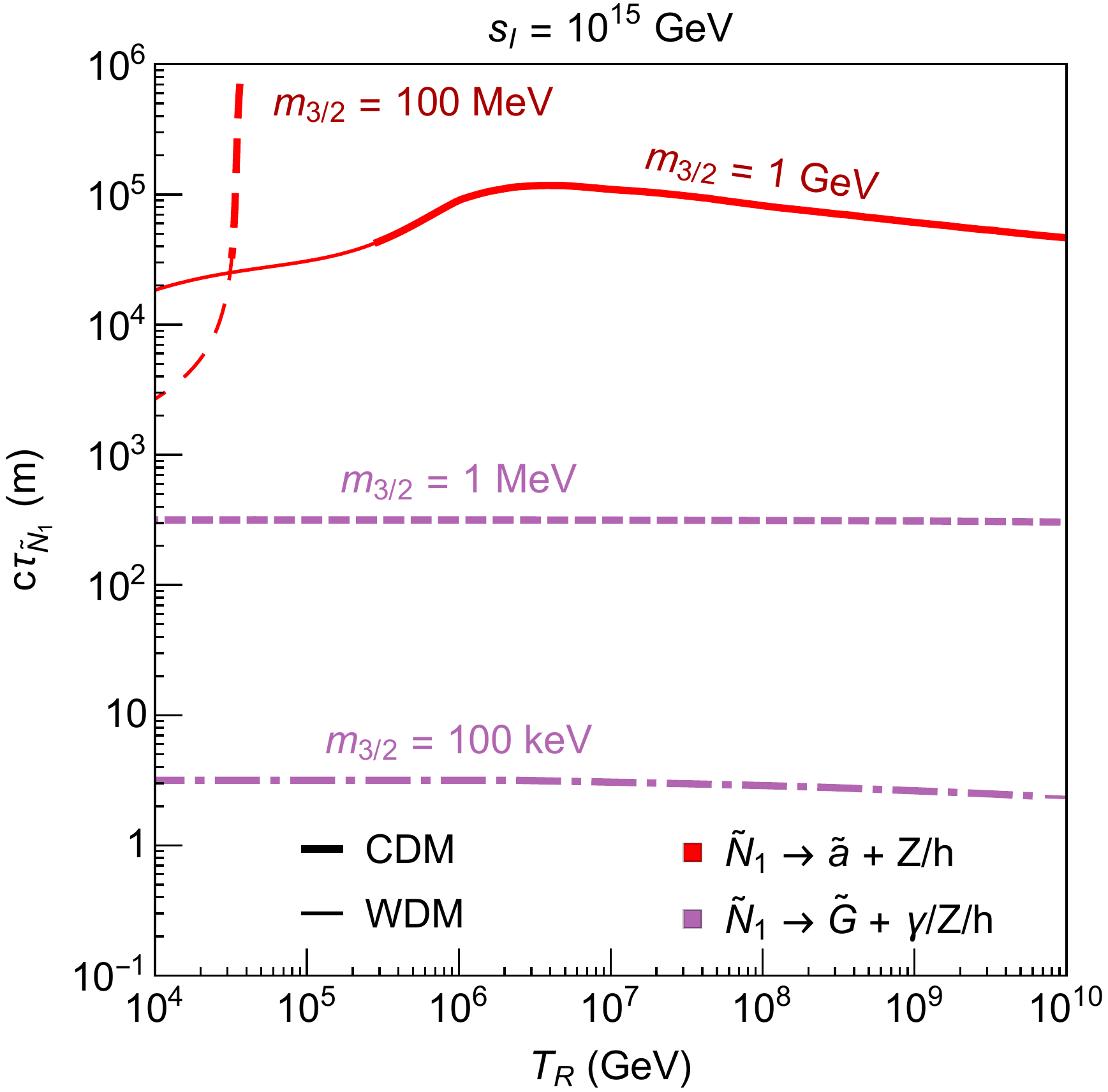} \includegraphics[width=0.495\linewidth]{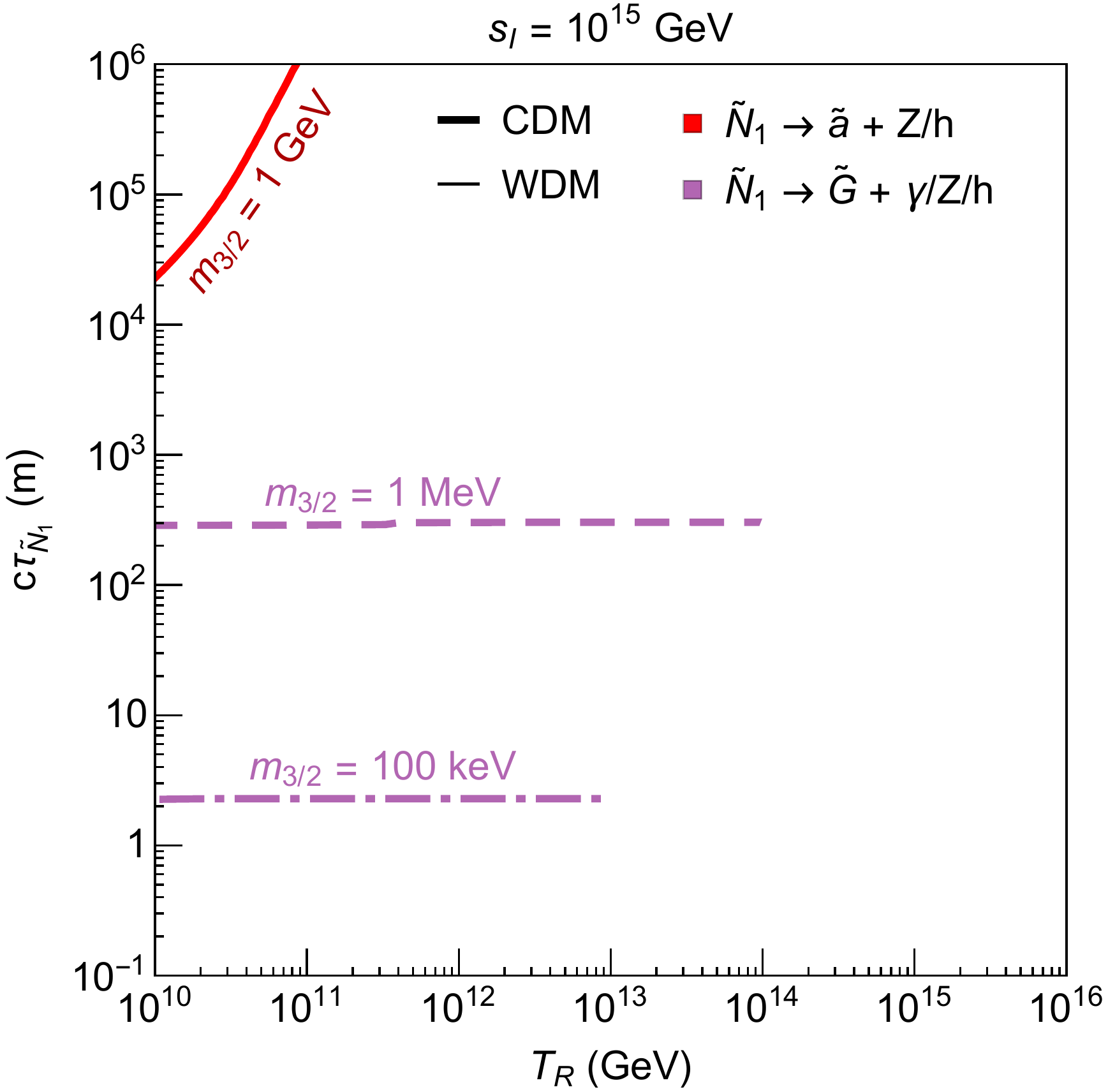}
\end{center}
\caption{The lifetime of the neutralino LOSP, decaying dominantly to $\tilde{a} + h/Z$ (red) or to $ \tilde{G}+\gamma/Z$(pink), predicted from the $\Omega h^2 = 0.11$ contours of \Fig{fig:LowDFSZ+}. We fix $q_\mu=2$, $\mathcal{D}=4$, $\tan\beta=2$, and $M_2/2 = M_1 = \mu = 1$ TeV and $m_s = 600$ GeV. The left (right) panel is for the cosmology with $T_R \lsim (\gsim) 10^{10}$ GeV discussed in \Sec{subsec:lowTR} (\Sec{subsec:highTR}).}
\label{fig:ctauLowDFSZPlus}
\end{figure}


The dark matter abundance is shown in \Fig{fig:LowDFSZ+} as contours of $\Omega h^2 = 0.11$. 
In the high $T_R$ regimes axino FI is always subdominant, so that in discussing the upper panels of \Fig{fig:LowDFSZ+}, we only need to identify axino UV and gravitino UV. Since the thick parts of the curves indicate gravitino UV domination, the features of these parts are identical to those in \Fig{fig:LowDFSZ0}. The new features are in the thin parts. In the upper panels, the vertical thin curves correspond to axino UV thermal abundance given by \Eq{eq:Yeq}. In the lower left panel, since the dilution factor becomes unity in regions where $T_R \lsim 10^{5-6}$ GeV and $V_{PQ} \lsim 10^{14-15}$ GeV, the gravitino UV abundance decreases with $T_R$ and the curve becomes thin and vertical, corresponding to axino FI. As axino FI and UV production in \Eqs{eq:nDecay}{eq:cDecay} and \Eq{eq:Yaxino} are independent of $m_{\tilde{a}}$, the result generically applies for any $m_{\tilde{a}}$ between $650 \GeV - 1 \TeV$ as required by the free-streaming length for cold dark matter and neutralino decay kinematics, respectively.

Finally, we also make predictions of neutralino LOSP lifetimes in \Fig{fig:ctauLowDFSZPlus}. Similar to \Fig{fig:ctauLowDFSZ0}, the relevant decay modes are ${\widetilde{N}}_1 \rightarrow \tilde{a}$ (red) and ${\widetilde{N}}_1 \rightarrow \tilde{G}$ (purple), with the latter dominating for low $m_{3/2}$. The prediction of the total decay rate, when dominated by $\Gamma_{{\widetilde{N}}_1 \rightarrow \tilde{G}}$, depends on $m_{3/2}$ and is insensitive to dark matter production and dilution. On the other hand, once $m_{3/2}$ is sufficiently large, ${\widetilde{N}}_1 \rightarrow \tilde{a}$ dominates and the lifetime prediction is set by the value of $V_{PQ}$ that gives $\Omega h^2 =0.11$ in \Fig{fig:LowDFSZ+}. The thin (thick) lines label warm (cold) dark matter. The left (right) panel corresponds to the low (high) $T_R$ cosmology studied in \Sec{subsec:lowTR} (\Sec{subsec:highTR}). In the right panel, the purple curves are truncated on the right at the values of $T_R$ where the curves of the upper right panel of \Fig{fig:LowDFSZ+} enter the light gray excluded region. It is worth noting that for clarity we only show the prediction for one value of $s_I$ and there is a large set of parameters that lead to collider signals with viable cosmology.

\section*{Acknowledgments}
We acknowledge useful conversations with Marcin Badziak, Kimberly Boddy, and Keisuke Harigaya.  This work was supported in part by the Director, Office of Science, Office of High Energy and Nuclear Physics, of the US Department of Energy under Contract DE-AC02-05CH11231 and by the National Science Foundation under grants PHY-1002399 and PHY-1316783. R.C. was supported in part by the National Science Foundation Graduate Research Fellowship under Grant No. DGE 1106400. F.D. is supported by the U.S. Department of Energy grant number DE-SC0010107.  The work of L. J. Hall was performed in part at the Institute for Theoretical Studies ETH Zurich and at the Aspen Center for Physics, which is supported by National Science Foundation grant PHY-1066293.

\appendix
\section{Axion Supermultiplet Interactions}
\label{app:Interactions}

In this Appendix we develop a general framework to describe effective interactions of the axion supermultiplet. We consider both self-interactions and couplings to MSSM fields. In the next Appendix, we use these results to compute the decay widths used in this work.

At energies below the PQ breaking scale the theory is still approximately supersymmetric. We assume the PQ symmetry to be broken by a set of chiral superfields $\Phi_i$, which we expand around their vacuum expectation values (vev)
\be
\Phi_i = v_i \exp\left[ q_i \frac{A}{V_{PQ}} \right] \ .
\label{eq:Phi}
\ee
The axion $a$ fills the supermultiplet $A$ as explicitly given in \Eq{eq:axionsupermultiplet}, together with its superpartners, the saxion $s$ and the axino $\tilde a$. We normalize the PQ charges $q_i$ of the PQ breaking fields in such a way that they are all integers and $|q_i|$ as small as possible. Within this convention, the effective PQ breaking scale is defined as follows
\be
V^2_{PQ} = \sum_i q_i^2 v_i^2 \ .
\label{eq:VPQdef}
\ee
The effect of a PQ rotation with an angle $\alpha$ on the axion superfield is the following
\be
A \; \rightarrow \; A + i  \alpha V_{\rm PQ} \ .
\label{eq:shiftofA}
\ee
The interactions of the axion supermultiplet are significantly constrained by this shift invariance.

\subsection{Color Anomaly}

The PQ symmetry is broken by a color anomaly, which generates the low-energy interaction
\be
{\cal L}_{A W W} = - \frac{g_3^2 N_{\rm DW}}{32\pi^2 v_{\rm PQ}} 
\int d^2 \theta A \, W^\alpha W_\alpha + {\rm h.c.} 
\label{eq:LagAWW}
\ee
between the axion superfield $A$ and the supersymmetric QCD field strength $W^\alpha$. The domain wall number $N_{\rm DW}$ appearing in the above expression is the QCD anomaly coefficient of the PQ symmetry. Finally, we define the axion decay constant
\be
f_a  \; =  \; \frac{\sqrt{2} }{ N_{\rm DW}} \; V_{PQ} \ .
\label{eq:VPQdef2}
\ee
Upon expanding in component fields the supersymmetric expression in \Eq{eq:LagAWW}, we identify the effective interaction between the axion $a$ and the QCD field strength $G$
\be
{\cal L}_{aG\tilde{G}} = \frac{g_3^2}{32\pi^2} \, \frac{a}{f_a} \, G^{\mu\nu} \tilde{G}_{\mu\nu}  \ .
\label{eq:f}
\ee

\subsection{Supersymmetric Interactions}

We assume the fields in \Eq{eq:Phi} to be canonically normalized, and the K\"ahler potential reads
\be
K = \sum_i \Phi_i^\dag \Phi_i =  \sum_i v_i^2 \exp\left[ q_i \left( \frac{A + A^\dag}{V_{PQ}} \right) \right] =  A^\dag A + \frac{1}{2} \sum_i \frac{q^3_i v_i^2}{V_{PQ}^3} A^\dag A  \, (A + A^\dag) + \ldots \ . \ .
\label{eq:Kaler}
\ee
This function only depends on the PQ invariant combination $A + A^\dag$, consistently with the invariance under the shift in \Eq{eq:shiftofA}. The axion superfield $A$ is canonically normalized. 

Holomorphy and PQ invariance forbid superpotential self-interactions for $A$. However, the axion superfield $A$ can appear in the superpotential of DFSZ theories, where the $\mu$ term is PQ charged. We are allowed to write the PQ invariant operator
\be
W = \mu \exp\left[ q_\mu \frac{A}{V_{PQ}} \right] H_u H_d =  \mu H_u H_d  + q_\mu \frac{\mu}{V_{PQ}} \, A \, H_u H_d  + \ldots \ .
\label{eq:Wdefinition}
\ee
Here, we denote $q_\mu$ as the model-dependent PQ charge of the $\mu$ term. 

\subsection{SUSY Breaking Interactions}

Finally, we account for SUSY breaking. We find it convenient to employ the non-linear field
\be
X = \left( \theta + \frac{\tilde \eta}{\sqrt{2} F} \right)^2 F \ ,
\label{eq:Xdef}
\ee
where $F$ is the SUSY breaking scale and $\tilde \eta$ is the associated goldstino. 

SUSY breaking is transmitted to the PQ sector through the higher dimensional operator
\be
K_{\rm \cancel{SUSY}} = c_{AAX} \, \frac{\left( A + A^\dag\right)^2 \left(X + X^\dag \right)}{M_{\rm mess}} \ ,
\label{eq:LagMplSupprAxino}
\ee
with $M_{\rm mess}$ the mass scale of the particles coupling the two sectors. We expect this operator to be generated by Planck scale dynamics~\cite{Cheung:2011mg}, and therefore we have the upper limit on the mediation scale $M_{\rm mess} \lesssim M_{\rm Pl}$. A consequence of this operator is a contribution to the axino mass 
\be
\int d^4 \theta K_{\rm \cancel{SUSY}} = - c_{AAX} \frac{F}{M_{\rm mess}} \tilde a \tilde a =
- \frac{1}{2} (2 \sqrt{3} \, c_{AAX} \, \frac{M_{\rm Pl}}{M_{\rm mess}} \, m_{3/2}) \tilde a \tilde a  + \ldots \ ,
\ee
where we have used the known relation $m_{3/2} = F / (\sqrt{3} M_{\rm Pl})$. As discussed in the main text of this work, it is not natural to have an axino much lighter than the gravitino, unless there is some good reason to suppress the size of the coefficient $c_{AAX}$ (e.g. sequestering).

Furthermore, we also require the presence of an effective $B \mu$ term necessary for a successful electroweak symmetry breaking
\be
W_{\rm \cancel{SUSY}} = - c_B \, X \, \exp\left[ q_\mu \frac{A}{V_{PQ}} \right] H_u H_d \ ,
\label{eq:WSUSYbreaking}
\ee
with $c_B F = B \mu$. 

\section{Decay Widths}
\label{app:Decay}

Saxion decays are responsible for reheating the universe and producing dark radiation. Decays and inverse decays of neutralinos and charginos generate an axino freeze-in abundance. The axino decay to the gravitino creates a warm DM population and vice versa. Finally, neutralino decays to the axino or gravitino lead to displaced collider events. We compute the decay widths for all these processes, using the interactions derived above.

\subsection{Saxion Decays}
\label{app:SaxDecays}

The saxion can decay to three possible final states: axions, axinos and Higgs bosons. The first two processes are mediated by the axion multiplet self-interactions, more specifically the cubic K\"ahler potential term in \Eq{eq:Kaler}. Once we expand in component fields, we find the operators
\be
\label{eq:Lsaa} 
\mathcal{L}_{s a a, s \tilde{a} \tilde{a}} \, = \, - \frac{\kappa}{\sqrt{2} V_{\rm PQ}} \, s \, \partial^\mu a \partial_\mu a +
\frac{\kappa}{\sqrt{2} V_{\rm PQ}} \, s \; m_{\tilde{a}} \left( \tilde{a} \tilde{a} + \tilde{a}^\dag \tilde{a}^\dag \right)  \ .
\ee
Here, we define the dimensionless parameter $\kappa = \sum_i q_i^3  \, v_i^2/V_{PQ}^2$. For models with only a single PQ breaking field, or theories with more than one but all with the same PQ charge, we have $\kappa = 1$. In more general cases $\kappa$ is a free parameter. If $\kappa$ is non-zero, axinos can be copiously produced from saxion decays. Neglecting the final state masses, the decay widths are
\begin{align}
\label{eq:stoaa}
\Gamma_{s \, \rightarrow \, a a } = & \, \frac{\kappa^2 \, m_s^3}{64 \pi V_{\rm PQ}^2} \ , \\
\label{eq:stoaxinos}
\Gamma_{s \, \rightarrow \, \tilde{a} \tilde{a} } = & \,  \frac{\kappa^2 \, m^2_{\tilde{a}} m_s}{8 \pi V_{\rm PQ}^2}
 \ .
\end{align}

More importantly for the reheating process, the saxion can decay to Higgs bosons and longitudinal electroweak gauge bosons. These decays are mediated from the supersymmetric interactions arising from the superpotential in \Eq{eq:Wdefinition} as well as the SUSY breaking interactions arising from \Eq{eq:WSUSYbreaking}. The resulting scalar potential interactions are
\be
V_{s H H} = \sqrt{2} \, q_\mu  \frac{\mu^2}{V_{\rm PQ}} \, s \left( H_u^\dag H_u + H_d^\dag H_d \right) + 
q_\mu \,  \frac{B}{V_{\rm PQ}} \, \frac{s}{\sqrt{2}} \left( H_u H_d + {\rm h.c.}  \right) \ .
\label{eq:VsHH}
\ee
The Higgs doublets $H_u$ and $H_d$ contain three Goldstones ($G^\pm$ and $G^0$) providing electroweak gauge bosons longitudinal modes, two CP-even ($h$ and $H$) and one CP-odd ($A$) neutral scalars and one charged scalar ($H^\pm$). We consider the decoupling limit where the non-SM fields are heavy. In such a limit, we find it convenient to introduce the doublets
\be
H_{\rm SM} =  \left( \begin{array}{c} G^+ \\ v + \frac{h + i G^0}{\sqrt{2}}  \end{array} \right) \ , \qquad \qquad \qquad
H_{\rm Heavy} = \left( \begin{array}{l} \frac{H + i A}{\sqrt{2}}   \\   \, H^- \end{array} \right) \ .
\ee
The SM-like Higgs boson $h$ lives within the multiplet $H_{\rm SM}$, which takes the electroweak symmetry breaking vev. The decoupling limit holds as long as $m_A \gg m_Z$, and in such a limit the connection between gauge eigenstates and mass eigenstates reads
\begin{align}
\label{eq:Hudecoupling} H_u =  & \, \sin\beta \, H_{\rm SM} \, + \cos\beta \, H^{(c)}_{\rm Heavy} \ , \\
\label{eq:Hddecoupling}  H_d = & \, \cos\beta \, H^{(c)}_{\rm SM} \, +  \sin\beta \, H_{\rm Heavy}\ ,
\end{align}
where we introduce the ratio between the two vevs $\tan\beta = v_u / v_d$ and with $ H^{(c)}_i = i \sigma^2 H_i^*$. We report here the decay width in such a decoupling limit and for large $\tan\beta$ (for more general expressions see App.~A of Ref.~\cite{Co:2016vsi}) 
\be
\Gamma_{s \, \rightarrow \, {\rm visible}} = \mathcal{D} \, \frac{q^2_\mu \mu^4}{16 \pi m_s V_{\rm PQ}^2} \ .
\label{eq:Gammastovisible}
\ee
Here, $\mathcal{D}$ counts the number of kinematically allowed final state particles. For a SM Higgs sector we have $\mathcal{D} = 4$, whereas $\mathcal{D} = 8$ for the full MSSM Higgs sector.

\subsection{Neutralino and Chargino Decays to Axinos}

We start by defining the mixing matrices for the MSSM neutralinos and charginos. We collect the neutralinos into the four-dimensional array 
\be
\widetilde{\chi} = \left( \begin{array}{cccc} \widetilde{B} & \widetilde{W}^3 & \widetilde{h}_d^0 & \widetilde{h}_u^0 \end{array}   \right) \ ,
\ee
and the connection with mass eigenstates $\widetilde{N_i}$ is achieved through the rotation
\be
\widetilde{\chi} = R \, \widetilde{N}   \ , \qquad \qquad \qquad \qquad (R^\dag R = R R^\dag = 1) \ .
\label{eq:NvsChi}
\ee
Likewise, we group the charginos into the two different two-dimensional arrays
\be
\widetilde{\psi}^+ = \left( \begin{array}{c} \widetilde{W}^+ \\ \widetilde{h}_u^+ \end{array} \right) \ , \qquad \qquad \qquad 
\widetilde{\psi}^- = \left( \begin{array}{c} \widetilde{W}^-  \\ \widetilde{h}_d^-  \end{array}   \right) \ ,
\ee
containing positively and negatively charged states, respectively. We separately rotate the two chargino arrays 
\begin{align}
\label{eq:C+def} \widetilde{C}^+ = & \, V^\dag \, \widetilde{\psi}^+ \ , \qquad \qquad \qquad (V^\dag V = V V^\dag = 1) \ , \\
\label{eq:C-def} \widetilde{C}^- = & \, U^\dag \, \widetilde{\psi}^- \ , \qquad \qquad \qquad (U^\dag U = U U^\dag = 1) \ ,
\end{align}
and as expected we find that the mass eigenvalues for the two arrays are the same. The positively and negatively charged mass eigenstates fill a Dirac fermion.

For neutralinos decays and inverse decays we have
\begin{align}
\label{eq:Gammaitoa} 
\Gamma_{\widetilde{N}_i \; \rightarrow \; \tilde{a} H} = & \,   
\left(\frac{q_\mu \mu}{V_{\rm PQ}} \right)^2 \frac{m_{\widetilde{N}_i}}{16 \pi} \left( \squared{s_\beta R_{3 i} } + \squared{c_\beta R_{4 i} } \right) \ , \\
\label{eq:Gammaatoi} 
\Gamma_{\tilde{a} \; \rightarrow \;  \widetilde{N}_i H} =  & \, 
\left(\frac{q_\mu \mu}{V_{\rm PQ}} \right)^2 \frac{m_{\tilde{a}}}{16 \pi} \left( \squared{s_\beta R_{3 i} } + \squared{c_\beta R_{4 i} } \right) \ ,
\end{align}
where $R$ is the rotation matrix defined in \Eq{eq:NvsChi}. Likewise, for charginos
\begin{align}
\label{eq:Gammaitoacharged} 
\Gamma_{\widetilde{C}^\pm_i  \; \rightarrow \; \tilde{a} \phi_+} = & \,
\left(\frac{q_\mu \mu}{V_{\rm PQ}} \right)^2 \frac{m_{\widetilde{C}^\pm_i}}{32 \pi} \left( \squared{s_\beta \, U_{2 i}} + \squared{c_\beta \, V_{2 i}} \right)  \ , \\
\label{eq:Gammaatoicharged} 
\Gamma_{\tilde{a} \; \rightarrow \; \widetilde{C}^\pm_i \phi_+} = & \, 
\left(\frac{q_\mu \mu}{V_{\rm PQ}} \right)^2 \frac{m_{\tilde{a}}}{32 \pi} \left( \squared{s_\beta \, U_{2 i}} + \squared{c_\beta \, V_{2 i}} \right)  \ ,
\end{align}
with $U$ and $V$ defined in \Eqs{eq:C+def}{eq:C-def}.

\subsection{Gravitino Decays}
\label{app:gravdecays}

Gravitino decays are mediated by the Planck suppressed supergravity operators~\cite{Cremmer:1982en}
\begin{align}
 \label{eq:gravitinodecayop} 
 \mathcal{L}_{\tilde G} = & \, - \frac{i}{2 M_{\rm Pl}} \mathcal{J}^\mu \tilde{G}_\mu  + {\rm h.c.} \ ,\\
\label{eq:Jsugra} 
\mathcal{J}^\mu = & \, 
 \sqrt{2} \, \sigma^\nu \sigmabar^\mu \chi \, \partial_\nu \phi^\dag 
- \sigma^{\alpha \beta} \sigma^\mu \lambda^{a \dag} F^a_{\alpha \beta}  \ .
\end{align}
The supercurrent $\mathcal{J}^\mu$ contains both chiral superfields with components $(\phi, \chi)$ as well as gauge superfields with components $(V, \lambda)$. Here, $F$ is the field strength of the vector field $V$. The gravitino decay width to chiral multiplet components reads
\be
\Gamma_{\tilde{G} \rightarrow \phi \chi} = N_c^{(\phi)} \frac{m_{3/2}^3}{384 \pi \, M_{\rm Pl}^2} \ ,
\label{eq:gravdecchiral}
\ee
where we neglect the final state masses and the color factor accounts for a decay to a quark/squark pair ($N_c^{(\text{quarks})} = 3$). Likewise, the decay width to gauge multiplet components results in
\be
\Gamma_{\tilde{G} \rightarrow V \lambda} = N_c^{(V)} \frac{m_{3/2}^3}{32 \pi \, M_{\rm Pl}^2} \ .
\ee
Here, the multiplicity factor is $(8,3,1)$ for the final state (gluino, wino, bino).

\subsection{Axino and Neutralino Decays to Gravitinos}
\label{app:togravdecays}

In the last part of this Appendix, we consider decays to final states involving gravitino. We assume the gravitino to be much lighter than the decaying particle, so that we can approximate the process with decays to longitudinal gravitinos, in accordance with the equivalence theorem. The process is correctly described by the effective Lagrangian~\footnote{An important exception is when the main source for the soft masses is Anomaly Mediation~\cite{D'Eramo:2012qd,D'Eramo:2013mya}. For the light gravitinos considered here, these corrections are very suppressed.}
\begin{align}
\mathcal{L}_{\tilde \eta} = & \, - \frac{1}{F} \tilde{\eta} \; \partial_\mu J^\mu + {\rm h.c.} \ , \label{eq:gravitinodecayop2}\\
J^\mu = & \, \sigma^\nu \sigmabar^\mu \chi \, \partial_\nu \phi^\dag 
- \frac{1}{\sqrt{2}} \sigma^{\alpha \beta} \sigma^\mu \lambda^{a \dag} F^a_{\alpha \beta} \ ,
\end{align}
where we consider again both chiral $(\phi, \chi)$ and vector $(V, \lambda)$ supermultiplets. We note that the above interactions can also be derived from the full supergravity result in \Eq{eq:gravitinodecayop}, by identifying the longitudinal component of the gravitino
\be
\tilde{G}_\mu \; \rightarrow \; i \, \sqrt{\frac{2}{3}} \frac{\partial_\mu \tilde \eta}{m_{3/2}} \ .
\ee
Upon using the Goldstino equations of motion~\cite{Cheung:2010mc,Cheung:2011jq} and the relation $m_{3/2} = F / (\sqrt{3} M_{\rm Pl})$, we recover the interaction between the Goldstino and the flat-space global SUSY supercurrent. 

The first case we discuss is the axino decay to axion and gravitino. The associated axino decay to saxion and gravitino is assumed to be kinematically forbidden. Upon using the general supercurrent result, and accounting for only the axion final state, we find 
\be
\Gamma_{\tilde a \; \rightarrow \; \tilde{G}  \, a} = \frac{m^5_{\tilde a}}{96 \pi \, m_{3/2}^2 M_{\rm Pl}^2} .
\label{eq:axinodecaygrav2}
\ee

The last cases we discuss are the ones relevant for displaced collider signatures. We only consider decays of the lightest neutralino $\widetilde{N}_1$, since all other R-odd particles produced at collider will promptly decay to it. The mass eigenstate $\widetilde{N}_1$ is related to the gauge eigenstates through the rotation in \Eq{eq:NvsChi}. For decays to photons we are only sensitive to Goldstino interactions with the neutral gauginos, and we find
\be
\label{eq:graviToGamma}
\Gamma_{\widetilde{N}_1 \; \rightarrow \; \tilde G\, \gamma} = 
\left| R_{11} c_w + R_{21} s_w \right|^2 
\frac{m_{\widetilde{N}_1}^5}{48 \pi \, m_{3/2}^2 M_{\rm Pl}^2} \ .
\ee
For decays to Z bosons, we have both longitudinal and transverse final states
\begin{align}
\label{eq:graviToZT}
\Gamma_{\widetilde{N}_1 \; \rightarrow \; \tilde G \, Z_T} = 
& \, \left| R_{11} s_w - R_{21} c_w \right|^2 \frac{m_{\widetilde{N}_1}^5}{48 \pi \, m_{3/2}^2 M_{\rm Pl}^2} \ , \\
\label{eq:graviToZL}
\Gamma_{\widetilde{N}_1 \; \rightarrow \; \tilde G \, Z_L} = 
& \, \left| R_{41} s_\beta - R_{31} c_\beta \right|^2  \frac{m_{\widetilde{N}_1}^5}{96 \pi \, m_{3/2}^2 M_{\rm Pl}^2} \ .
\end{align}
Finally, for decays to Higgs bosons we have
\be
\label{eq:graviToh}
\Gamma_{\widetilde{N}_1 \; \rightarrow \; \tilde G \, h} = 
\left| R_{41} s_\beta + R_{31} c_\beta \right|^2
\frac{m_{\widetilde{N}_1}^5}{96 \pi \, m_{3/2}^2 M_{\rm Pl}^2} \ ,
\ee
where we identify the SM-like Higgs boson $h$ in the decoupling limit as in Eqs.~\eqref{eq:Hudecoupling} and \eqref{eq:Hddecoupling}

\section{Free Streaming of Warm DM Component}
\label{app:FreeStreaming}

Our framework provides a warm dark matter source through production of NLSP particles and subsequent decay ${\rm NLSP}  \rightarrow  {\rm LSP} \, a$, where $a$ is a (nearly) massless axion. We always assume that if the LSP is the gravitino (axino), then the NLSP is the axino (gravitino). In this Appendix, we provide the calculation of the free streaming length for such a warm component  
\be
\lambda_{\rm FS} = \int_{\tau_{\rm NLSP}}^{t_{\rm eq}} \frac{v_{\rm LSP}(t)}{a(t)} \, dt =
\frac{2 \, t_{\rm eq}}{a_{\rm eq}^2} \int_{a_\tau}^{a_{\rm eq}} v(a) \, da \ .
\label{eq:lambdaFSdef}
\ee
Here, $\tau_{\rm NLSP}$ is the NLSP lifetime, whereas $t_{\rm eq}$ and $a_{\rm eq}$ are the time and scale factor at matter-radiation equality, respectively. In the second equality, we defined $a_\tau = a (\tau_{\rm NLSP})$ and we changed integration variable by using the relation for a radiation dominated universe $a \propto t^{1/2}$, justified if NLSP decays happen after BBN (i.e. $\tau_{\rm NLSP} \gtrsim 1 \, {\rm sec}$).

The LSP velocity $v(a)$ after decays is just a consequence of free streaming. The initial energy and momentum at the decay time $\tau_{\rm NLSP}$ follow from four-momentum conservation 
\be
\left\{ E^\tau_{\rm LSP}, p^\tau_{\rm LSP} \right\} =  \left\{ E_{\rm LSP}(\tau_{\rm NLSP}), p_{\rm LSP}(\tau_{\rm NLSP}) \right\} = \left\{ \frac{m_{\rm NLSP}^2 + m_{\rm LSP}^2}{2 m_{\rm NLSP}} , \frac{m_{\rm NLSP}^2 - m_{\rm LSP}^2}{2 m_{\rm NLSP}} \right\}  \ .
\ee
The LSP momentum red-shifts with the Hubble expansion
\be
v_{\rm LSP}(a) = \frac{p_{\rm LSP}(a)}{E_{\rm LSP}(a)} = \left[ 1 + \left(\frac{m_{\rm LSP}}{p_{\rm LSP}^\tau} \frac{a}{a_\tau} \right)^2 \right]^{ - \scalebox{1.01}{$\frac{1}{2}$} } \ ,
\label{eq:vLSPfree}
\ee
and the free streaming scale defined in \Eq{eq:lambdaFSdef} results in 
\begin{align}
\lambda_{\rm FS} = & \, 2 \, \frac{t_{\rm eq}}{a_{\rm eq}} \frac{p_{\rm LSP}(a_{\rm eq})}{m_{\rm LSP}} \; 
\left[ \mathcal{F}(a_{\rm eq}) - \mathcal{F}(a_\tau) \right] \ , \\ 
\mathcal{F}(a) = & \, \log\left[ \frac{m_{\rm LSP}}{p_{\rm LSP}(a)}  + \sqrt{1 + \left( \frac{m_{\rm LSP}}{p_{\rm LSP}(a)}\right)^2} \right] \ . 
\end{align}

It is convenient to derive an approximate expression for $\lambda_{\rm FS}$ by identifying the scale factor value $a_{\rm NR}$, correspondent to the time when the free streaming LSP enters the non-relativistic regime. The LSP velocity in \Eq{eq:vLSPfree} can be approximated as follows
\be
v(a) \simeq \left\{ \begin{array}{cccccl}
1 & & & & & a < a_{\rm NR} \\
a_{\rm NR} / a & & & & & a \geq a_{\rm NR} 
\end{array}
\right. \ .
\ee
We find $a_{\rm NR}$ by imposing $p_{\rm LSP}(a_{\rm NR}) \simeq m_{\rm LSP}$ and we find $a_{\rm NR} \simeq a_\tau p^\tau_{\rm LSP} / m_{\rm LSP}$. The free streaming length, as defined in \Eq{eq:lambdaFSdef}, approximately reads
\be
\lambda_{\rm FS} \simeq 
\frac{2 \, t_{\rm eq}}{a_{\rm eq}} \frac{a_\tau}{a_{\rm eq}} \frac{p^\tau_{\rm LSP}}{m_{\rm LSP}} 
\left[1 +  \log\left(\frac{a_{\rm eq}}{a_\tau} \frac{m_{\rm LSP}}{p_{\rm LSP}^\tau} \right) \right] \ .
\ee
We evaluate this expression by using the known values $t_{\rm eq} / a_{\rm eq} \simeq 93 \, {\rm Mpc}$,  $t_{\rm eq} \simeq 2.2 \times 10^{12} \, {\rm sec}$, and the time dependence of the scale factor $a(t) = a_{\rm eq} (t / t_{\rm eq})^{1/2}$. In the $m_{\rm NLSP} \gg m_{\rm LSP}$ limit, the free streaming length reads
\be
\lambda_{\rm FS} \simeq 0.6 \, {\rm Mpc} \, 
\left( \frac{m_{\rm NLSP}}{10 \, m_{\rm LSP}}\right) \left( \frac{\tau_{\rm NSLP}}{10^4 \, {\rm sec}} \right)^{1/2} 
\left[ 1 +  0.1 \log\left(  \frac{10 \, m_{\rm LSP}}{m_{\rm NLSP}} \left( \frac{10^4 \, {\rm sec}}{\tau_{\rm NSLP}} \right)^{1/2} \right)\right] \ .
\label{eq:lambdaFSpprox}
\ee


\begin{thebibliography}{0}  

\bibitem{Peccei:1977hh} 
  R.~D.~Peccei and H.~R.~Quinn,
  Phys.\ Rev.\ Lett.\  {\bf 38}, 1440 (1977).
  
\bibitem{Peccei:1977ur} 
  R.~D.~Peccei and H.~R.~Quinn,
  Phys.\ Rev.\ D {\bf 16}, 1791 (1977).
  
\bibitem{Weinberg:1977ma} 
  S.~Weinberg,
  Phys.\ Rev.\ Lett.\  {\bf 40}, 223 (1978).
  
\bibitem{Wilczek:1977pj} 
  F.~Wilczek,
  Phys.\ Rev.\ Lett.\  {\bf 40}, 279 (1978).
  
\bibitem{Dine:1981rt} 
  M.~Dine, W.~Fischler and M.~Srednicki,
  Phys.\ Lett.\ B {\bf 104}, 199 (1981).

\bibitem{Zhitnitsky:1980tq} 
  A.~R.~Zhitnitsky,
  Sov.\ J.\ Nucl.\ Phys.\  {\bf 31}, 260 (1980)
  [Yad.\ Fiz.\  {\bf 31}, 497 (1980)].
  
\bibitem{Chun:2011zd} 
  E.~J.~Chun,
  Phys.\ Rev.\ D {\bf 84}, 043509 (2011)
  [arXiv:1104.2219 [hep-ph]].
  
\bibitem{Co:2015pka} 
  R.~T.~Co, F.~D'Eramo, L.~J.~Hall and D.~Pappadopulo,
  JCAP {\bf 1512}, no. 12, 024 (2015)
  [arXiv:1506.07532 [hep-ph]].
  
\bibitem{Hall:2009bx} 
  L.~J.~Hall, K.~Jedamzik, J.~March-Russell and S.~M.~West,
  JHEP {\bf 1003}, 080 (2010)
  [arXiv:0911.1120 [hep-ph]].
  
\bibitem{Covi:2001nw} 
  L.~Covi, H.~B.~Kim, J.~E.~Kim and L.~Roszkowski,
  JHEP {\bf 0105}, 033 (2001)
  [hep-ph/0101009].
  
\bibitem{Strumia:2010aa} 
  A.~Strumia,
  JHEP {\bf 1006}, 036 (2010)
  [arXiv:1003.5847 [hep-ph]].

\bibitem{Nanopoulos:1983up} 
  S.~Weinberg, unpublished.
  D.~V.~Nanopoulos, K.~A.~Olive and M.~Srednicki,
  Phys.\ Lett.\  {\bf 127B}, 30 (1983).
   M.~Y.~Khlopov, A.~D.~Linde,
  Phys.\ Lett.\  {\bf B138}, 265-268 (1984).  
  
\bibitem{Bae:2011jb} 
  K.~J.~Bae, K.~Choi and S.~H.~Im,
  JHEP {\bf 1108}, 065 (2011)
  [arXiv:1106.2452 [hep-ph]].
  
\bibitem{Cheung:2011mg} 
  C.~Cheung, G.~Elor and L.~J.~Hall,
  Phys.\ Rev.\ D {\bf 85}, 015008 (2012)
  [arXiv:1104.0692 [hep-ph]].
  
\bibitem{Randall:1994fr} 
  L.~Randall and S.~D.~Thomas,
  Nucl.\ Phys.\ B {\bf 449}, 229 (1995)
  [hep-ph/9407248]. 
\bibitem{Hashimoto:1998ua} 
  M.~Hashimoto, K.~I.~Izawa, M.~Yamaguchi and T.~Yanagida,
  Phys.\ Lett.\ B {\bf 437}, 44 (1998)
  [hep-ph/9803263].
  
\bibitem{Kim:1979if} 
  J.~E.~Kim,
  Phys.\ Rev.\ Lett.\  {\bf 43}, 103 (1979).
  
\bibitem{Shifman:1979if} 
  M.~A.~Shifman, A.~I.~Vainshtein and V.~I.~Zakharov,
  Nucl.\ Phys.\ B {\bf 166}, 493 (1980).
  
\bibitem{Graf:2013xpe} 
  P.~Graf and F.~D.~Steffen,
  JCAP {\bf 1312}, 047 (2013)
  [arXiv:1302.2143 [hep-ph]].
  
\bibitem{Chung:1998rq} 
  D.~J.~H.~Chung, E.~W.~Kolb and A.~Riotto,
  Phys.\ Rev.\ D {\bf 60}, 063504 (1999)
  [hep-ph/9809453].
 
\bibitem{Giudice:2000ex} 
  G.~F.~Giudice, E.~W.~Kolb and A.~Riotto,
  Phys.\ Rev.\ D {\bf 64}, 023508 (2001)
  [hep-ph/0005123].
  
\bibitem{Kawasaki:1995vt} 
  M.~Kawasaki, T.~Moroi and T.~Yanagida,
  Phys.\ Lett.\ B {\bf 383}, 313 (1996)
  [hep-ph/9510461].
  
\bibitem{Co:2016vsi} 
  R.~T.~Co, F.~D'Eramo and L.~J.~Hall,
  Phys.\ Rev.\ D {\bf 94}, no. 7, 075001 (2016)
  [arXiv:1603.04439 [hep-ph]].
 
\bibitem{Dimopoulos:1996vz} 
  S.~Dimopoulos, M.~Dine, S.~Raby and S.~D.~Thomas,
  Phys.\ Rev.\ Lett.\  {\bf 76}, 3494 (1996)
  [hep-ph/9601367].
  
\bibitem{Martin:2000eq} 
  S.~P.~Martin,
  Phys.\ Rev.\ D {\bf 62}, 095008 (2000)
  [hep-ph/0005116].
    
\bibitem{Weinberg:1982zq} 
  S.~Weinberg,
  Phys.\ Rev.\ Lett.\  {\bf 48}, 1303 (1982).
  
\bibitem{Kawasaki:2008jc} 
  M.~Kawasaki and K.~Nakayama,
  Phys.\ Rev.\ D {\bf 77}, 123524 (2008)
  [arXiv:0802.2487 [hep-ph]].
  
\bibitem{Kawasaki:2011ym} 
  M.~Kawasaki, N.~Kitajima and K.~Nakayama,
  Phys.\ Rev.\ D {\bf 83}, 123521 (2011)
  [arXiv:1104.1262 [hep-ph]].
  
\bibitem{Asaka:2000ew} 
  T.~Asaka and T.~Yanagida,
  Phys.\ Lett.\ B {\bf 494}, 297 (2000)
  [hep-ph/0006211].
  
\bibitem{Baer:2010gr} 
  H.~Baer, S.~Kraml, A.~Lessa and S.~Sekmen,
  JCAP {\bf 1104}, 039 (2011)
  [arXiv:1012.3769 [hep-ph]].
  
\bibitem{Baer:2011eca} 
  H.~Baer and A.~Lessa,
  JHEP {\bf 1106}, 027 (2011)
  [arXiv:1104.4807 [hep-ph]].
  
\bibitem{Bae:2014rfa} 
  K.~J.~Bae, H.~Baer, A.~Lessa and H.~Serce,
  JCAP {\bf 1410}, no. 10, 082 (2014)
  [arXiv:1406.4138 [hep-ph]].
  
  \bibitem{Hasenkamp:2010if} J.~Hasenkamp and J.~Kersten, 
  Phys.\ Rev.\ D {\bf 82}, 115029 (2010)
  [arXiv:1008.1740 [hep-ph]].
  
\bibitem{Moroi:1993mb} 
  T.~Moroi, H.~Murayama and M.~Yamaguchi,
  Phys.\ Lett.\ B {\bf 303}, 289 (1993).
  V.~S.~Rychkov and A.~Strumia,
  Phys.\ Rev.\ D {\bf 75}, 075011 (2007)
  [hep-ph/0701104].
   
\bibitem{Cheung:2011nn} 
  C.~Cheung, G.~Elor and L.~Hall,
  Phys.\ Rev.\ D {\bf 84}, 115021 (2011)
  [arXiv:1103.4394 [hep-ph]].
  
\bibitem{Jedamzik:2004er} 
  K.~Jedamzik,
  Phys.\ Rev.\ D {\bf 70}, 063524 (2004)
  [astro-ph/0402344].
  
\bibitem{Kawasaki:2004yh} 
  M.~Kawasaki, K.~Kohri and T.~Moroi,
  Phys.\ Lett.\ B {\bf 625}, 7 (2005)
  [astro-ph/0402490].
 
\bibitem{Ellis:2005ii} 
  J.~R.~Ellis, K.~A.~Olive and E.~Vangioni,
  Phys.\ Lett.\ B {\bf 619}, 30 (2005)
  [astro-ph/0503023].
  
\bibitem{Boyarsky:2008xj} 
  A.~Boyarsky, J.~Lesgourgues, O.~Ruchayskiy and M.~Viel,
  JCAP {\bf 0905}, 012 (2009)
  [arXiv:0812.0010 [astro-ph]].
  
\bibitem{Harada:2014lma} 
  A.~Harada and A.~Kamada,
  JCAP {\bf 1601}, no. 01, 031 (2016)
  [arXiv:1412.1592 [astro-ph.CO]].
 
\bibitem{Kamada:2016vsc} 
  A.~Kamada, K.~T.~Inoue and T.~Takahashi,
  Phys.\ Rev.\ D {\bf 94}, no. 2, 023522 (2016)
  [arXiv:1604.01489 [astro-ph.CO]].
  
\bibitem{Weinberg:2013aya} 
  D.~H.~Weinberg, J.~S.~Bullock, F.~Governato, R.~Kuzio de Naray and A.~H.~G.~Peter,
  Proc.\ Nat.\ Acad.\ Sci.\  {\bf 112}, 12249 (2014)
  [arXiv:1306.0913 [astro-ph.CO]].
  
\bibitem{El-Badry15}
K.~El-Badry, A.~Wetzel, M.~Geha , P.~F.~Hopkins, D.~Kere\v{s}, T.~K.~Chan, C.~A.~Faucher-Gigu\`ere, 
Astrophys.\ J.\  {\bf 820}, 131 (2015),
[arXiv: 1512.01235 [astro-ph.GA]] 

\bibitem{Wetzel:2016wro} 
  A.~R.~Wetzel, P.~F.~Hopkins, J.~h.~Kim, C.~A.~Faucher-Giguere, D.~Keres and E.~Quataert,
  Astrophys.\ J.\  {\bf 827}, no. 2, L23 (2016)
  [arXiv:1602.05957 [astro-ph.GA]].
  
\bibitem{Oman:2015xda} 
  K.~A.~Oman {\it et al.},
  Mon.\ Not.\ Roy.\ Astron.\ Soc.\  {\bf 452}, no. 4, 3650 (2015)
  [arXiv:1504.01437 [astro-ph.GA]].
  
\bibitem{Papastergis:2015} 
  E.~Papastergis and F.~Shankar,
  A$\&$A 591, A58 (2016)
  [arXiv:1511.08741 [astro-ph.GA]].

\bibitem{Kaplinghat:2005sy} 
  M.~Kaplinghat,
  Phys.\ Rev.\ D {\bf 72}, 063510 (2005)
  [astro-ph/0507300].
  
\bibitem{Cembranos:2005us} 
  J.~A.~R.~Cembranos, J.~L.~Feng, A.~Rajaraman and F.~Takayama,
  Phys.\ Rev.\ Lett.\  {\bf 95}, 181301 (2005)
  [hep-ph/0507150].
  
\bibitem{Borschensky:2014cia} 
  C.~Borschensky, M.~Kr�mer, A.~Kulesza, M.~Mangano, S.~Padhi, T.~Plehn and X.~Portell,
  Eur.\ Phys.\ J.\ C {\bf 74}, no. 12, 3174 (2014)
  [arXiv:1407.5066 [hep-ph]].
    
\bibitem{Chou:2016lxi} 
  J.~P.~Chou, D.~Curtin and H.~J.~Lubatti,
  arXiv:1606.06298 [hep-ph].
  
\bibitem{Tang:2015qga} 
  J.~Tang {\it et al.},
  arXiv:1507.03224 [physics.acc-ph].
  A. Ball, M. Benedikt, L. Bottura, O. Dominguez, F. Gianotti, B. Goddard, P. Lebrun, M. Mangano, D. Schulte, E. Shaposhnikova, R. Tomas, and F. Zimmermann (FCC-hh), Future Circular Collider Study Hadron Collider Parameters, Tech. Rep.
(CERN, Geneva, 2014).
  
\bibitem{Ade:2015xua} 
  P.~A.~R.~Ade {\it et al.} [Planck Collaboration],
  arXiv:1502.01589 [astro-ph.CO].

\bibitem{Abazajian:2016yjj} 
  K.~N.~Abazajian {\it et al.} [CMB-S4 Collaboration],
  arXiv:1610.02743 [astro-ph.CO].
  
\bibitem{Cremmer:1982en} 
  E.~Cremmer, S.~Ferrara, L.~Girardello and A.~Van Proeyen,
  Nucl.\ Phys.\ B {\bf 212}, 413 (1983).
  
\bibitem{D'Eramo:2012qd} 
  F.~D'Eramo, J.~Thaler and Z.~Thomas,
  JHEP {\bf 1206}, 151 (2012)
  [arXiv:1202.1280 [hep-ph]].
  
\bibitem{D'Eramo:2013mya} 
  F.~D'Eramo, J.~Thaler and Z.~Thomas,
  JHEP {\bf 1309}, 125 (2013)
  [arXiv:1307.3251 [hep-ph]].

\bibitem{Cheung:2010mc} 
  C.~Cheung, Y.~Nomura and J.~Thaler,
  JHEP {\bf 1003}, 073 (2010)
  [arXiv:1002.1967 [hep-ph]].

\bibitem{Cheung:2011jq} 
  C.~Cheung, F.~D'Eramo and J.~Thaler,
  JHEP {\bf 1108}, 115 (2011)
  [arXiv:1104.2600 [hep-ph]].
  
\end{thebibliography}
\end{document}